\documentclass[aps,prd,preprintnumbers,nofootinbib,superscriptaddress,preprint]{revtex4-1}
\usepackage{graphicx}
\usepackage{amssymb}
\usepackage{amsmath}
\usepackage{color}
\usepackage{booktabs}
\usepackage{dcolumn}

\graphicspath{{fig/}}

\newcommand{\beq}{\begin{eqnarray}}
\newcommand{\eeq}{\end{eqnarray}}

\def\fsl#1{\setbox0=\hbox{$#1$}
   \dimen0=\wd0
   \setbox1=\hbox{/} \dimen1=\wd1
   \ifdim\dimen0>\dimen1
      \rlap{\hbox to \dimen0{\hfil/\hfil}}
      #1
   \else
      \rlap{\hbox to \dimen1{\hfil$#1$\hfil}}
      /
   \fi}

\begin{document}

\preprint{UTHEP-770, UTCCS-P-144, HUPD-2207, KUNS-2930}

\title{$K_{\ell 3}$ form factors at the physical point: Toward the continuum limit}

\date{ 
October  6 2022 
}



\author{Ken-ichi~Ishikawa}
\affiliation{Graduate School of Advanced Science and Engineering, Hiroshima University, Higashi-Hiroshima, Hiroshima 739-8526, Japan}
\affiliation{Core of Research for the Energetic Universe, Graduate School of Advanced Science and Engineering, Hiroshima University, Higashi-Hiroshima, Hiroshima 739-8526, Japan}

\author{Naruhito~Ishizuka}
\affiliation{Center for Computational Sciences, University of Tsukuba, \\ Tsukuba, Ibaraki 305-8577, Japan}

\author{Yoshinobu~Kuramashi}
\affiliation{Center for Computational Sciences, University of Tsukuba, \\ Tsukuba, Ibaraki 305-8577, Japan}

\author{Yusuke~Namekawa}
\affiliation{Department of Physics, Kyoto University, Kitashirakawa-Oiwake-cho, Kyoto 606-8502, Japan}

\author{Yusuke~Taniguchi}
\affiliation{Center for Computational Sciences, University of Tsukuba, \\ Tsukuba, Ibaraki 305-8577, Japan}

\author{Naoya~Ukita}
\affiliation{Center for Computational Sciences, University of Tsukuba, \\ Tsukuba, Ibaraki 305-8577, Japan}

\author{Takeshi~Yamazaki}
\affiliation{Faculty of Pure and Applied Sciences, University of Tsukuba, \\ Tsukuba, Ibaraki 305-8571, Japan}
\affiliation{Center for Computational Sciences, University of Tsukuba, \\ Tsukuba, Ibaraki 305-8577, Japan}

\author{Tomoteru~Yoshi\'e}
\affiliation{Center for Computational Sciences, University of Tsukuba, \\ Tsukuba, Ibaraki 305-8577, Japan}

\collaboration{PACS Collaboration}

\begin{abstract}

We present updated results for the form factors of 
the kaon semileptonic $(K_{\ell 3})$ 
decay process calculated with $N_f = 2 + 1$ nonperturbatively $O(a)$-improved 
Wilson quark action and Iwasaki gauge action at the physical point
on large volumes of more than (10 fm)$^4$.
In addition to our previous calculation at the 
lattice spacing $a = 0.085$ fm,
we perform a calculation at the second lattice spacing of $0.063$ fm.
Using the results for the form factors extracted from
3-point functions with the local and also conserved vector currents 
at the two lattice spacings,
continuum extrapolation and interpolation of the momentum transfer 
are carried out simultaneously to obtain the value of the form factor $f_+(0)$
at the zero momentum transfer in the continuum limit.
After investigation of stability of $f_+(0)$ against several fit forms
and different data, we obtain $f_+(0) = 0.9615(10)(^{+47}_{\ -3})(5)$,
where the first, second, and third errors are
statistical, systematic errors from choice of the fit forms
and isospin breaking effect, respectively.
Furthermore, we obtain the slope and curvature of the form factors,
and the phase space integral from the momentum transfer dependence of 
the form factors.
Combining our value of $f_+(0)$ and experimental input of 
the $K_{\ell 3}$ decay, one of the Cabibbo-Kobayashi-Maskawa matrix 
elements $|V_{us}|$ is determined as $|V_{us}| = 0.2252(^{\ +5}_{-12})$,
whose error contains the experimental one as well as that 
in the lattice calculation.
This value is reasonably consistent with the ones determined 
from recent lattice QCD results of $f_+(0)$ and also the one determined 
through the kaon leptonic decay process.
We observe some tension between our value and 
$|V_{us}|$ evaluated from the unitarity of the CKM
matrix with $|V_{ud}|$, while it depends on the size of the error of $|V_{ud}|$.
It is also found that $|V_{us}|$ determined with our phase space integrals 
through six $K_{\ell 3}$ decay processes
is consistent with the above one using $f_+(0)$.

\end{abstract}

\maketitle

\section{Introduction}

Unitarity of the Cabibbo-Kobayashi-Maskawa (CKM) matrix is important
in search for signals beyond the standard model (BSM).
Since it should be satisfied in the standard model,
its violation indirectly suggests existence of a BSM physics.
The unitarity of the first row of the CKM matrix gives a condition
for the three matrix elements, $|V_{ud}|$, $|V_{us}|$, and $|V_{ub}|$,
as $|V_{ud}|^2 + |V_{us}|^2 + |V_{ub}|^2 = 1$.
In the current values of the matrix elements reviewed in
PDG20~\cite{ParticleDataGroup:2020ssz},
$|V_{ud}| = 0.97370(14)$, $|V_{us}| = 0.2245(8)$, and $|V_{ub}| = 0.00382(24)$,
the tension of 3.3 $\sigma$ from unity is observed as
\begin{equation}
1-(|V_{ud}|^2 + |V_{us}|^2 + |V_{ub}|^2) = 0.00149(45) .
\end{equation}
It could be a BSM signal, though the value of $|V_{us}|$ is
obtained from an average of slightly different values,
$|V_{us}| = 0.2252(5)$ and 0.2231(7) determined from 
the kaon leptonic ($K_{\ell 2}$) and semileptonic ($K_{\ell 3}$)
decay processes, respectively.
The error reductions, especially in the $K_{\ell 3}$ decay, are desirable 
to clarify the tension.
In both the determinations, lattice QCD calculation plays an essential role.

For the $K_{\ell 2}$ decay, $|V_{us}|$ is determined by
combining the experimental values
$|V_{us}|F_K/ \linebreak |V_{ud}|F_\pi = 0.27599(37)$~\cite{Moulson:2017ive}
and $|V_{ud}|$, and the lattice QCD result of 
the ratio of the decay constants for the kaon and pion, $F_K/F_\pi$.
In lattice QCD, its calculation is relatively simpler, and 
it is well determined in various calculations 
as reviewed in Ref.~\cite{Aoki:2021kgd}. 

In the $K_{\ell 3}$ decay,
$|V_{us}|$ is related to the kaon decay rate 
$\Gamma_{K_{\ell 3}}$ as
\begin{equation}
\Gamma_{K_{\ell 3}} = C_{K_{\ell 3}} ( |V_{us}| f_+(0) )^2 I^\ell_K,
\label{eq:decay_width_kl3}
\end{equation}
where $f_+(0)$ is the value of the $K_{\ell 3}$ form factor at $q^2 = 0$
with $q$ being the momentum transfer,
$C_{K_{\ell 3}}$ is a known factor including the electromagnetic 
correction and the SU(2) breaking effect.
$I^\ell_K$ is the phase space integral for $\ell = e, \mu$, 
which is calculated from the shape of
the experimental $K_{\ell 3}$ form factors.
The value of $f_+(0)$ is needed to be computed in a nonperturbative QCD
calculation, such as lattice QCD. In lattice QCD,
the calculation of $f_+(0)$ is not as simple as the one of $F_K/F_\pi$,
although several lattice QCD calculations~\cite{Bazavov:2012cd,Boyle:2007qe,Boyle:2013gsa,Boyle:2015hfa,Aoki:2017spo,Bazavov:2013maa,Carrasco:2016kpy,Bazavov:2018kjg} provided precise results of $f_+(0)$.
The recent results of $f_+(0)$ tend to yield a smaller $|V_{us}|$ than
the ones determined from the $K_{\ell 2}$ decay and also the unitarity
of the CKM matrix.

For the BSM search, it is important to confirm the discrepancies of $|V_{us}|$
from several calculations by different groups using independent setups
with a similar size of error to the most precise 
result~\cite{Bazavov:2018kjg}.
For this purpose we calculated the $K_{\ell 3}$ form factors
using a part of the PACS10 configurations~\cite{PACS:2019hxd}.
The PACS10 configurations were generated on 
the physical volume of more than (10 fm)$^4$ at the physical point, where
the pion mass $m_\pi$ and kaon mass $m_K$ are the physical ones.
This ensemble is suitable for a precise calculation of $f_+(0)$,
because systematic errors for the chiral extrapolation and 
the finite volume effect are considered to be negligible.
Furthermore, the data of the $K_{\ell 3}$ form factors near $q^2 = 0$
can be calculated even in the periodic boundary condition 
in the spatial directions thanks to the huge volume,
so that a stable $q^2$ interpolation of the $K_{\ell 3}$ form factors is
performed.
In our previous calculation
the obtained value of $|V_{us}|$ with $f_+(0)$ was consistent with
the one from the $K_{\ell 2}$ decay, while our error was larger than
the one in Ref.~\cite{Bazavov:2018kjg}.
Since our calculation was carried out at one lattice spacing, $a = 0.085$ fm,
the largest uncertainty of $f_+(0)$ comes from a systematic error of 
a finite lattice spacing effect.
Thus, the purpose of this study is to reduce the large systematic uncertainty
by adding data calculated with another set of the PACS10 configurations
at the finer lattice spacing, $a = 0.063$ fm.
We also aim to perform continuum extrapolation and $q^2$ interpolation
simultaneously to estimate the value of $f_+(0)$ in the continuum limit 
using the form factors at the two lattice spacings.

In this calculation, not only local but also conserved vector currents 
are employed in the calculation of 3-point functions,
from which the matrix elements of the $K_{\ell 3}$ form factors are extracted.
We newly calculate the conserved vector current data at $a = 0.085$ fm,
because in our previous calculation only the local vector current is employed
at this lattice spacing.
The results of $f_+(0)$ obtained from the two currents
have a different dependence on the lattice spacing.
It enables us to carry out a continuum extrapolation
of $f_+(0)$ using only the two lattice spacings.
The extrapolated result of $f_+(0)$ in the continuum limit agrees with
the one in our previous work, and also it has much smaller 
statistical and lower systematic errors than the previous one.
The upper systematic error, however, is still a similar size to
that in the previous work, which is mainly caused by a fit form
dependence in the continuum extrapolation.
This result reasonably agrees with the previous lattice QCD results
within 1.6 $\sigma$ in the total error, where the statistical
and systematic errors are added in quadrature.
As in the previous work, $|V_{us}|$ determined from our $f_+(0)$ 
is well consistent with the one from the $K_{\ell 2}$ decay.
A tension is seen in a comparison of our value of $|V_{us}|$ 
with the one evaluated from the unitarity of the CKM matrix, 
while its significance depends on the size of the error of $|V_{ud}|$.
The slope and curvature of the form factors are also evaluated from
the continuum limit results for the form factors.
Although the systematic errors coming from a fit function
dependence are large in those results, except for the slope of $f_+(q^2)$,
the results are reasonably consistent with the experimental ones and
also the previous lattice QCD calculations.
Furthermore, it is found that the phase space integrals calculated from
our result of the $q^2$ dependent form factors
agree with the experimental values.
Moreover, the values of $|V_{us}|$ are determined from 
our phase space integrals through six kaon decay processes and
their average. Those results also agree with that using $f_+(0)$.
A part of the preliminary results in this work was already reported 
in Ref.~\cite{Yamazaki:2021zxz}.

This paper is organized as follows.
Section~\ref{sec:methods} describes the definition of the $K_{\ell 3}$ 
form factors, simulation parameters, and our calculation methods
to extract the matrix elements.
The results of the form factors using the local and conserved currents
are presented in Sec.~\ref{sec:result_finite_a}
including their $q^2$ interpolations to $q^2 = 0$ at each lattice spacing.
Continuum extrapolations of the form factors are discussed 
in Sec.~\ref{sec:continuum_limit},
where the results for $f_+(0)$, their slope and curvature,
the phase space integrals, and determination of $|V_{us}|$ 
are also presented.
Section~\ref{sec:conclusions} is devoted to conclusion.
The appendix summarizes various fit results of continuum extrapolations.

All dimensionful quantities are expressed in units of 
the lattice spacing throughout this paper, 
unless otherwise explicitly specified.

\section{Calculation methods}
\label{sec:methods}

We calculate the two form factors of the $K_{\ell 3}$ decay
$f_+(q^2)$ and $f_0(q^2)$ defined below.
In this section we first give the definitions of the form factors 
from the matrix element.
After that, the simulation parameters, calculation method for
the correlators, and analysis method to extract the matrix elements
from the correlators are described.

\subsection{Definition of $K_{\ell 3}$ form factors}

The $K_{\ell 3}$ form factors $f_+(q^2)$ and $f_-(q^2)$ are defined by the matrix 
element of the weak vector current $V_{\mu}$ as,
\begin{eqnarray}
\langle \pi (\vec{p}_{\pi}) \left | V_{\mu} \right | K(\vec{p}_{K}) \rangle = ({p}_{K}+{p}_{\pi})_{\mu}f_{+}(q^2)+ ({p}_{K}-{p}_{\pi})_{\mu}f_{-}(q^2),
\label{eq:def_matrix_element}
\end{eqnarray}
where $q=p_{K}-p_{\pi}$ is the four-dimensional momentum transfer.
The scalar form factor $f_0(q^2)$ is defined by $f_+(q^2)$ and $f_-(q^2)$
as,
\begin{eqnarray}
f_{0}(q^2) =f_{+}(q^2) + \frac{-q^2}{{m^2_{K}}-{m^2_{\pi}}}f_{-}(q^2)
= f_{+}(q^2)\left(1+ \frac{-q^2}{{m^2_{K}}-{m^2_{\pi}}}\xi(q^2)\right), 
\label{eq:f0}
\end{eqnarray}
where $\xi(q^2)=f_{-}(q^2)/f_{+}(q^2)$.
At $q^2=0$, the two form factors $f_+(q^2)$ and $f_0(q^2)$ satisfy
the condition $f_+(0)=f_0(0)$.

\subsection{Simulation parameters and setup}

We employ a subset of the PACS10 configurations at
the two bare couplings $\beta = 2.00$ and 1.82, which 
are generated at the physical point on more than $(10\ {\rm fm})^4$ volume.
The lattice size and lattice cutoff determined from the $\Xi$ baryon mass input
are tabulated in Table~\ref{tab:mpi_mk_conf}.
The configuration generations were performed using 
the nonperturbatively improved Wilson quark action
with the six-stout link smearing~\cite{Morningstar:2003gk} 
and the Iwasaki gauge action~\cite{Iwasaki:2011jk}.
The simulation parameters of the configuration generation
at $\beta = 2.00$ and $\beta = 1.82$ are summarized in 
Refs.~\cite{Shintani:2019wai} and ~\cite{Ishikawa:2018jee}, respectively.
The number of the configuration used in this calculation is 20
at both the lattice spacings.
The separations between each configuration are 5 and 10 molecular dynamics
trajectories at $\beta = 2.00$ and 1.82, respectively.

The same quark action as in the configuration generation is 
adopted in the measurement of the $K_{\ell 3}$ form factors.
The parameters for the quark actions are summarized in 
Table~\ref{tab:quark_param}.
The parameters at $\beta = 1.82$ are also found in our previous
paper~\cite{PACS:2019hxd}.
The coefficient of the clover term is nonperturbatively 
determined in the Schr\"odinger functional scheme.
The stout smearing parameter is $\rho = 0.1$ at both the lattice spacings.
The statistical error of observables is evaluated by the jackknife method
with the bin size of 5 and 10 trajectories at $\beta = 2.00$ and 1.82,
respectively.
The measured masses for the pion and kaon, $m_\pi$ and $m_K$, are
presented in Table~\ref{tab:mpi_mk_conf}.
It is noted that the measured masses at each $\beta$ are
slightly different from the physical ones, $m_{\pi^-} = 0.13957061$ GeV 
and $m_{K^0} = 0.497611$ GeV, for the $K_{\ell 3}$ decay in this calculation.
In a later section, we will explain that the tiny differences are corrected 
using the next-to-leading order (NLO) SU(3) chiral perturbation theory (ChPT),
and the corrections in $f_+(0)$ are as small as or less than 
the statistical error.

\begin{table}[t!]
\caption{Parameters for the PACS10 gauge configurations.
$L$, $T$, $a$, and $N_{\rm conf}$ represent
the spatial and temporal extents, the lattice spacing, and
the number of the configurations, respectively.
The spatial extent in units of fm and the measured $m_\pi$ and $m_K$ are
also tabulated.}
\label{tab:mpi_mk_conf}
\begin{tabular}{cccccccc}\hline\hline
$\beta$ & $L^3\times T$ & $a^{-1}$[GeV] & $a$ [fm] & $L$[fm] & $m_\pi$[GeV] & $m_K$[GeV] & $N_{\rm conf}$ \\\hline
2.00 & 160$^3\times$160 & 3.1108(70) & 0.063 & 10.1 & 0.13777(20) & 0.50481(19) & 20 \\
1.82 & 128$^3\times$128 & 2.3162(44) & 0.085 & 10.9 & 0.13511(72) & 0.49709(35) & 20 \\
\hline\hline
\end{tabular}
\end{table}

\begin{table}[t!]
\caption{Parameters in the quark action at each $\beta$.
$\kappa_l$, $\kappa_s$, $c_{\rm SW}$, $\rho$, $N_{\rm stout}$ express
the hopping parameter for the light and strange quarks,
the coefficient of the clover term, the stout smearing parameter, 
and the number of the stout smearing, respectively.}
\label{tab:quark_param}
\begin{tabular}{cccccc}\hline\hline
$\beta$ & $\kappa_l$ & $\kappa_s$ & $c_{\rm SW}$ & $\rho$ & $N_{\rm stout}$
\\\hline
2.00 & 0.125814 & 0.124925 & 1.02 & 0.1 & 6 \\
1.82 & 0.126117 & 0.124902 & 1.11 & 0.1 & 6 \\
\hline\hline
\end{tabular}
\end{table}

For the calculation method of the form factors at $\beta = 2.00$,
we follow the one in our previous calculation~\cite{PACS:2019hxd}.
The matrix element in Eq.(\ref{eq:def_matrix_element}) is extracted 
from the ratio of 2- and 3-point functions given as
\begin{eqnarray}
C^{\pi}_L(\vec{p},t-t_i)&=&
\langle 0 | O_{\pi,L}( \vec{p},t) \left(O^s_{\pi,L}( \vec{p},t_i)\right)^\dagger| 0 \rangle,
\label{eq:def_2-pt_pi}\\
C^{K}_L(\vec{p},t-t_i)&=&
\langle 0 | O_{K,L}( \vec{p},t) \left(O^s_{K,L}( \vec{p},t_i)\right)^\dagger| 0 \rangle ,
\label{eq:def_2-pt_K}
\\
C_{\mu,L}^{K\pi}(\vec{p }, t-t_i)&=&
\langle 0 | O_{K,L}( \vec{0},t_f) V_\mu(\vec{p},t) \left(O^s_{\pi,L}( \vec{p},t_i)\right)^\dagger| 0 \rangle
\label{eq:def_3-pt}.
\end{eqnarray}
In the 3-point function $C_{\mu,L}^{K\pi}$, $t_i \le t \le t_f$ is assumed.
The pion, kaon, and weak vector current operators are defined as
\begin{eqnarray}
O_{\pi,L}(\vec{p},t) &=& \sum_{\vec{x}}
\overline{u}(\vec{x},t)\gamma_5 d(\vec{x},t) e^{i\vec{p}\cdot\vec{x}} ,
\label{eq:op_pi}\\
O_{K,L}(\vec{p},t) &=& \sum_{\vec{x}}
\overline{s}(\vec{x},t)\gamma_5 d(\vec{x},t) e^{i\vec{p}\cdot\vec{x}} ,
\label{eq:op_K}\\
V_\mu(\vec{p},t) &=& \sum_{\vec{x}}
\overline{u}(\vec{x},t)\gamma_\mu s(\vec{x},t) e^{i\vec{p}\cdot\vec{x}} 
\label{eq:op_vec} .
\end{eqnarray}
$q(\vec{x},t)$ for $q = u,d,s$ represents a local quark field, 
where the color and Dirac indices are omitted.
The source operator $O^s_{H,L}$ for $H = \pi, K$ at the time slice $t_i$
in Eqs.~(\ref{eq:def_2-pt_pi})--(\ref{eq:def_3-pt})
is calculated by the $Z(2)\otimes Z(2)$ random source~\cite{Boyle:2008yd},
whose random numbers are spread in all the color, spin, and 
spatial spaces.
For example $O^s_{\pi,L}$ is defined as,
\begin{equation}
O^s_{\pi,L}(\vec{p},t) = \frac{1}{N_r}\sum_j
\left[ \sum_{\vec{x}}\overline{u}(\vec{x},t)\eta_j^\dagger(\vec{x})e^{i\vec{p}\cdot\vec{x}} \right]
\gamma_5
\left[ \sum_{\vec{y}}d(\vec{y},t)\eta_j(\vec{y}) \right] ,
\label{eq:random_source}
\end{equation}
where $N_r$ is the number of the random source, and
the color and spin indices are omitted.
The $Z(2)\otimes Z(2)$ random source $\eta_j(\vec{x})$ 
satisfies the following condition as,
\begin{equation}
\frac{1}{N_r}\sum_j \eta_j^\dagger(\vec{x})\eta_j(\vec{y}) \xrightarrow[N_r \to \infty]{} \delta(\vec{x}-\vec{y}) 
\label{eq:rand_source}.
\end{equation}
Since this source reduces to the local operator when $N_r$ is large enough,
we shall call it as the local source operator.

At $\beta = 2.00$, we also employ the exponentially smeared quark operator 
with the Coulomb gauge fixed configuration
in the calculations for the 2- and 3-point functions, defined as,
\begin{equation}
q_S(\vec{x},t) = \sum_{\vec{y}} \phi(|\vec{y}-\vec{x}|) q(\vec{x},t) .
\label{eq:smeared_quark}
\end{equation}
The smearing function $\phi(r)$ is given with the spatial extent of $L$ as
\begin{equation}
\phi(r) = \left\{
\begin{array}{ll}
1 & (r=0)\\
A e^{-Br} & (r < L/2)\\
0 & (r \ge L/2)
\end{array}
\right. .
\end{equation}
The smearing parameters for the light and strange quarks are chosen as 
$(A, B) = (1.2,0.14)$ and (1.2,0.22),
respectively, to obtain earlier plateaus in the effective meson masses.
In the smeared operator calculation, all the quark fields $u, d, s$ 
in the meson operators in the 2- and 3-point functions, {\it i.e.,}
$O_{H,L}$ and $O^s_{H,L}$ for $H = \pi, K$, are replaced by $u_S, d_S, s_S$.
On the other hand, the quark fields in the vector current are unchanged.
In the following we will denote the meson operators and 
the correlators with the exponentially smeared operator as
$O_{H, S}$, $O^s_{H, S}$, $C^{\pi}_S$, $C^{K}_S$, and $C_{\mu,S}^{K\pi}$.

In order to estimate the form factors in the continuum limit,
another 3-point function is calculated with
the point splitting conserved vector current, which is given by
\begin{eqnarray}
\widetilde{V}_\mu (\vec{p},t)
= \sum_{\vec{x}}
\overline{u}(\vec{x},t)\widetilde{\gamma}_\mu(\vec{x},t) s(\vec{x},t) e^{i\vec{p}\cdot\vec{x}} 
\label{eq:op_con_vec} ,
\end{eqnarray}
where
\begin{equation}
\overline{f}(x) \widetilde{\gamma}_\mu(x) g(x) = \frac{1}{2}\left[
\overline{f}(x+\hat{\mu})(1+\gamma_\mu)U_\mu^\dagger(x)g(x)-
\overline{f}(x)(1-\gamma_\mu)U_\mu(x)g(x+\hat{\mu})\right] 
\end{equation}
with $x$ expresses a four-dimensional vector of $(\vec{x},t)$.
The 3-point function with the conserved current is represented by
$\widetilde{C}_{\mu,X}^{K\pi}$ for $X = L, S$.
In all the cases the 3-point functions are measured
using the sequential source method at the sink time slice $t_f$.

The conserved current data at $\beta = 1.82$ is newly added in this work, 
which was not calculated in our previous work~\cite{PACS:2019hxd}.
It is noted that only the local operator is employed in the
measurements for the 2- and 3-point functions at this lattice spacing.

The periodic boundary condition in the spatial directions
is employed in the calculation of quark propagators.
As in our previous work~\cite{PACS:2019hxd},
we average the correlators with the periodic and 
anti-periodic boundary conditions
in the temporal direction.
This average reduces unwanted wrapping around effects in the 3-point 
functions~\cite{Aoki:2008ss}, and also makes the periodicity of 
the 2-point functions effectively double.
In the correlators with the anti-periodic boundary condition, 
one of the quark propagators in Eqs.~(\ref{eq:def_2-pt_pi})--(\ref{eq:def_3-pt})
is computed with the temporal anti-periodic boundary condition
as explained in Ref.~\cite{PACS:2019hxd}.
All the correlators are regarded as the averaged ones 
in the following discussion, unless otherwise explicitly stated.

The momentum is labeled by an integer $n_p$
as $p^2 = \vec{p}^2 = (2\pi/L)^2 n_p$.
We employ $0 \le n_p \le 6$ to obtain the form factors in the vicinity of $q^2 = 0$.
For a finite momentum case, we compute the correlators with
several momentum assignments to increase statistics.
$q^2$ is evaluated with
\begin{equation}
q^2 = -( m_K - E_\pi(p) )^2 + p^2
\end{equation}
where $E_\pi(p) = \sqrt{m_\pi^2 + (2\pi/L)^2 n_p}$ with the measured $m_\pi$.
At each $\beta$ the values of $q^2_{n_p}$ denoting $q^2$ with $n_p$ 
are tabulated in Table~\ref{tab:q2} 
as well as the number of the momentum assignment.

We investigate the current time dependence of 
the 3-point function using several 
time separations between the source and sink operators, 
$t_{\rm sep}= t_f - t_i$ in Eq.~(\ref{eq:def_3-pt}).
The values of $t_{\rm sep}$ employed in our calculation are 
tabulated in Table~\ref{tab:tsep_nmeas}.
The same set of $t_{\rm sep}$ is used in the 3-point functions with
both the local and conserved vector currents.
It is noted that we employ shorter $t_{\rm sep}$ in the smeared operator
calculation than the one in the local case at $\beta = 2.00$, because 
excited state effects are well suppressed by the smearing.

Using the same techniques in Ref.~\cite{PACS:2019hxd},
the signal of the correlators is improved by repeating the calculation
on each configuration with 4 temporal directions by rotating the configuration
thanks to the hypercube symmetry,
and averaging a backward 3-point function
in $t_f \le t \le t_i$ with the same $t_{\rm sep}$.
We also adopt calculations with different 8 source time slices in 
equally separated 20 and 16 time intervals at 
$\beta = 2.00$ and 1.82, respectively, 
except for $t_{\rm sep}=36$ with the smeared operator 
calculation at $\beta = 2.00$,
where different 4 source time slices with 40 separations are employed.
The number of the random source $N_r$ in Eq.~(\ref{eq:rand_source})
is unity in almost all the cases, while $N_r = 2$ is used in 
larger $t_{\rm sep}$ cases: the local and smeared operator calculations
with $t_{\rm sep} = 58$ and 48 at $\beta = 2.00$, respectively,
and $t_{\rm sep} = 42, 48$ at $\beta = 1.82$.
The total numbers of measurement are summarized in 
Table~\ref{tab:tsep_nmeas}.

\begin{table}[t!]
\caption{The value of the momentum transfer squared $q^2_{n_p}$[GeV$^2$] 
at each $\beta$ and $n_p = (L/2\pi)^2 p^2$.
$\nu_p$ is the number of the momentum assignment in the calculation
of the correlators.}
\label{tab:q2}
\begin{tabular}{cccccccc}\hline\hline
$\beta$ & $q_0^2$ & $q^2_1$ & $q^2_2$ & $q^2_3$ & $q^2_4$ & $q^2_5$ & $q^2_6$ \\\hline
2.00 & $-0.13471(23)$ & $-0.08791(18)$ & $-0.05072(15)$ & $-0.01889(13)$ &   0.00938(11)  & 0.03507(10) & 0.05878(8)\\
1.82 & $-0.13103(48)$ & $-0.08980(33)$ & $-0.05656(25)$ & $-0.02792(20)$ & $-0.00239(17)$ & 0.02087(15) & 0.04239(13)\\
$\nu_p$ & 1 & 6 & 12 & 8 & 6 & 9 & 9 \\
\hline\hline
\end{tabular}
\end{table}

\begin{table}[t!]
\caption{The source-sink separation $t_{\rm sep} = t_f - t_i$ of 
the 3-point function in lattice and physical units at 
each $\beta$ for local and smeared operators.
The numbers of measurements $N_{\rm meas}$\footnote{We adopt a different counting rule of measurement from that in the previous paper~\cite{PACS:2019hxd}. Thus, $N_{\rm meas}$ at $\beta = 1.82$ becomes half of the previous paper, though it means the same.} for the correlators are also presented.}
\label{tab:tsep_nmeas}
\begin{tabular}{cccc}\hline\hline
$\beta$ & \multicolumn{2}{c}{2.00} & 1.82 \\
operator & local & smeared & local \\\hline
$t_{\rm sep}$ & (50, 58, 64) & (36, 42, 48, 54) & (36, 42, 48) \\
$t_{\rm sep}$[fm] & (3.2, 3.7, 4.1) & (2.3, 2.7, 3.0, 3.4) & (3.1, 3.6, 4.1) \\
$N_{\rm meas}$ & (1280, 2560, 1280) & (640, 1280, 2560, 1280)
& (1280, 2560, 2560) \\
\hline\hline
\end{tabular}
\end{table}

\subsection{Extraction of matrix element}

First, we focus on the calculation with the local vector current,
and the analysis method to extract the matrix element
from the data at $\beta = 2.00$ is explained below.

The matrix element in Eq.~(\ref{eq:def_matrix_element})
is obtained from the ground state contribution in the 3-point function.
Its time dependence differs from excited state contributions.
To extract the matrix element, we define a ratio $R_{\mu,X}(t;n_p)$ as,
\begin{eqnarray}
R_{\mu,X}(t;n_p) &=& Z_X^\pi(p)Z_X^K(0)
\frac{N_\mu (\vec{p}) C_{\mu,X}^{K\pi}(\vec{p},t)}
{C^\pi_X(\vec{p},t)C^K_X(\vec{0},t_{\rm sep}-t)} ,
\label{eq:def_Rmu}
\end{eqnarray}
at each $p^2 = (2\pi/L)^2 n_p$ for $X = L,S$.
In the equation it is assumed $t_i = 0$, which is also employed in
the following discussion.
The coefficient $N_\mu$ is defined by $N_4(\vec{p}) = 1$ and 
$N_i(\vec{p}) = 1/p_i$ with $i = 1,2,3$,
and $Z_X^\pi(p)$ and $Z_X^K(0)$ are determined from fitting of 
the 2-point functions in a large $t$ region using a fit form given by
\begin{eqnarray}
C^{H}_X(\vec{p},t)= \frac{\left(Z_X^H(p)\right)^2}{2E_H(p)} ( e^{-E_H(p) t}+e^{-E_H(p) (2T - t)} ),
\label{eq:asym_form_two-pt}
\end{eqnarray}
with $E_H(p) = \sqrt{m_H^2 + p^2}$ for $H = \pi, K$.
Noted that $C^{H}_X$ and $C_{\mu,X}^{K\pi}$ have the $2T$ periodicity,
since they are calculated by average of periodic and anti-periodic correlators
in the temporal direction as explained in the last subsection.
In the local operator calculation ($X = L$), 
we use the relation $Z_L^\pi(p) = Z_L^\pi(0)$,
which is confirmed statistically.

In a middle time region, $0 \ll t \ll t_{\rm sep}$, $R_{\mu,X}$ behaves as,
\begin{eqnarray}
R_{\mu,X}(t;n_p) 
&=&
R_{\mu}(p) + \Delta A_{\mu,X}(p,t),
\label{eq:Rmu_time_dep}
\end{eqnarray}
where 
the constant part corresponds to the matrix element 
$R_\mu(p) = \frac{N_\mu (\vec{p})}{Z_V}
\langle \pi (\vec{p}) \left | V_{\mu} \right | K(\vec{0}) \rangle$
with $p^2 = (2\pi/L)^2 n_p$.
The last term on the right-hand side represents the leading 
contribution from excited states.
In a later section, its functional form is discussed 
to extract the matrix elements.
The renormalization factor of the local vector current $Z_V$ is 
determined from a ratio with 3-point functions for $\pi$ and $K$
electromagnetic form factors at $q^2 = 0$ as,
\begin{equation}
R_{Z_V}(t) = \sqrt{\frac{C^\pi_X(\vec{0},t_{\rm sep})C^K_X(\vec{0},t_{\rm sep})}
{C^{\pi\pi}_{4,X}(t)C^{KK}_{4,X}(t)}}.
\label{eq:zv}
\end{equation}
The correlators are defined by
\begin{eqnarray}
C^{\pi\pi}_{4,X}(t) &=& \left\langle 0 \left| O_{\pi,X}(\vec{0},t_{\rm sep})V^{\rm em}_4(\vec{0},t)\left(O^s_{\pi,X}(\vec{0},0)\right)^\dagger\right|0\right\rangle ,
\label{eq:3-pt_em_pi}\\
C^{KK}_{4,X}(t) &=& \left\langle 0 \left| O_{K,X}(\vec{0},t_{\rm sep}) V^{\rm em}_4(\vec{0},t)\left(O^s_{K,X}(\vec{0},0)\right)^\dagger\right|0\right\rangle ,
\label{eq:3-pt_em_K}
\end{eqnarray}
where $V_4^{\rm em}(t)$ is the temporal component of the local electromagnetic 
current.

\begin{figure}[ht!]
\includegraphics*[scale=.45]{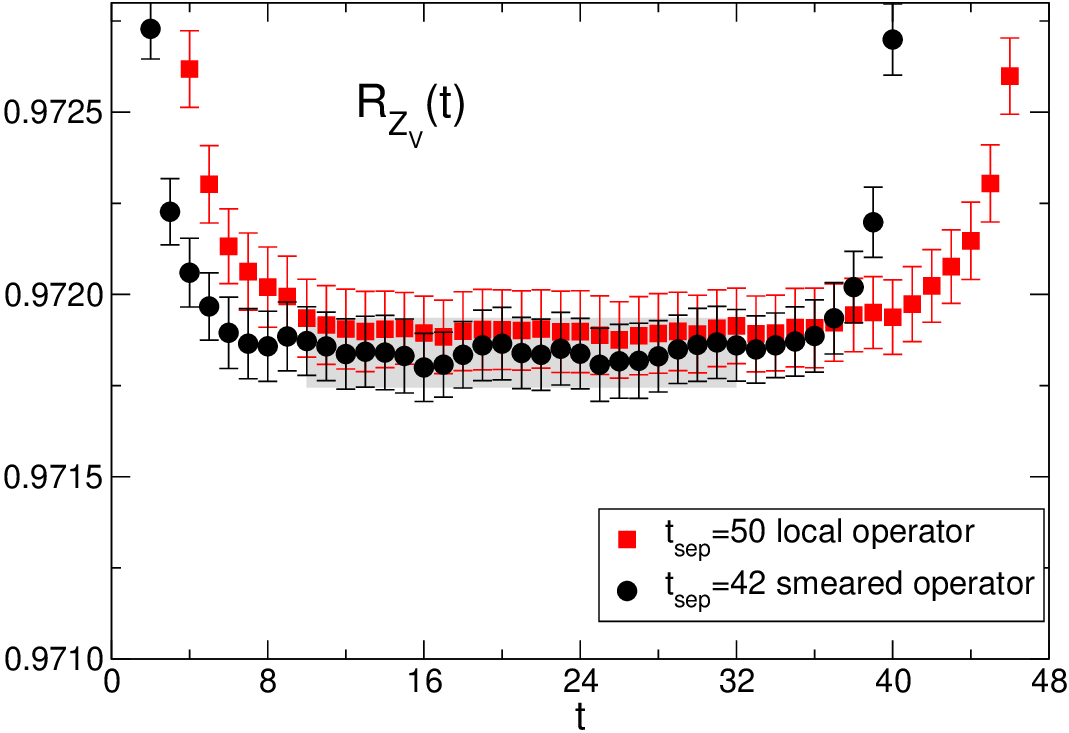}
\caption{Ratios defined in Eq.~(\ref{eq:zv}) for the local (square) and 
smeared (circle) data in $t_{\rm sep} = 50$ and 42, respectively, as a
function of $t$. The shaded band represents a constant fit result
of the smeared data with one standard deviation,
which corresponds to the value of $Z_V$.}
\label{fig:zv-b200}
\end{figure}

Typical data of $R_{Z_V}$ at $\beta = 2.00$ using the local and smeared 
operators in $t_{\rm sep} = 50$ and 42, respectively, 
are shown in Fig.~\ref{fig:zv-b200}.
From a constant fit in a flat region of the smeared operator data,
we obtain $Z_V = 0.971840(96)$.
This value agrees with the one obtained from the other operator
calculation in the figure, $Z_V = 0.97190(11)$.
Those are also statistically consistent with results
obtained from other $t_{\rm sep}$ data using the local and smeared operators.
The value of the smeared operator is adopted in the following analysis 
at $\beta = 2.00$.

At $\beta = 1.82$ we use $Z_V$ from the local operator calculation 
to adopt the common setups as in $\beta = 2.00$, 
although the difference of $Z_V$ obtained 
from the local operator calculation and the Schr\"odinger functional 
scheme~\cite{Taniguchi:2012kk} was discussed in Ref.~\cite{PACS:2019hxd}.

In the case of the 3-point function with the conserved vector current,
the calculation strategy is basically the same as above.
The difference is only that $Z_V = 1$ is employed.
We confirm that $Z_V$ of the conserved vector current statistically 
agrees with unity by the same calculation as in Eq.~(\ref{eq:zv}), but using
$\widetilde{C}_{4,X}^{\pi\pi}$ and $\widetilde{C}_{4,X}^{\pi\pi}$
calculated with the conserved electromagnetic vector current.

\section{Result at finite lattice spacings}
\label{sec:result_finite_a}

In this section we will present results of the form factors 
at $\beta = 2.00$ and 1.82.
First, the results using the local vector current at only $\beta = 2.00$ are
described, because those at $\beta = 1.82$ are already reported 
in Ref.~\cite{PACS:2019hxd}.
And then the results from the conserved vector current at both 
the lattice spacings are discussed.

\subsection{Local current result at $\beta = 2.00$}
\label{sec:loc_cur_ana}

In this subsection the data calculated with the local vector current
at $\beta = 2.00$ are presented.
They are calculated using the local and smeared operators.

\begin{figure}[ht!]
\includegraphics*[scale=.45]{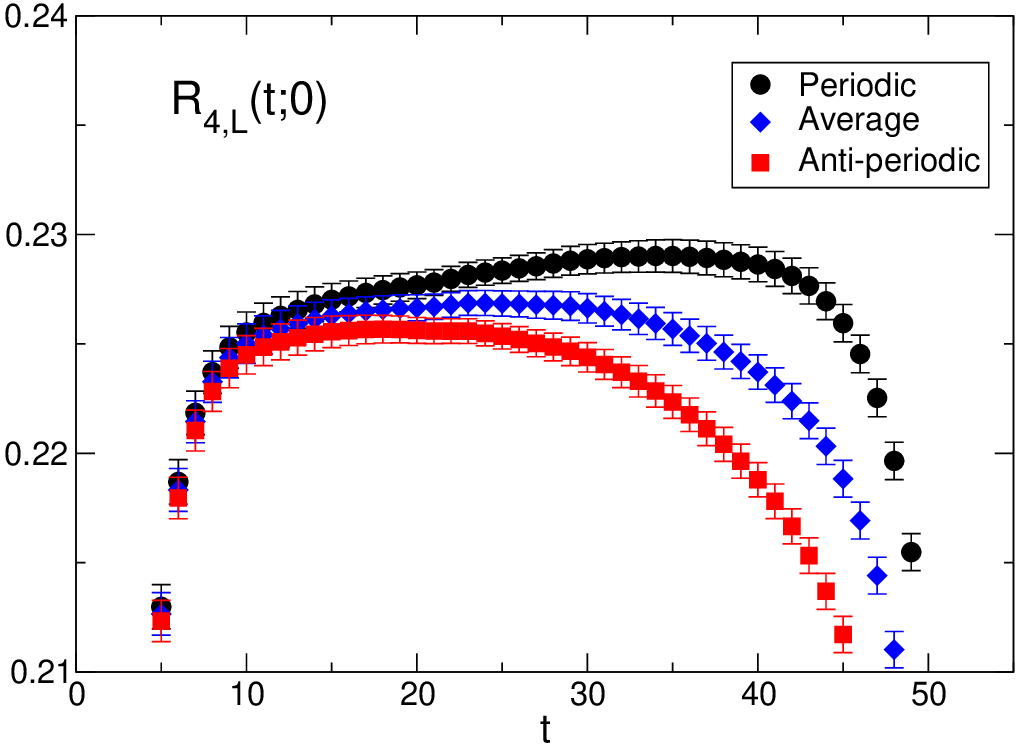}
\includegraphics*[scale=.45]{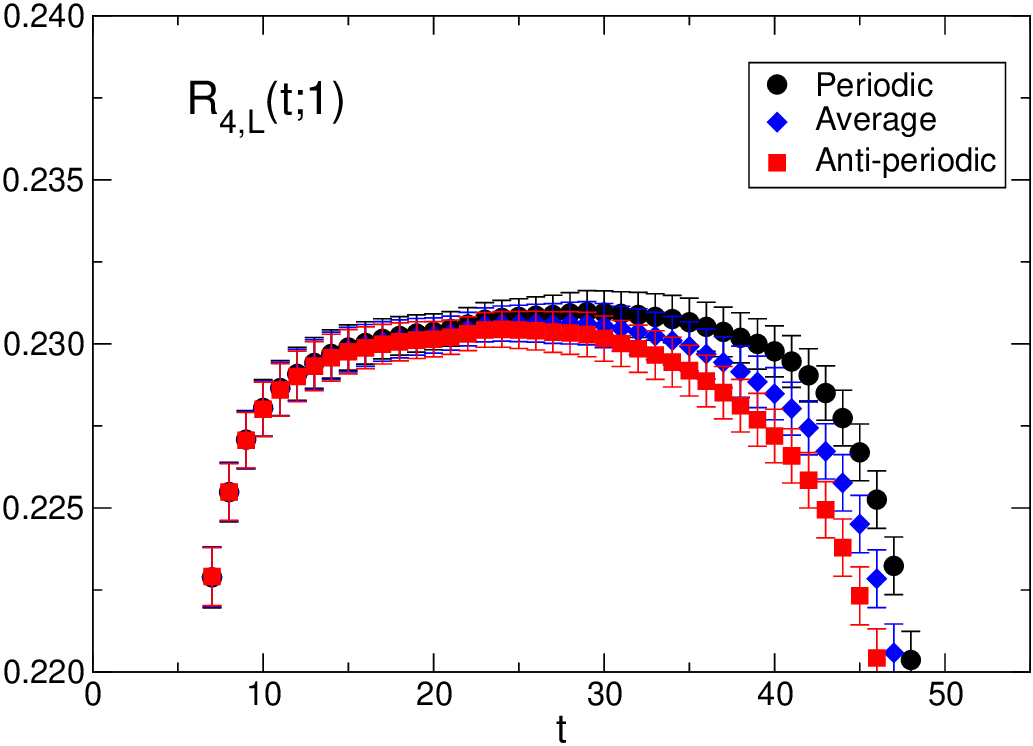}
\caption{Ratios $R_{4,L}(t;n_p)$ with $t_{\rm sep} = 58$
defined in Eq.(\ref{eq:def_Rmu}) for 
$n_p = 0$ and 1 correspond to the left and right panels, respectively.
The ratios calculated from the 3-point function with the periodic and
anti-periodic boundary conditions in the temporal direction are expressed
by circle and square symbols. Their average is also shown by diamond symbol.}
\label{fig:v4.0-1-lcl}
\end{figure}

\begin{figure}[ht!]
\includegraphics*[scale=.45]{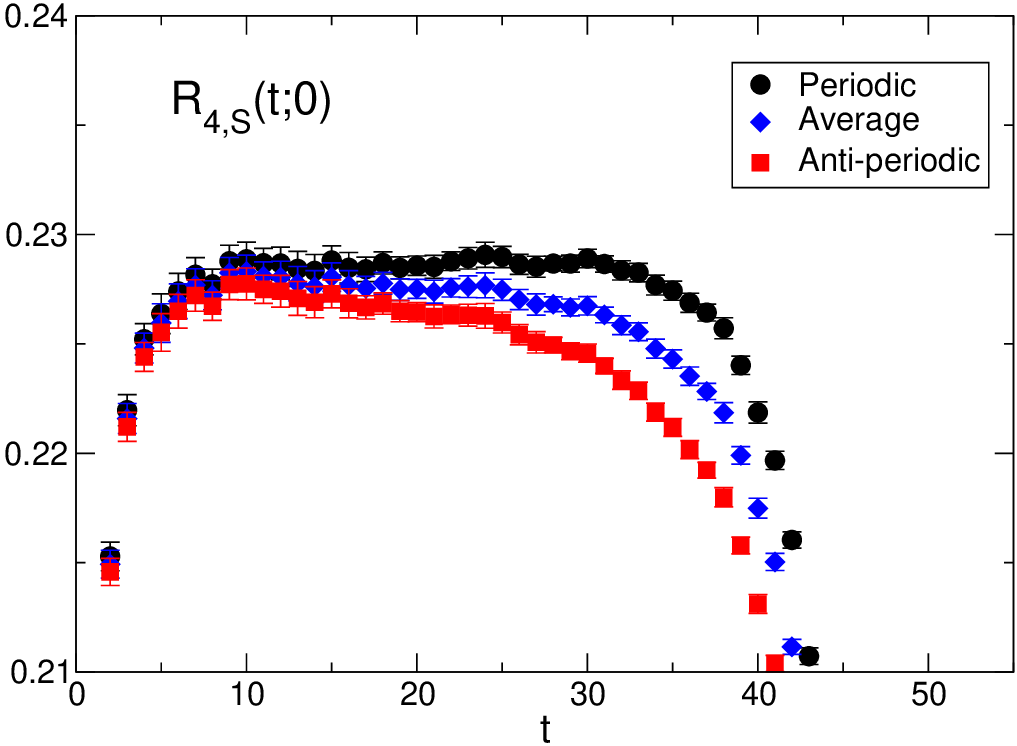}
\includegraphics*[scale=.45]{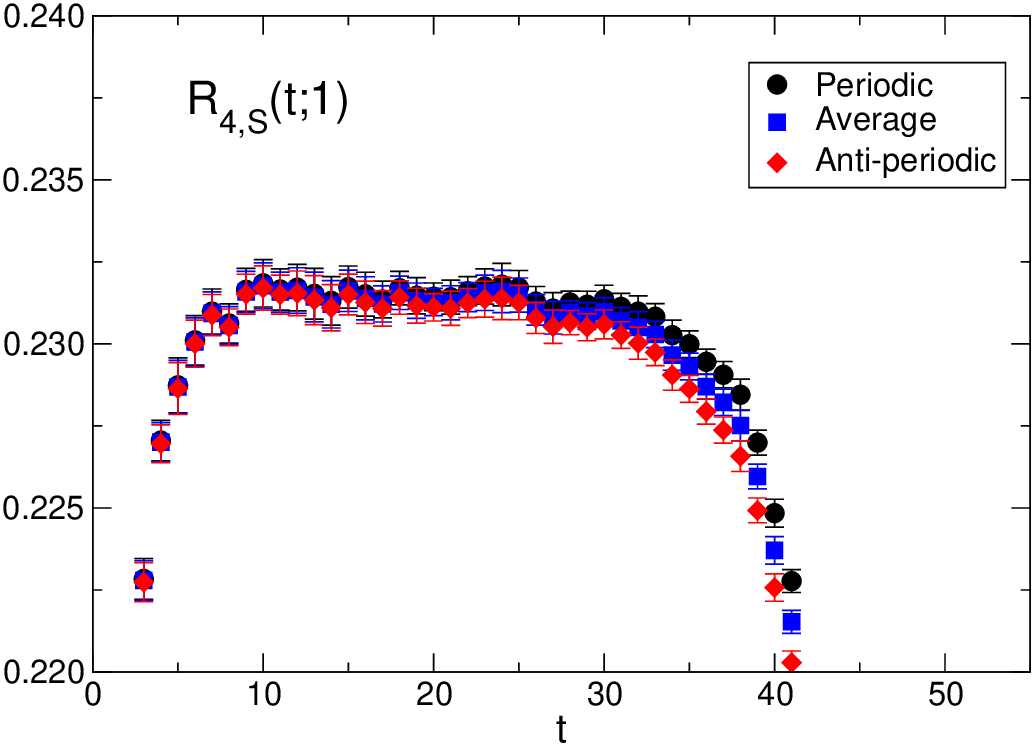}
\caption{The same figure as Fig.~\ref{fig:v4.0-1-lcl}, but for $R_{4,S}$
with $t_{\rm sep} = 48$.}
\label{fig:v4.0-1-smr}
\end{figure}

The wrapping around effects in the 3-point functions are
suppressed by averaging the correlators with two boundary conditions,
as discussed in our previous work~\cite{PACS:2019hxd}.
A similar effect of the suppression is seen in this calculation.
The left panel of Fig.~\ref{fig:v4.0-1-lcl} shows
the two ratios $R_{4,L}(t;0)$ using the local operator with $t_{\rm sep} = 58$ 
defined in Eq.~(\ref{eq:def_Rmu}),
but using $C_{4,L}^{K\pi}$ with the periodic and anti-periodic
boundary conditions in the temporal direction represented by
circle and square symbols, respectively.
The two data have different time dependence in the middle region, 
$0 \ll t \ll t_{\rm sep}$.
It is considered to be caused by wrapping around effects, which
is similar to the ones in Ref.~\cite{Aoki:2008ss}.
The average with the two data denoted by diamonds gives a mild $t$ dependence
in the same region, because the effects are effectively canceled out in the average.
The wrapping around effects become smaller as the momentum increases
shown in the right panel of Fig.~\ref{fig:v4.0-1-lcl}.
The same trend was seen in our previous work~\cite{PACS:2019hxd}.
Figure~\ref{fig:v4.0-1-smr} presents that a similar suppression
is observed in the smeared operator calculation with $t_{\rm sep} = 48$,
though the wrapping around effects seem relatively smaller than
the local operator case.
As explained in the previous section, the averaged data are employed
in the following analyses.

\begin{figure}[ht!]
\includegraphics*[scale=.45]{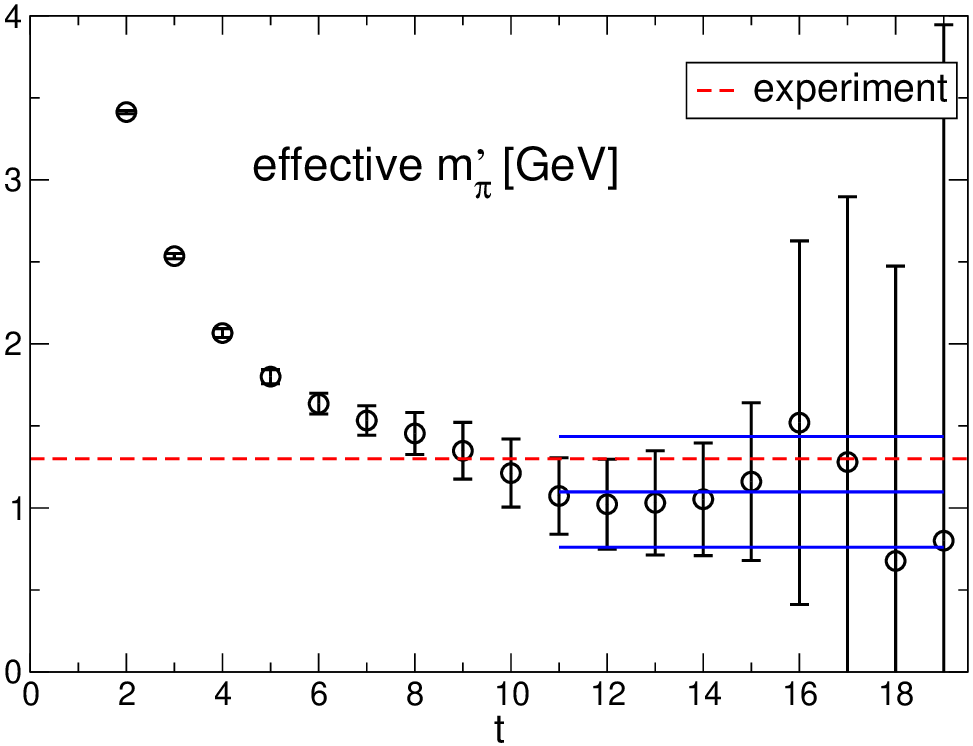}
\includegraphics*[scale=.45]{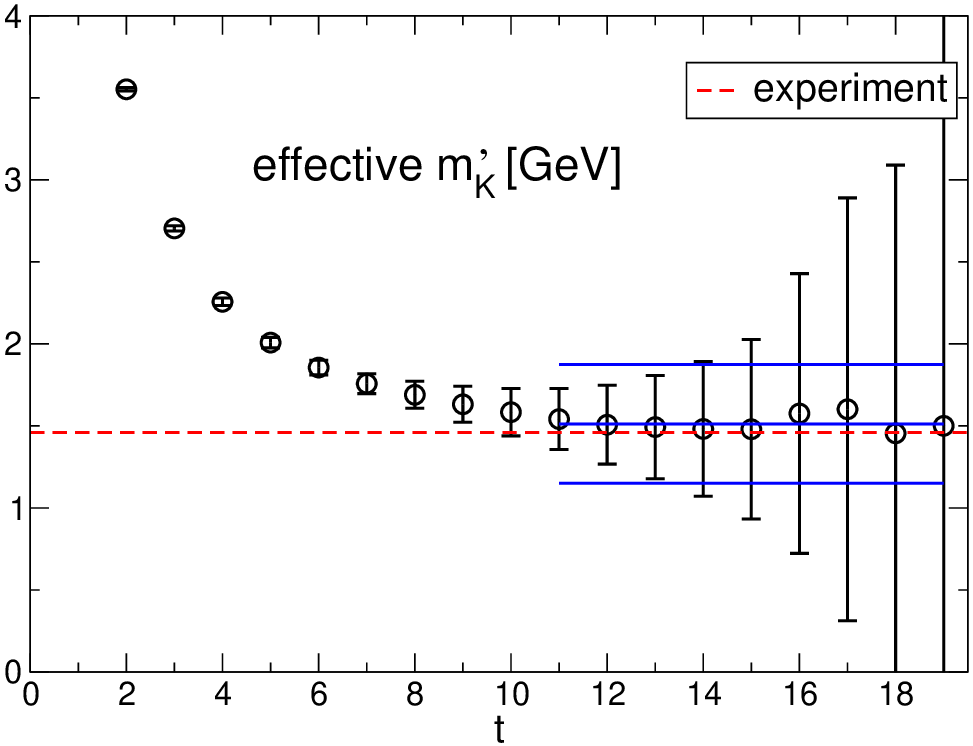}
\caption{Effective masses of the first excited state for
$\pi$ (left) and $K$ (right) in GeV units.
The dashed lines correspond to their experimental values.
The solid lines express a fit result with one standard deviation,
and their lengths represent the fit range.}
\label{fig:excited_meson_masses}
\end{figure}

In order to extract the constant part in $R_{\mu,X}$, which is 
$R_\mu(p)$ in Eq.~(\ref{eq:Rmu_time_dep}) corresponding to the matrix element,
$R_{\mu,X}$ in a middle $t$ region is fitted by a fitting form given in 
Eq.~(\ref{eq:Rmu_time_dep}) with an appropriate leading
excited state term $\Delta A_{\mu, X}$.
We adopt the same form of $\Delta A_{\mu, X}$
as in our previous calculation~\cite{PACS:2019hxd} given by
\begin{equation}
\Delta A_{\mu,X}(p,t) =
A_{\mu,X}^\pi(p) e^{-\Delta_\pi(p)t} +
A_{\mu,X}^K(p) e^{-\Delta_K(t_f-t)} .
\label{eq:def_radial_excited_effect}
\end{equation}
In the local operator case we choose
\begin{eqnarray}
\Delta_\pi(p) &=& \sqrt{(m_\pi^\prime)^2+p^2} - E_\pi(p), \\
\Delta_K &=& m_K^\prime - m_K .
\end{eqnarray}
The value of $m_H^\prime$ is fixed by the experimental value of 
the first excited state mass of the $H$ meson,
$m_\pi^\prime = 1.3$ GeV and $m_K^\prime = 1.46$ GeV 
in PDG20~\cite{ParticleDataGroup:2020ssz}.
This is because the first excited masses estimated from our 2-point functions
with the local operator presented in Fig.~\ref{fig:excited_meson_masses} are 
well consistent with those values, albeit the errors are large.
The analysis method to extract the first excited state mass from the 
2-point function with the local operator is the same as 
explained in the previous work~\cite{PACS:2019hxd}.
On the other hand, $\Delta_\pi(p)$ and $\Delta_K$ are chosen as 
free parameters in the smeared operator data.

\begin{figure}[ht!]
\includegraphics*[scale=.45]{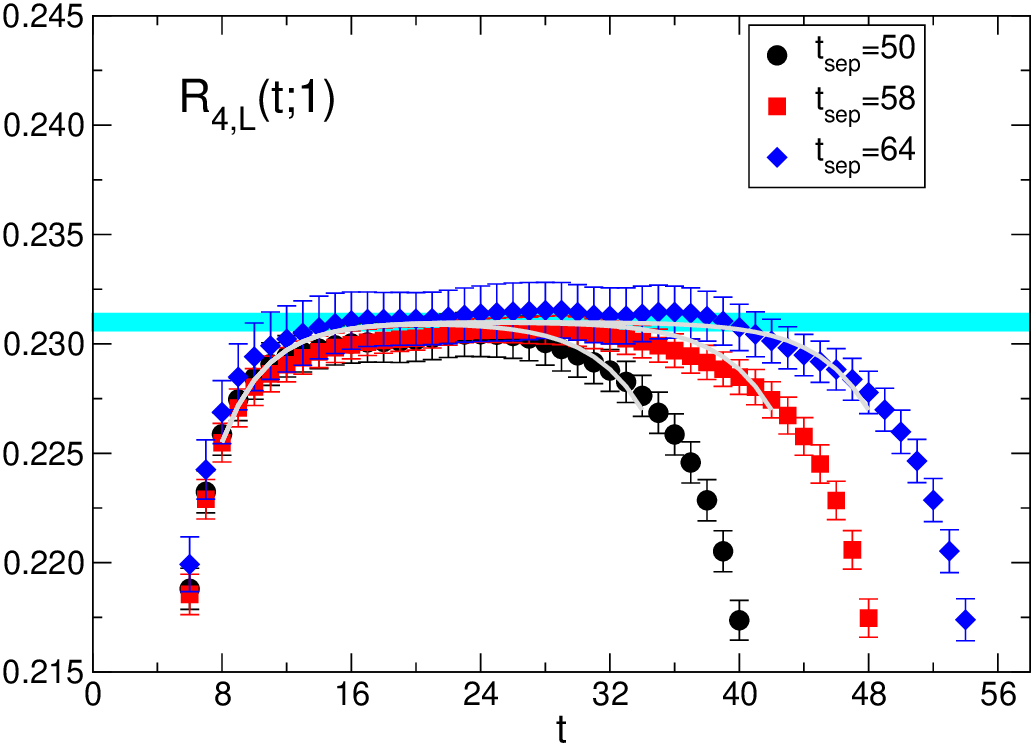}
\includegraphics*[scale=.45]{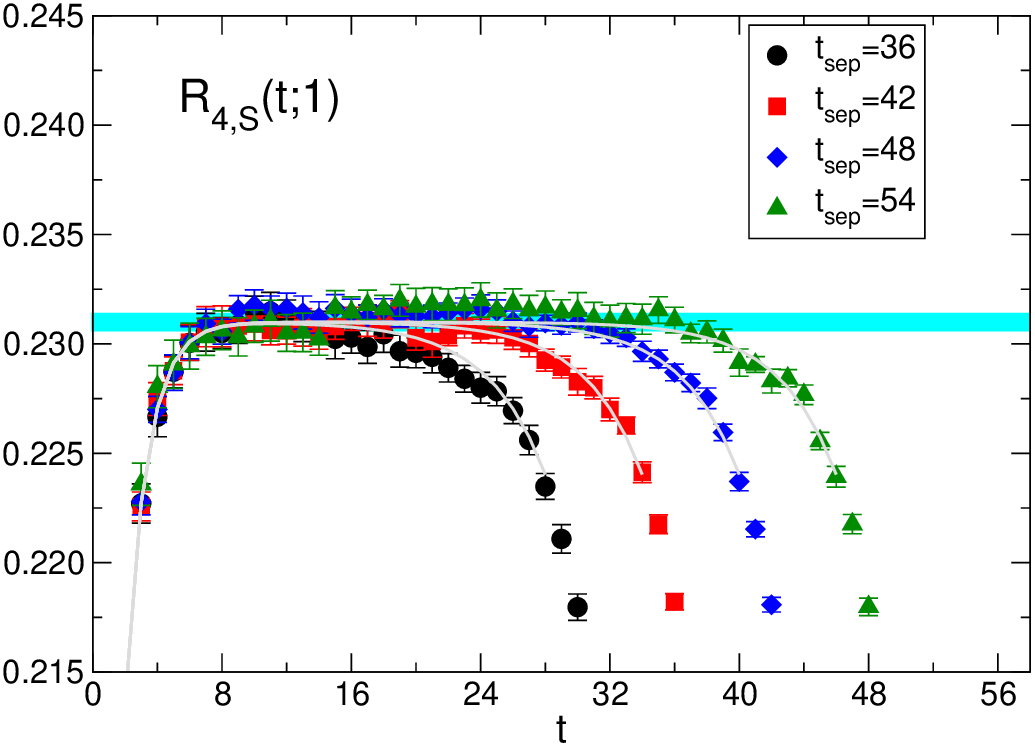}
\caption{Ratios $R_{4,X}(t;1)$ defined in Eq.(\ref{eq:def_Rmu}) for 
the local $X=L$ (left) and smeared $X=S$ (right) operators.
Different symbols express different $t_{\rm sep}$ data in each panel.
The gray curves correspond to results from a simultaneous fit.
The light blue band is the result of the matrix element $R_4(p)$.}
\label{fig:v4.1-fit-lcl-smr}
\end{figure}

We carry out a simultaneous fit using
all the $t_{\rm sep}$ data with both the operators 
to obtain a common $R_\mu(p)$.
As a typical example of the fit result,
the results from $R_{4,L}(t;1)$ and $R_{4,S}(t;1)$ are plotted in 
Fig.~\ref{fig:v4.1-fit-lcl-smr}.
The fit ranges are chosen such that the value of 
the uncorrelated $\chi^2/$dof is less than unity.
The minimum time $t_{\min}$ of the fit range is fixed,
while the maximum is shifted by $t_{\rm max} = t_{\rm sep} - t_{\rm fit}$.
The choice of $t_{\rm min}$ and $t_{\rm max}$ depends on the operator.
In the figure, we employ $(t_{\rm min},t_{\rm fit}) = (8,16)$
and (2,8) for the local and smeared operators, respectively.
The fit curves well explain our data in the
middle $t$ region for both the operators.
From the fit we extract $R_4(p)$ denoted by the light blue band in the
figure, which agrees with the larger $t_{\rm sep}$ data in the middle region.
Noted that the value of $\chi^2/$dof is not so tiny even in uncorrelated fits,
because the smeared operator data do not behave as a smooth function of 
$t$ as shown in the right panel of Fig.~\ref{fig:v4.1-fit-lcl-smr}.
It seems to be caused by a weak correlation among the data in the different 
time slices.
Another reason of the large $\chi^2/$dof might be a weak correlation among 
the different operator data, which are calculated with different random 
sources in each source operator.

\begin{figure}[ht!]
\includegraphics*[scale=.45]{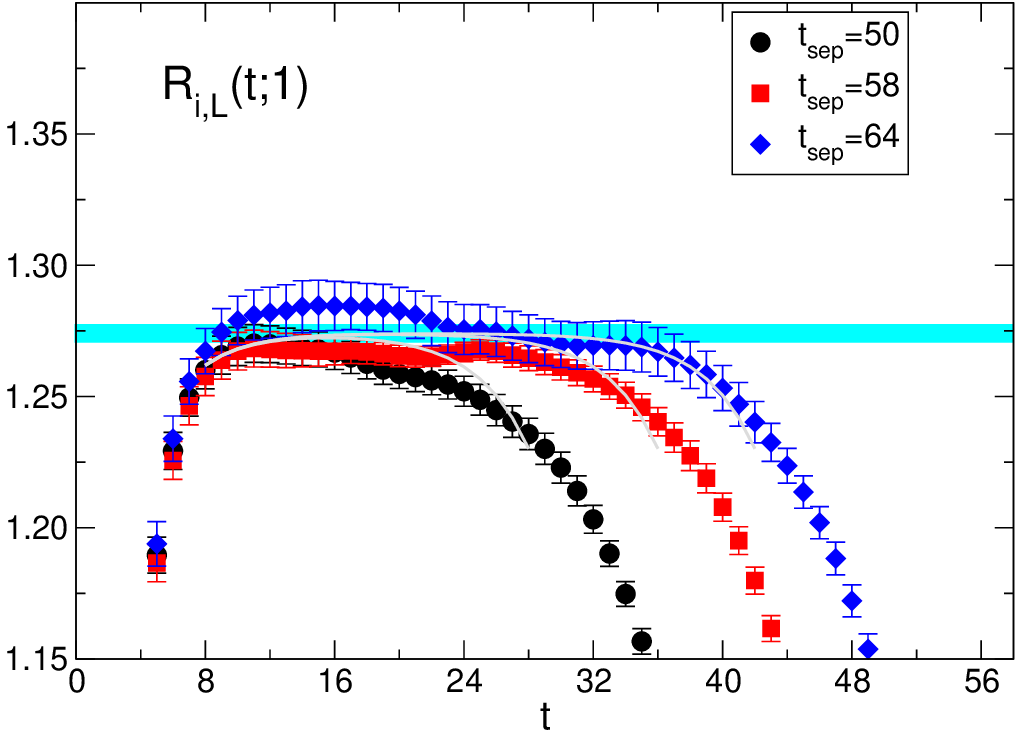}
\includegraphics*[scale=.45]{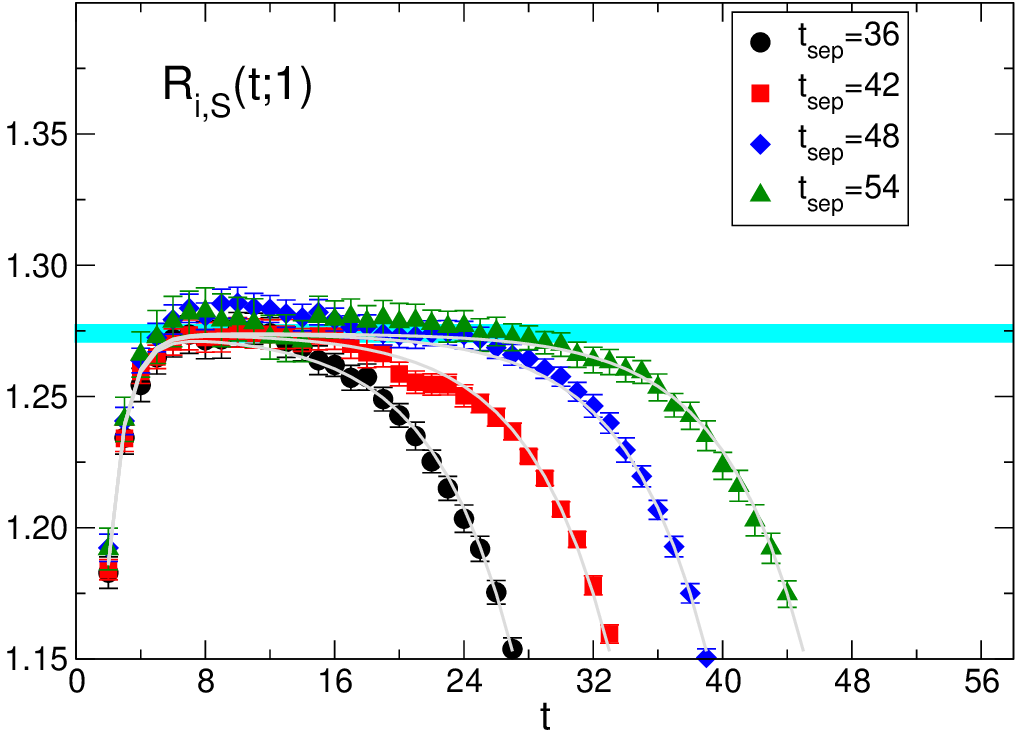}
\caption{The same figure as Fig.~\ref{fig:v4.1-fit-lcl-smr},
but for $R_{i,X}(t;1)$.}
\label{fig:vi.1-fit-lcl-smr}
\end{figure}

The spatial component of the matrix element $R_i(p)$ is extracted
from the same simultaneous fit using the two operator data.
The data of the spatial component are obtained by the average
of $R_{i,X}(t;n_p)$ in $i = 1, 2, 3$ at each $n_p$ and $X$.
Figure~\ref{fig:vi.1-fit-lcl-smr} shows that
the data for $R_{i,L}(t;1)$ and $R_{i,S}(t;1)$ have rather larger
statistical fluctuations than those for $\mu = 4$.
Using $(t_{\rm min},t_{\rm fit}) = (8,22)$
and (2,9) for the local and smeared operators,
a reasonable fit can be carried out represented by gray curves,
and the result of $R_i(p)$ is expressed by a light blue band in the figure.

In the $\mu = 4$ and $i$ cases, we also carry out another analysis
using free exponents in $\Delta A_{\mu,L}$ instead of the fixed ones.
The results from the two analyses 
agree with each other within the statistical error.
Furthermore, a two-exponential fit analysis is performed,
where another excited state term $\Delta B_{\mu,X}(p,t)$ 
with the same form as $\Delta A_{\mu,X}(p,t)$
in Eq.~(\ref{eq:def_radial_excited_effect})
is added in a fit function, which is given as
\begin{equation}
R_{\mu,X}(t;n_p) 
=
R_{\mu}(p) + \Delta A_{\mu,X}(p,t) + \Delta B_{\mu,X}(p,t) .
\end{equation}
Since in general a two-exponential fit is numerically unstable, 
in this analysis, the exponents in $\Delta A_{\mu,X}$
are fixed as above for both the operators $X = L, S$,
while those in $\Delta B_{\mu,X}$ are free parameters.
The results from the two-exponential analysis 
are statistically consistent with those from the above analyses,
although the statistical errors are larger.

\begin{figure}[ht!]
\includegraphics*[scale=.45]{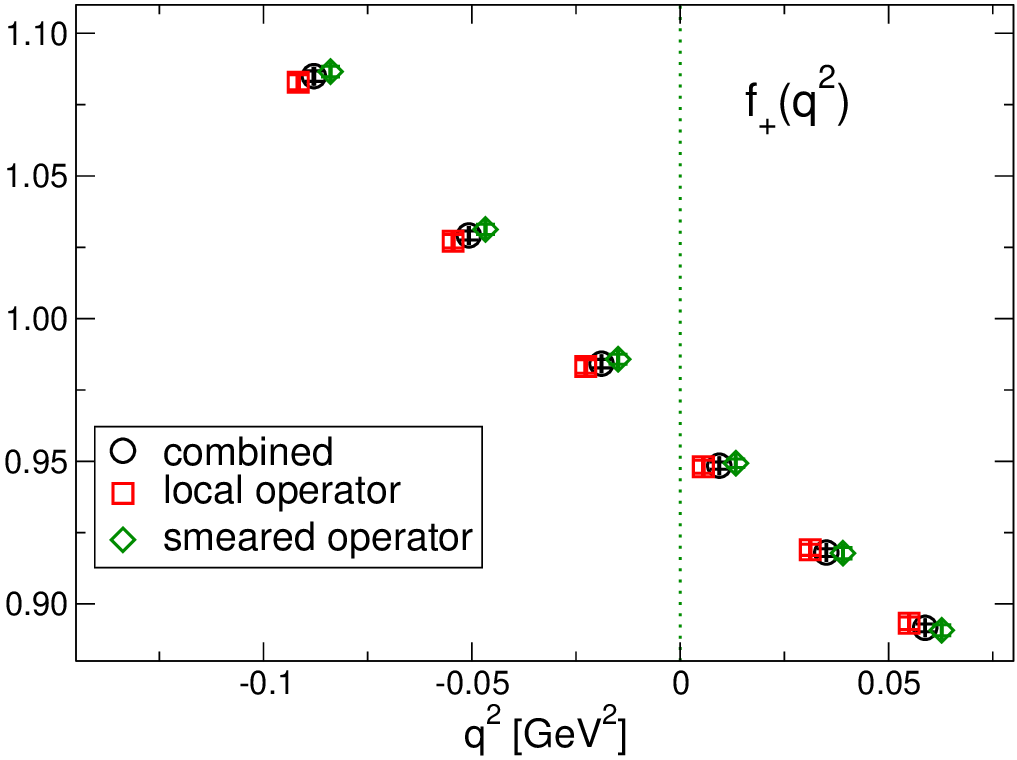}
\includegraphics*[scale=.45]{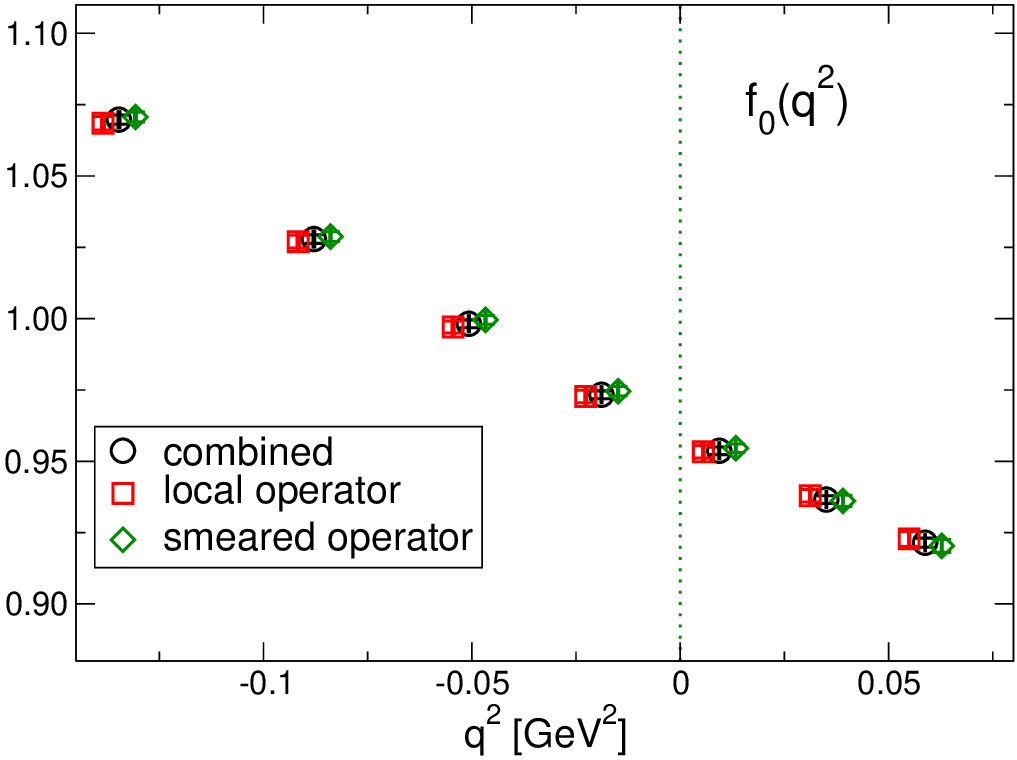}
\caption{$K_{\ell 3}$ form factors $f_+(q^2)$ (left) and $f_0(q^2)$ (right)
obtained from the local and smeared operators, and also 
combined analysis using both the operators. 
Those data are expressed by square, diamond, and circle symbols.
The results are obtained from the local current data.
The square and diamond symbols are slightly shifted to the negative and positive
$x$ direction, respectively, for clarity.
}
\label{fig:f+0-comb-lcl-smr}
\end{figure}

The bare matrix elements obtained from the above fits are renormalized with 
$Z_V$ discussed in the previous section.
Using the matrix elements in $\mu = 4$ and $i$, 
the form factors $f_+$ and $f_0$ are determined by solving a
linear equation at each $q^2_{n_p}$.
The results for the form factors are plotted in Fig.~\ref{fig:f+0-comb-lcl-smr}
as a function of $q^2$.
We also perform similar analyses explained above 
with only the local or smeared operator data
to study a stability of our result.
As shown in the figure, their results are in good agreement 
with those from the combined analysis using both the operator data 
discussed above.
The values for the form factors from all the analyses are tabulated 
in Table~\ref{tab:kl3_form_factor_b200}.
In the combined analysis, the values of $\xi(q^2) = f_-(q^2)/f_+(q^2)$
are also listed.
We note that the relative difference between the form factors
from the local and smeared operators is at most 1.6 $\sigma$.
It can be mainly caused by poor statistics in our calculation.
As presented in Table~\ref{tab:kl3_form_factor_b200}, the error of 
the combined analysis is smaller than those in the two operator
data, especially in the larger $q^2$. While we suspect a reason is that
the correlation between the local and smeared data becomes weaker 
as $q^2$ increase, we would need more detailed studies to clarify a reason 
of the error reduction.
In this paper, our main result is obtained from the combined 
analysis, while a systematic error is estimated 
using the other two results.

\begin{figure}[ht!]
\includegraphics*[scale=.45]{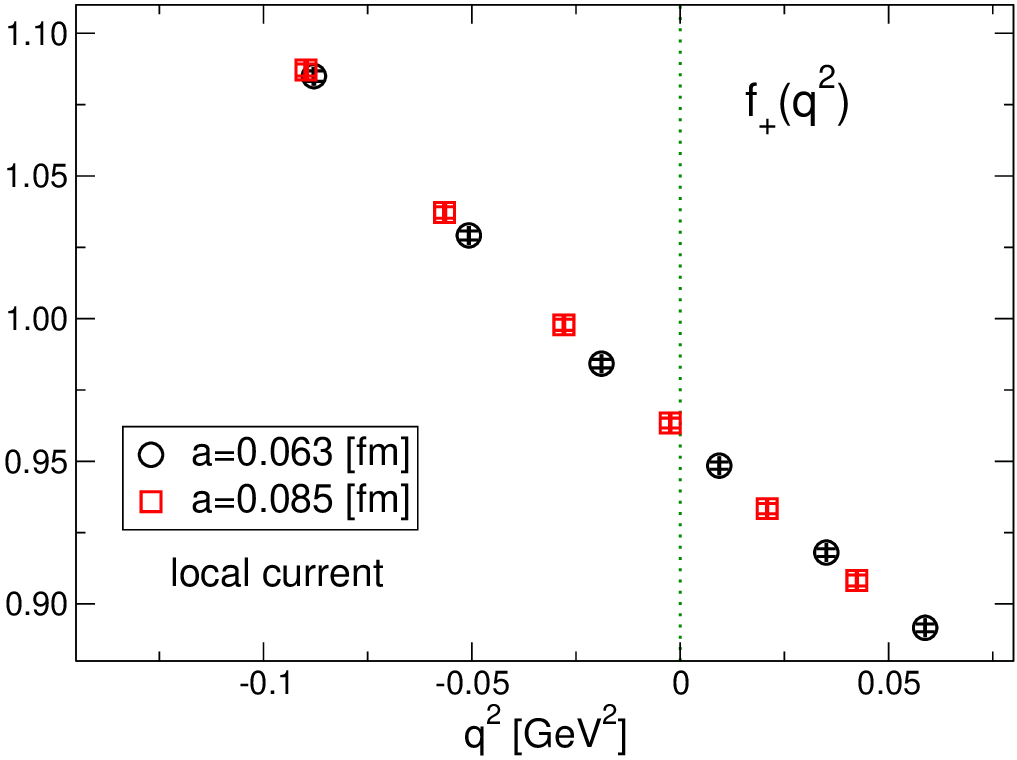}
\includegraphics*[scale=.45]{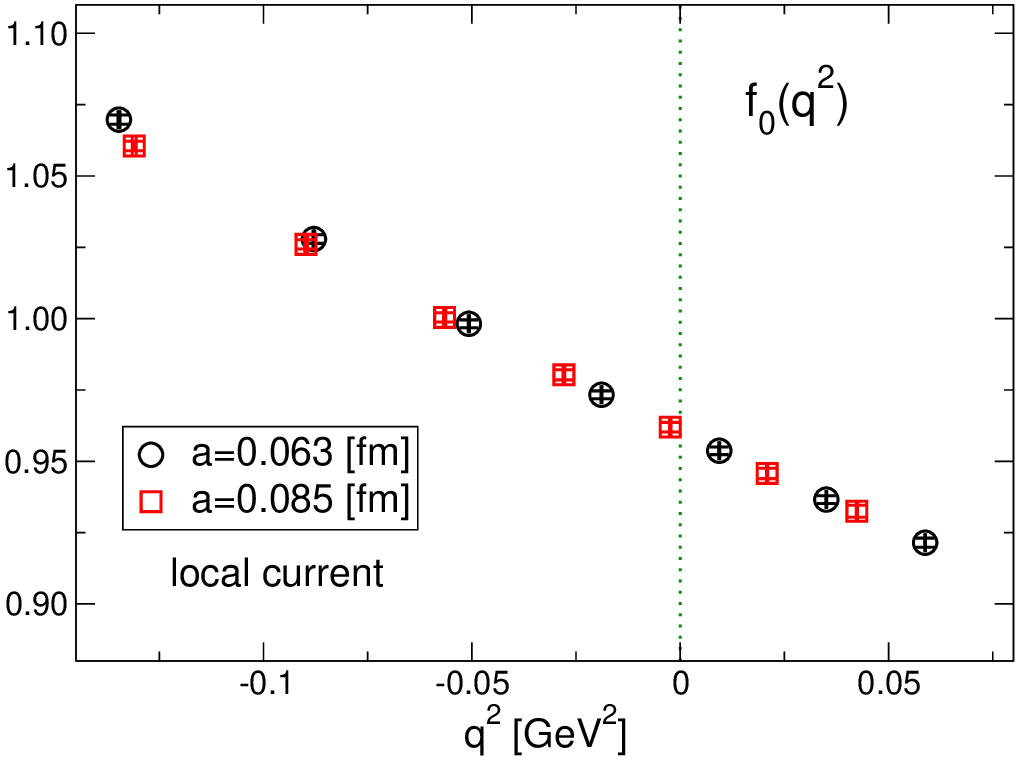}
\caption{Comparisons for $K_{\ell 3}$ form factors $f_+(q^2)$ (left)
and $f_0(q^2)$ (right) at the two lattice spacings obtained from
the local current data.
The data at the fine and coarse lattice spacings are expressed by
circle and square symbols, respectively.}
\label{fig:f+0-comp-lcl-cur}
\end{figure}

The form factors are compared with those at $\beta = 1.82$
obtained in our previous work~\cite{PACS:2019hxd}.
Figure~\ref{fig:f+0-comp-lcl-cur} shows that
two data of $f_+$ at the different lattice spacings are
well consistent with each other in all $q^2$ region we measured.
It means that finite lattice spacing effects in $f_+$ seem small.
A similar consistency is observed in $f_0$ near $q^2 = 0$,
while a little difference is seen in a negative $q^2$ region.
In a later section, we will chose a fit form of the finite lattice spacing
effects in a continuum extrapolation based on this tendency.

\begin{table}[ht!]
\caption{Results for the form factors $f_+(q^2)$ and $f_0(q^2)$
in each $q^2$ at $\beta = 2.00$ using the local vector current with
the combined analysis, local, and smeared operator data, respectively.
The ratio $\xi(q^2) = f_-(q^2)/f_+(q^2)$
is also presented in the combined analysis.}
\label{tab:kl3_form_factor_b200}
\begin{tabular}{cccccccccc}\hline\hline
& \multicolumn{3}{c}{combined} && \multicolumn{2}{c}{local}&& \multicolumn{2}{c}{smeared}\\
$q^2$ & $f_+(q^2)$   & $f_0(q^2)$ & $\xi(q^2)$    && $f_+(q^2)$ & $f_0(q^2)$ && $f_+(q^2)$ & $f_0(q^2)$ \\\hline
$q^2_0$ & $\cdots$   & 1.0697(17) & $\cdots$      && $\cdots$   & 1.0685(24) && $\cdots$   & 1.0707(17) \\
$q^2_1$ & 1.0850(18) & 1.0279(15) & $-0.1412(11)$ && 1.0829(27) & 1.0270(22) && 1.0866(18) & 1.0288(17) \\
$q^2_2$ & 1.0292(16) & 0.9982(14) & $-0.1400(12)$ && 1.0271(24) & 0.9971(21) && 1.0313(18) & 0.9996(15) \\
$q^2_3$ & 0.9842(15) & 0.9733(14) & $-0.1384(16)$ && 0.9833(24) & 0.9727(22) && 0.9858(18) & 0.9746(16) \\
$q^2_4$ & 0.9485(13) & 0.9537(13) & $-0.1380(33)$ && 0.9481(22) & 0.9533(22) && 0.9493(15) & 0.9546(15) \\
$q^2_5$ & 0.9180(13) & 0.9366(14) & $-0.1360(29)$ && 0.9190(20) & 0.9379(22) && 0.9178(20) & 0.9361(20) \\
$q^2_6$ & 0.8916(14) & 0.9215(16) & $-0.1342(46)$ && 0.8933(22) & 0.9228(26) && 0.8908(19) & 0.9203(23)
\\\hline\hline
\end{tabular}
\end{table}

\clearpage

\subsection{Conserved current result}
\label{sec:data-cns}

As in the local current data discussed in the above,
the form factors at $\beta = 2.00$ are extracted from 
the 3-point functions with the conserved vector current by the same analysis
method described in Sec.~\ref{sec:loc_cur_ana}.
The obtained results for the form factors from the combined analysis
are shown in Table~\ref{tab:kl3_form_factor_b200-cns-cur}
together with those from the local or smeared operator data.
As in the local current case, we confirm that the relative difference 
between the form factors from the local and smeared operators is less than
1.5 $\sigma$. The three results are in good agreement with each other.
The error reduction of the combined analysis compared to the two operator 
data is observed in large $q^2$ region, which might be caused by
a weak correlation between the local and smeared operator data as discussed
in the previous subsection.

At $\beta = 1.82$, the form factors with the conserved vector current
are calculated using only the local operator data as in our previous 
calculation~\cite{PACS:2019hxd}.
The analysis method is the same as that in the previous one,
which is also the same as the local operator data analysis
at $\beta = 2.00$, where $\Delta_\pi(p)$ and $\Delta_K$
in $\Delta A_{\mu,L}$ defined in Eq.~(\ref{eq:def_radial_excited_effect})
are fixed by the experimental values of the first excited meson masses.
We also perform an extra analysis with $\Delta_\pi(p)$ and $\Delta_K$
as free parameters.
In the following the two choices are called ``A1'' and ``A2,'' respectively, 
as in the previous work~\cite{PACS:2019hxd}.
The values for the form factors obtained from both the analyses 
are tabulated in Table~\ref{tab:kl3_form_factor_b182-cns-cur}.
The statistical errors in the A2 analysis are larger than those
in the A1 analysis especially at the largest $q^2$.
The same trend was observed in the local current data 
in Ref.~\cite{PACS:2019hxd}.
The A1 result is used in our main analysis,
while A2 is adopted for estimate of systematic error,
explained in a later section.

\begin{figure}[ht!]
\includegraphics*[scale=.45]{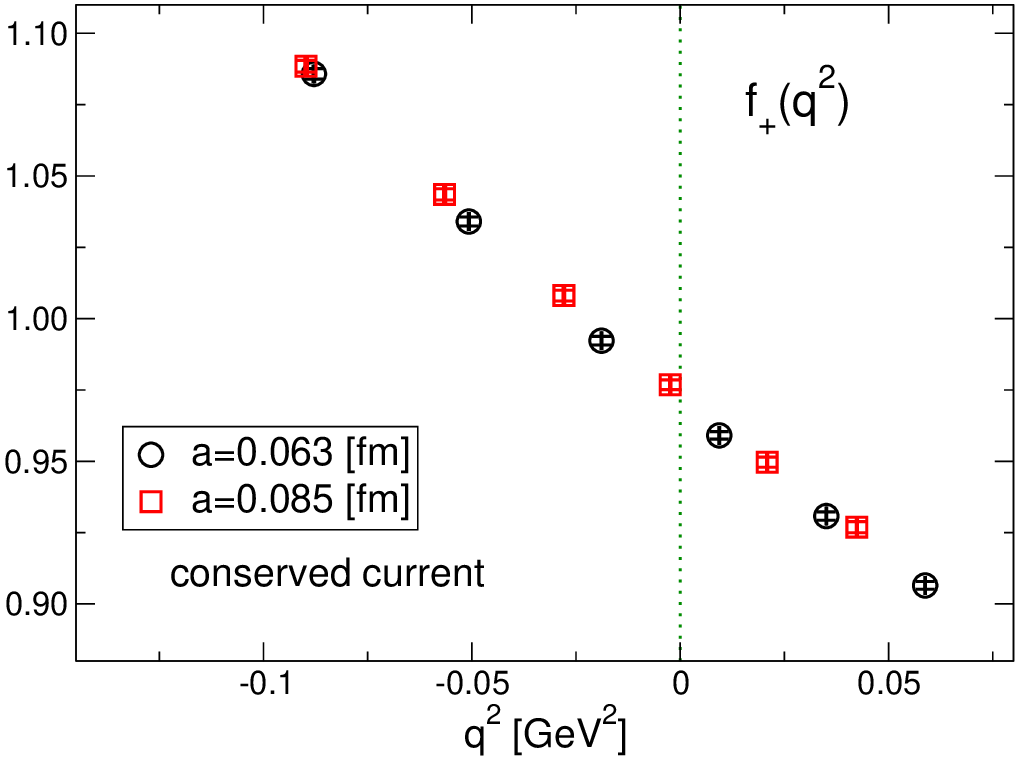}
\includegraphics*[scale=.45]{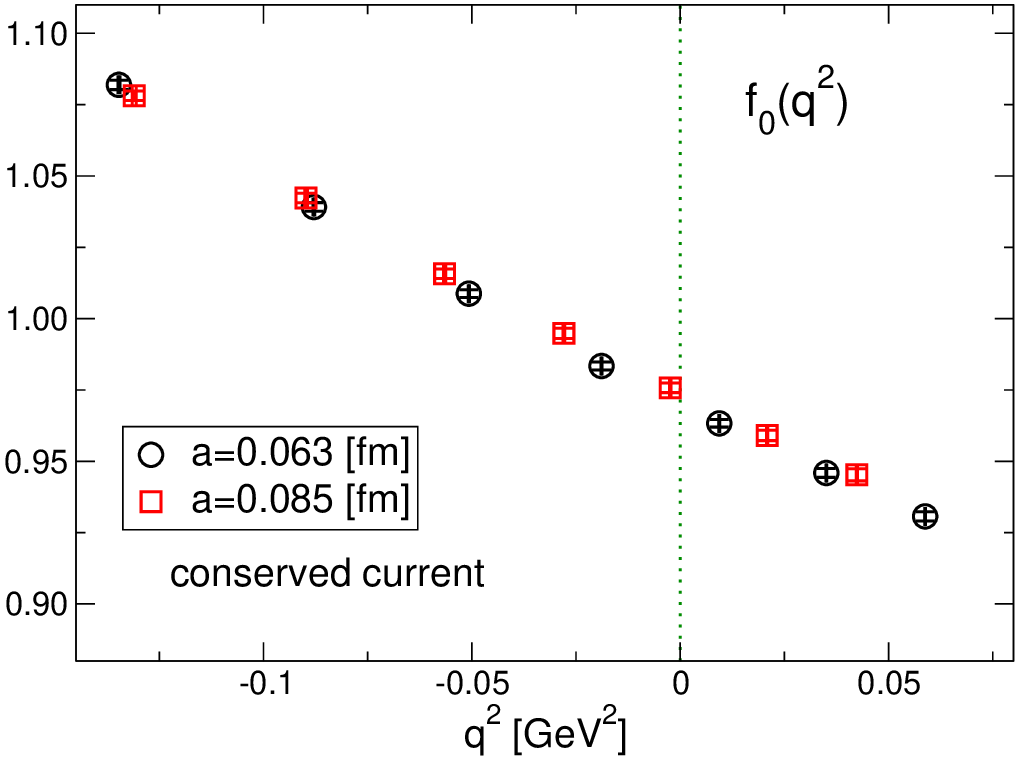}
\caption{The same figure as Fig.~\ref{fig:f+0-comp-lcl-cur},
but for the conserved vector current data.}
\label{fig:f+0-comp-cns-cur}
\end{figure}

The data at the different lattice spacings are compared 
in each form factor as shown in Fig.~\ref{fig:f+0-comp-cns-cur}.
In contrast to the local current data in Fig.~\ref{fig:f+0-comp-lcl-cur},
small finite lattice spacing effects are observed in the vicinity of
$q^2 = 0$ for both the form factors, and
these effects seem to increase with $q^2$ in both the form factors.

\begin{table}[ht!]
\caption{The same table as Table~\ref{tab:kl3_form_factor_b200},
but for the conserved vector current at $\beta = 2.00$.}
\label{tab:kl3_form_factor_b200-cns-cur}
\begin{tabular}{cccccccccc}\hline\hline
& \multicolumn{3}{c}{combined} && \multicolumn{2}{c}{local}&& \multicolumn{2}{c}{smeared}\\
$q^2$ & $f_+(q^2)$   & $f_0(q^2)$ & $\xi(q^2)$    && $f_+(q^2)$ & $f_0(q^2)$ && $f_+(q^2)$ & $f_0(q^2)$ \\\hline
$q^2_0$ & $\cdots$   & 1.0819(17) & $\cdots$      && $\cdots$   & 1.0807(24) && $\cdots$   & 1.0828(17) \\
$q^2_1$ & 1.0858(18) & 1.0391(15) & $-0.1153(11)$ && 1.0837(26) & 1.0382(23) && 1.0873(18) & 1.0400(17) \\
$q^2_2$ & 1.0340(16) & 1.0088(14) & $-0.1137(13)$ && 1.0320(24) & 1.0077(21) && 1.0361(18) & 1.0101(15) \\
$q^2_3$ & 0.9923(15) & 0.9834(14) & $-0.1118(16)$ && 0.9913(24) & 0.9828(22) && 0.9938(18) & 0.9845(17) \\
$q^2_4$ & 0.9591(13) & 0.9633(13) & $-0.1105(33)$ && 0.9587(23) & 0.9629(23) && 0.9599(15) & 0.9642(15) \\
$q^2_5$ & 0.9308(15) & 0.9459(16) & $-0.1092(29)$ && 0.9317(20) & 0.9471(22) && 0.9306(20) & 0.9454(19) \\
$q^2_6$ & 0.9065(14) & 0.9307(16) & $-0.1072(46)$ && 0.9078(22) & 0.9317(27) && 0.9058(20) & 0.9298(24)
\\\hline\hline
\end{tabular}
\end{table}

\begin{table}[ht!]
\caption{Results for the form factors $f_+(q^2)$ and $f_0(q^2)$
in each $q^2$ at $\beta = 1.82$ using the conserved vector current.
The meaning for the A1 and A2 analyses are the same 
as in Ref.~\cite{PACS:2019hxd}.
The ratio $\xi(q^2) = f_-(q^2)/f_+(q^2)$
is also presented in the A1 analysis.}
\label{tab:kl3_form_factor_b182-cns-cur}
\begin{tabular}{ccccccc}\hline\hline
& \multicolumn{3}{c}{A1} && \multicolumn{2}{c}{A2}\\
$q^2$   & $f_+(q^2)$ & $f_0(q^2)$ & $\xi(q^2)$    && $f_+(q^2)$ & $f_0(q^2)$ \\\hline
$q^2_0$ & $\cdots$   & 1.0781(17) & $\cdots$      && $\cdots$   & 1.0785(16) \\
$q^2_1$ & 1.0885(21) & 1.0423(16) & $-0.1082(14)$ && 1.0893(24) & 1.0426(17) \\
$q^2_2$ & 1.0435(20) & 1.0157(17) & $-0.1078(20)$ && 1.0459(32) & 1.0167(20) \\
$q^2_3$ & 1.0081(19) & 0.9948(18) & $-0.1078(25)$ && 1.0108(36) & 0.9966(27) \\
$q^2_4$ & 0.9768(17) & 0.9758(17) & $-0.1059(32)$ && 0.9779(25) & 0.9767(25) \\
$q^2_5$ & 0.9497(18) & 0.9591(18) & $-0.1083(33)$ && 0.9506(22) & 0.9604(25) \\
$q^2_6$ & 0.9269(19) & 0.9453(21) & $-0.1070(46)$ && 0.9304(45) & 0.9521(76) 
\\\hline\hline
\end{tabular}
\end{table}

\subsection{Interpolation in finite lattice spacing}
\label{sec:interpolate-each}

Before discussing the continuum extrapolation 
using the data at the two lattice spacings, 
a $q^2$ interpolation at each lattice spacing is performed to investigate 
a lattice spacing dependence of each form factor in this subsection.
We analyze the data using the combined and A1 analyses at $\beta = 2.00$ 
and 1.82, respectively, in both the local and conserved currents.
Those values are presented above, while the local current data
at $\beta = 1.82$ are reported in Table~II of Ref.~\cite{PACS:2019hxd}.

An interpolation of the form factors to $q^2 = 0$ is performed
with the same fitting forms as in our previous work~\cite{PACS:2019hxd}.
They are based on the NLO SU(3) ChPT, and we add correction terms as,
\begin{eqnarray}
f_+(q^2) &=& 1 - \frac{4}{F_0^2} L_9 q^2 + K_+(q^2) + c_0 + c_2^+ q^4 ,
\label{eq:NLO_chpt_f+}\\
f_0(q^2) &=& 1 - \frac{8}{F_0^2} L_5 q^2 + K_0(q^2) + c_0 + c_2^0 q^4 ,
\label{eq:NLO_chpt_f0}
\end{eqnarray}
where $F_0$ is the pion decay constant\footnote{We adopt the normalization of $F_\pi \sim 132$ MeV at the physical point.} in the SU(3) chiral limit.
The NLO functions $K_+(q^2)$ and $K_0(q^2)$ are given 
in Refs.~\cite{Gasser:1984ux,Gasser:1984gg}, which 
depend on $m_\pi$, $m_K$, $q^2$, $F_0$, and the scale $\mu = 0.77$ GeV.
Their explicit forms are shown in Appendix A of Ref.~\cite{PACS:2019hxd},
and they satisfy $K_+(0) = K_0(0)$.
$L_9$, $L_5$, $c_0$, and $c^{+,0}_{2}$ are free parameters in a fit.
A common $c_0$ is employed to satisfy
the constraint of $f_+(0) = f_0(0)$.
The last two terms in both the equations 
may be considered as a part of the NNLO analytic terms in the SU(3) ChPT.

\begin{figure}[ht!]
\includegraphics*[scale=.45]{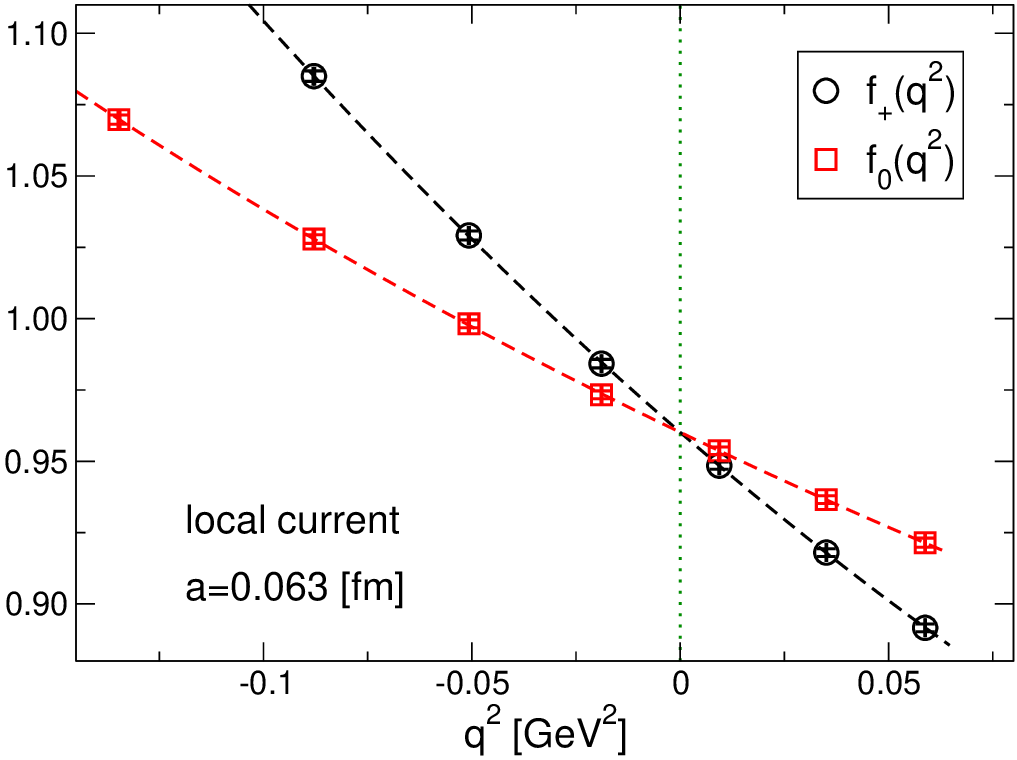}
\caption{Interpolation of $K_{\ell 3}$ form factors $f_+(q^2)$
and $f_0(q^2)$ to $q^2=0$ with fit forms
based on the NLO SU(3) ChPT formulas in 
Eqs.~(\ref{eq:NLO_chpt_f+}) and (\ref{eq:NLO_chpt_f0}) using
the local vector current data at $\beta = 2.00$.
The dashed curves express results from a simultaneous fit using 
$f_+(q^2)$ and $f_0(q^2)$ data.}
\label{fig:fit-f+0-b200-lcl-cur}
\end{figure}

\begin{figure}[ht!]
\includegraphics*[scale=.45]{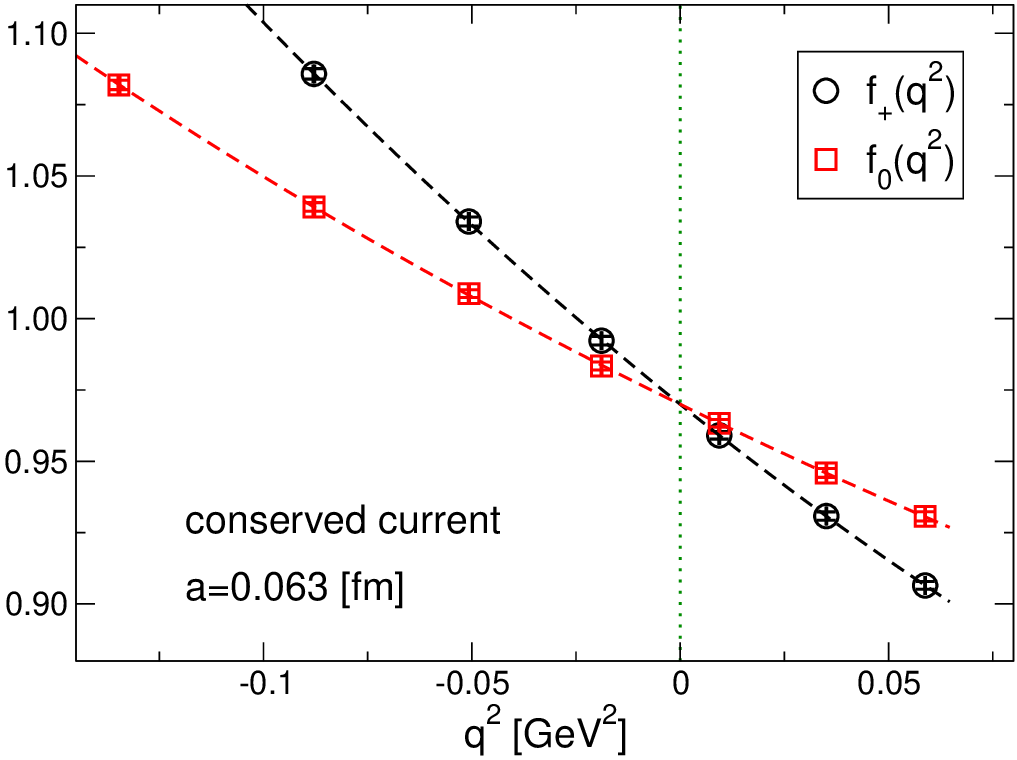}
\includegraphics*[scale=.45]{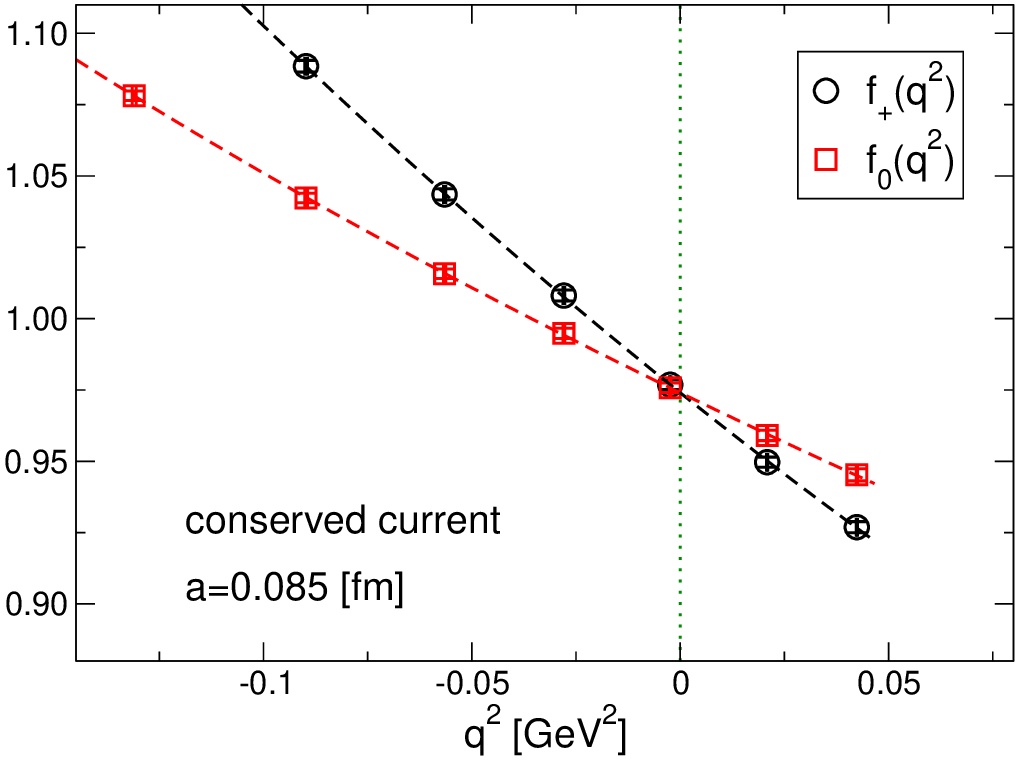}
\caption{The same figure as Fig.~\ref{fig:fit-f+0-b200-lcl-cur},
but using the conserved vector current data at $\beta = 2.00$ (left)
and 1.82 (right).}
\label{fig:fit-f+0-cns-cur}
\end{figure}

We employ a fixed value of $F_0 = 0.11205$ GeV~\footnote{$F_0$ is determined from the average of ratios, $F/F_0 = 1.229(59)$~\cite{Allton:2008pn} and $F/F_0 = 1.078(44)$~\cite{Aoki:2008sm}, and the pion decay constant of the SU(2) ChPT in the chiral limit $F = 0.12925$ GeV, which was calculated at $\beta = 1.82$~\cite{Ishikawa:2015rho,Kakazu:2017fhv}.} in interpolations as in Ref.~\cite{PACS:2019hxd}.
The fit results are presented in Figs.~\ref{fig:fit-f+0-b200-lcl-cur} 
and \ref{fig:fit-f+0-cns-cur} for the local current at $\beta = 2.00$ and 
the conserved current at both $\beta$, respectively.
All the fits well explain our data.
The values for the fit parameters and the uncorrelated $\chi^2/$dof 
are tabulated in Table~\ref{tab:fit_chpt-lcl-cur} together with the results
obtained from the same fit of the local current data at 
$\beta = 1.82$~\cite{PACS:2019hxd} for comparison.
In the local current, the results for $f_+(0)$ and $L_9$ are well
consistent at the different lattice spacings, which
is expected from the consistency of $f_+(q^2)$ at the different lattice spacings
presented in the left panel of Fig.~\ref{fig:f+0-comp-lcl-cur}.

Using the fit results a tiny chiral extrapolation to the physical
point, $m_{\pi^-} = 0.13957061$ GeV and $m_{K^0} = 0.497611$ GeV, 
is carried out at each data.
The value at the physical point
is evaluated by replacing $m_\pi$ and $m_K$ in the functions $K_+$ and $K_0$
in Eqs.~(\ref{eq:NLO_chpt_f+}) and (\ref{eq:NLO_chpt_f0})
by $m_{\pi^-}$ and $m_{K^0}$.
Due to the short extrapolation,
we neglect the mass dependences of the last two terms
in Eqs.~(\ref{eq:NLO_chpt_f+}) and (\ref{eq:NLO_chpt_f0}).
The evaluated value of $f_+(0)$ at the physical point is presented 
in Table~\ref{tab:f0-slope-curvature-each-beta}.
Comparing with the values of $f_+(0)$ at each $\beta$
in between Tables~\ref{tab:fit_chpt-lcl-cur} and 
\ref{tab:f0-slope-curvature-each-beta}, it is found that the corrections 
of the chiral extrapolation are as small as or less than the statistical error.
The slope and curvature of the form factors defined by
\begin{eqnarray}
\lambda^\prime_s &=& \frac{m_{\pi^-}^2}{f_+(0)} \left. \frac{d f_s(q^2)}{d (-q^2)}\right|_{q^2=0},\label{eq:f_slope}\\
\lambda^{\prime\prime}_s &=& \frac{m_{\pi^-}^4}{f_+(0)} \left. \frac{d^2 f_s(q^2)}{d (-q^2)^2}\right|_{q^2=0}, \label{eq:f_curvature}
\end{eqnarray}
for $s = +, 0$ are also calculated using the fit results
at the physical point.
Those results are compiled in Table~\ref{tab:f0-slope-curvature-each-beta}.

\begin{figure}[ht!]
\includegraphics*[scale=.45]{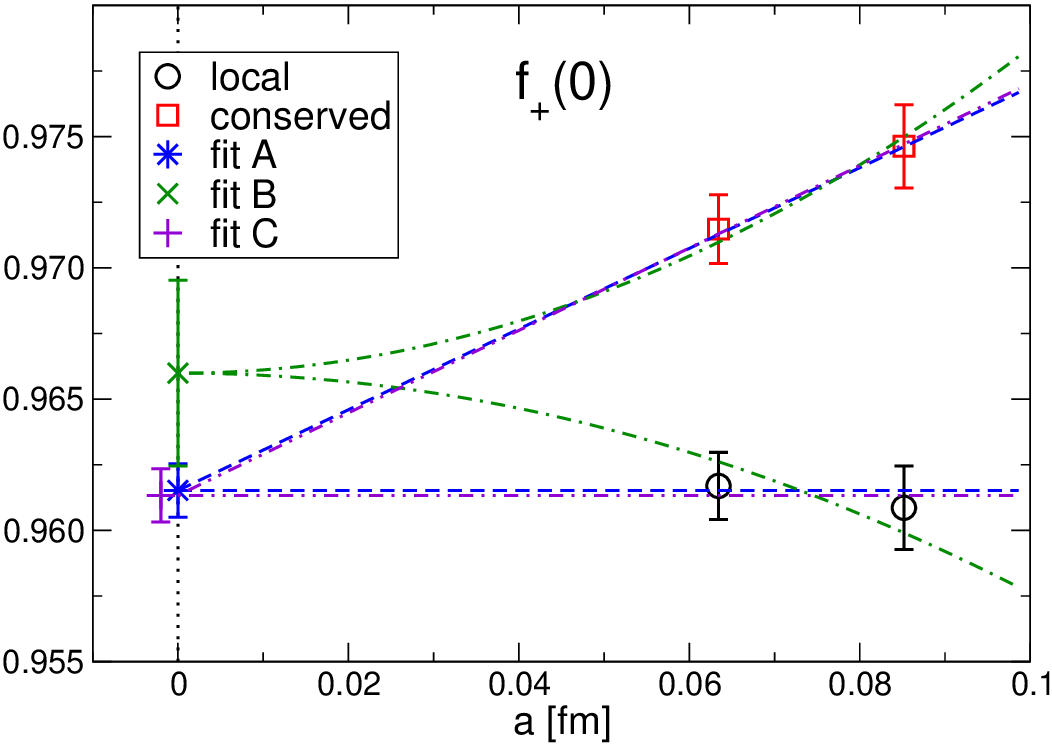}
\caption{Lattice spacing dependence of $f_+(0)$, whose values are 
evaluated from a fit result at each lattice spacing using
the local (circle) and conserved (square) vector current data.
As explained in Sec.~\ref{sec:continuum_extrapolation}, 
fit lines evaluated from 
continuum extrapolations for fit A, B, C are also presented.
}
\label{fig:fit-f+-q0-a0}
\end{figure}

Here, we discuss the lattice spacing dependence of the physical
quantities listed in Table~\ref{tab:f0-slope-curvature-each-beta}.
Each result of $f_+(0)$ is plotted in Fig.~\ref{fig:fit-f+-q0-a0} as 
a function of the lattice spacing $a$.
The result of the local current has a mild $a$ dependence,
while that of the conserved current almost behaves as a linear function.
A similar tendency was observed in 
the hadron vacuum polarization calculation with
the local and conserved vector currents on
the PACS10 configurations~\cite{Shintani:2019wai}.
It is expected that the value in the continuum limit could be
in between the two current data at the smaller lattice spacing,
and the two current data converge in the continuum limit.
Based on this observation, we will choose fitting forms of 
the form factor for the continuum extrapolation in the next section.

\begin{figure}[ht!]
\includegraphics*[scale=.45]{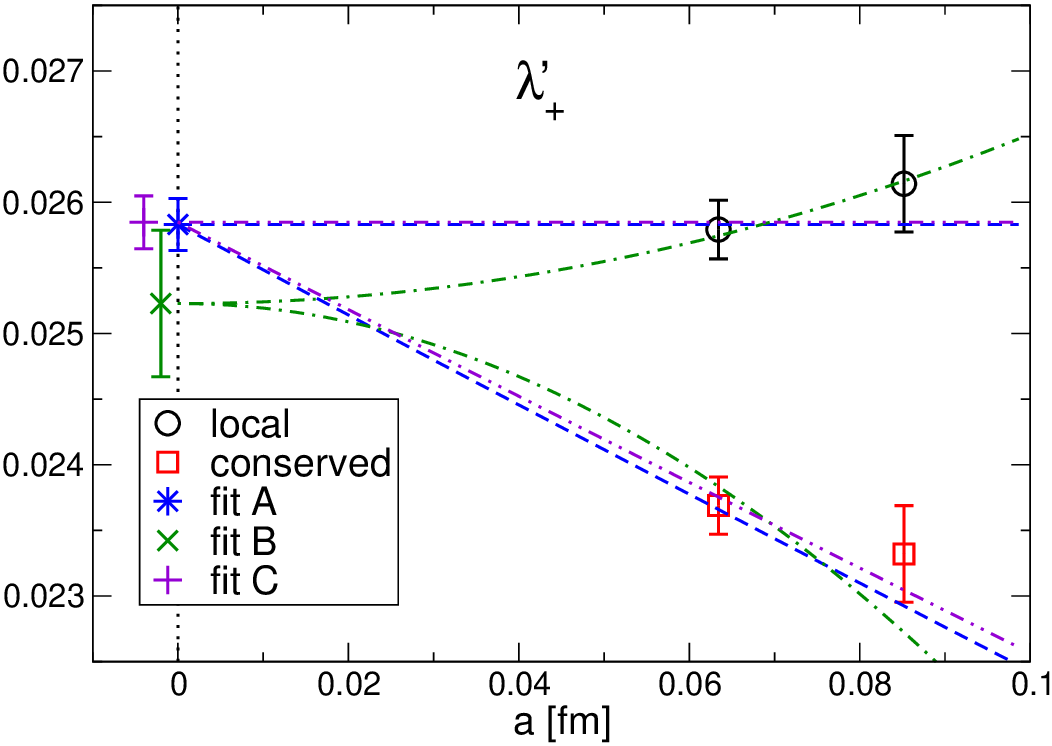}
\includegraphics*[scale=.45]{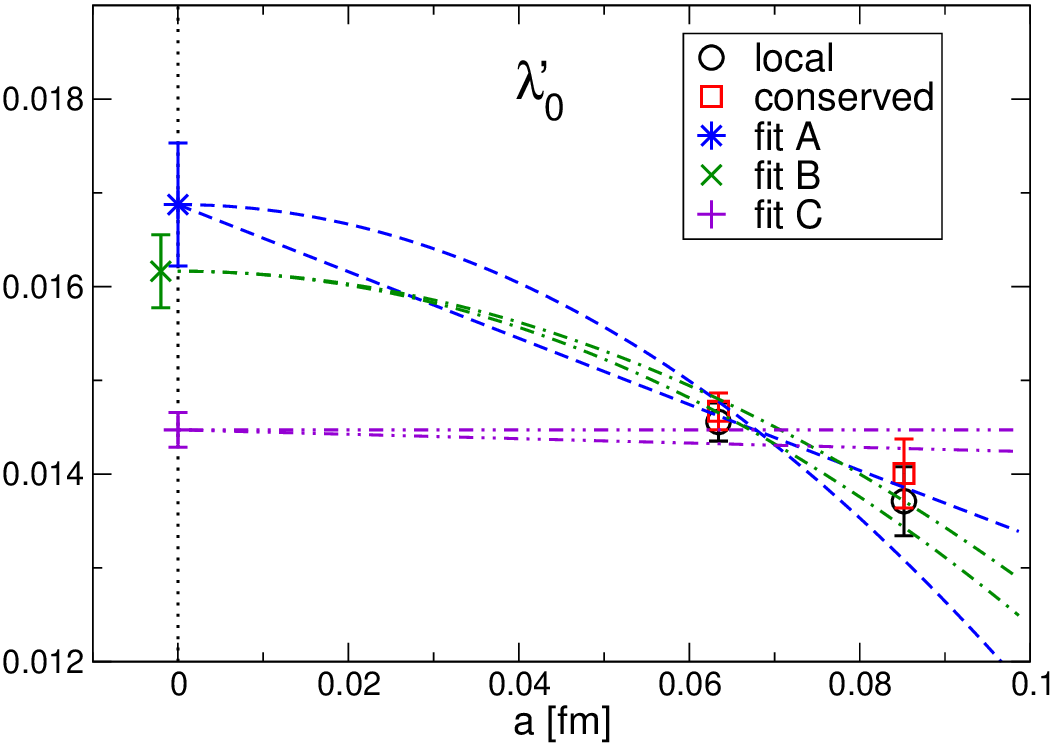}
\caption{The same figure as as Fig.~\ref{fig:fit-f+-q0-a0},
but for $\lambda^\prime_+$ (left) and $\lambda^\prime_0$ (right).}
\label{fig:lam+0-a0}
\end{figure}

\begin{figure}[ht!]
\includegraphics*[scale=.45]{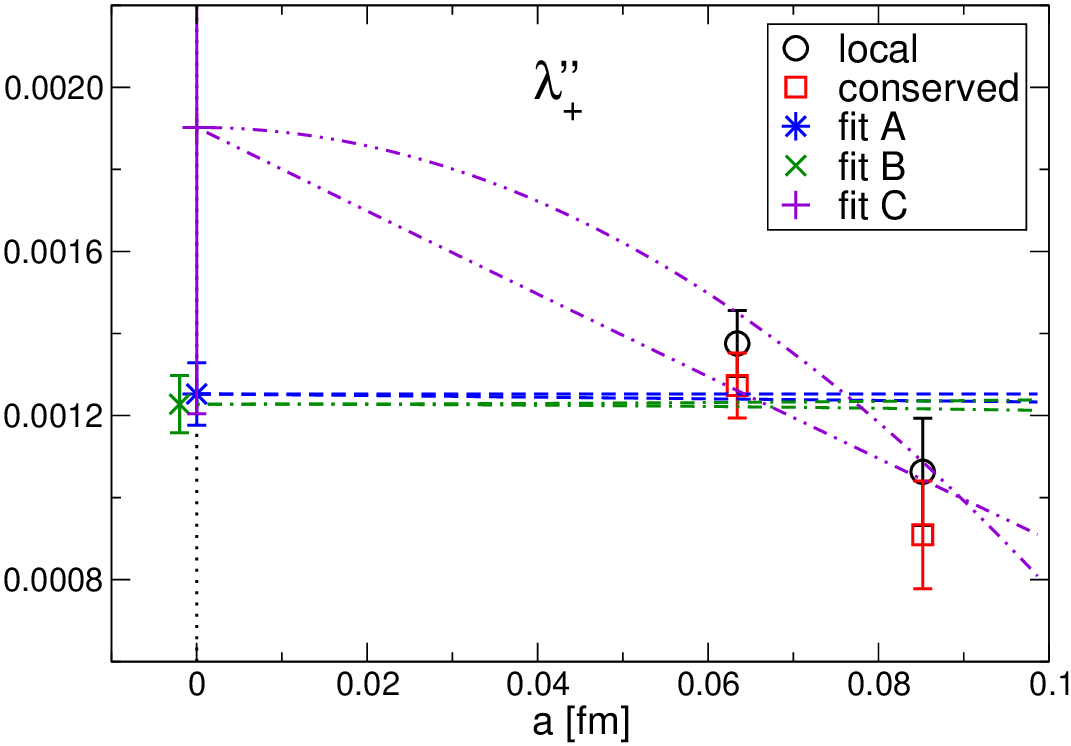}
\includegraphics*[scale=.45]{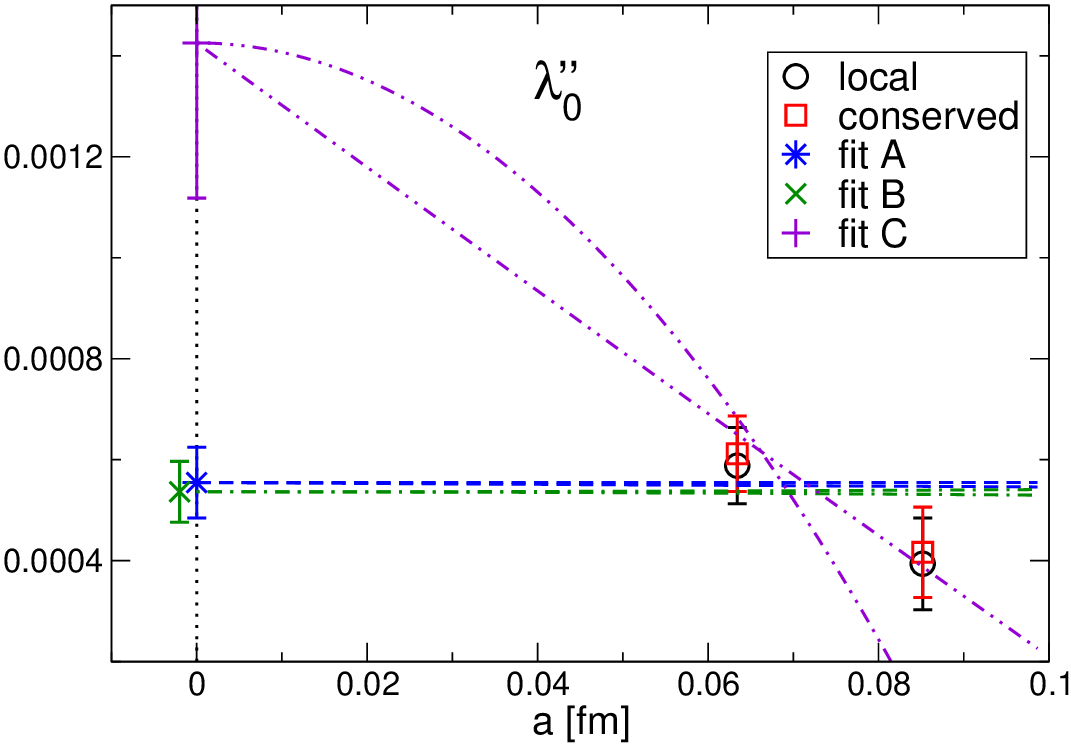}
\caption{The same figure as as Fig.~\ref{fig:fit-f+-q0-a0},
but for $\lambda^{\prime\prime}_+$ (left) and 
$\lambda^{\prime\prime}_0$ (right).}
\label{fig:dlam+0-a0}
\end{figure}

A similar split between the local and conserved current data 
is observed in the result of $\lambda_+^\prime$
shown in the left panel of Fig.~\ref{fig:lam+0-a0}.
In contrast to $f_+(0)$ and $\lambda_+'$, 
the difference between the local and conserved current results
is not clear at each lattice spacing as presented in
the right panel of Fig.~\ref{fig:lam+0-a0} for $\lambda_0^\prime$,
and Fig.~\ref{fig:dlam+0-a0} for $\lambda_+^{\prime\prime}$
and $\lambda_0^{\prime\prime}$.
These three quantities have large fitting form dependences
in the continuum extrapolation,
which will be discussed in the next section.

\begin{table}[ht!]
\caption{Fit results of $K_{\ell 3}$ form factors
based on the NLO SU(3) ChPT formulas in 
Eqs.~(\ref{eq:NLO_chpt_f+}) and (\ref{eq:NLO_chpt_f0}) 
with the local and conserved vector currents at each $\beta$
together with the value of the uncorrelated $\chi^2/$dof.
$F_0$ is fixed in all the fits.
The result with the local current at $\beta = 1.82$ was presented in Ref.~\cite{PACS:2019hxd}.
}
\label{tab:fit_chpt-lcl-cur}
\begin{tabular}{ccccc}\hline\hline
current& \multicolumn{2}{c}{local} & \multicolumn{2}{c}{conserved}\\
$\beta$              & 2.00          & 1.82          & 2.00         & 1.82 \\\hline 
$f_+(0)$             & 0.9602(13)    & 0.9603(16)    & 0.9700(13)   & 0.9740(16)   \\
$L_9$ [$10^{-3}$]    & 3.873(37)     & 3.924(57)     & 3.585(36)    & 3.539(58)    \\
$L_5$ [$10^{-4}$]    & 7.60(16)      & 6.94(28)      & 7.80(15)     & 7.32(29)     \\
$c_2^+$ [GeV$^{-4}$] & 1.59(10)      & 1.19(17)      & 1.48(10)     & 1.01(17)     \\
$c_2^0$ [GeV$^{-4}$] & $-0.152(96)$  & $-0.40(11)$   & $-0.114(96)$ & $-0.36(11)$  \\
$c_0$                & $-0.0068(13)$ & $-0.0077(16)$ & $0.0029(13)$ & $0.0061(16)$ \\
$\chi^2/$dof         & 0.05          & 0.05          & 0.05         & 0.05         
\\\hline\hline
\end{tabular}
\end{table}

\begin{table}[ht!]
\caption{Results for $f_+(0)$, $\lambda_s^\prime$,
and $\lambda_s^{\prime\prime}$ for $s = +, 0$ at the physical point
at each $\beta$, evaluated from the fit results shown in Table~\ref{tab:fit_chpt-lcl-cur}.
}
\label{tab:f0-slope-curvature-each-beta}
\begin{tabular}{ccccc}\hline\hline
current& \multicolumn{2}{c}{local} & \multicolumn{2}{c}{conserved}\\
$\beta$                                & 2.00       & 1.82       & 2.00       & 1.82 \\\hline 
$f_+(0)$                               & 0.9617(13) & 0.9609(16) & 0.9715(13) & 0.9746(16) \\
$\lambda_+^\prime$ [$10^{-2}$]         & 2.579(22)  & 2.614(37)  & 2.369(22)  & 2.332(37)  \\
$\lambda_0^\prime$ [$10^{-2}$]         & 1.456(20)  & 1.371(37)  & 1.467(20)  & 1.401(37)  \\
$\lambda_+^{\prime\prime}$ [$10^{-3}$] & 1.376(80)  & 1.06(13)   & 1.273(79)  & 0.91(13)   \\
$\lambda_0^{\prime\prime}$ [$10^{-3}$] & 0.588(76)  & 0.394(91)  & 0.612(75)  & 0.417(89) 
\\\hline\hline
\end{tabular}
\end{table}

\clearpage

\section{Result in the continuum limit}
\label{sec:continuum_limit}

In this section continuum extrapolations of the form factors are
discussed with the results from the local and conserved vector currents
at the two lattice spacings.
In our calculation the main result is obtained from the data using the combined
and A1 analyses at $\beta = 2.00$ and 1.82, respectively,
whose values are explained in the last section.
The data obtained from only the local or smeared operator at $\beta = 2.00$,
and from the A2 analysis at $\beta = 1.82$ are used for a systematic
error estimation discussed below.

\subsection{Continuum extrapolation}
\label{sec:continuum_extrapolation}

A continuum extrapolation is carried out using all the data
we calculated: the two form factors $f_+(q^2)$ and $f_0(q^2)$ with
the local and conserved currents at the two lattice spacings.
We adopt fit forms based on the NLO SU(3) ChPT formula
given in Eqs.~(\ref{eq:NLO_chpt_f+}) and (\ref{eq:NLO_chpt_f0}),
and add further correction terms corresponding to 
finite lattice spacing effects.
As discussed in Sec.~\ref{sec:result_finite_a},
the effect is different in $f_+(q^2)$ and $f_0(q^2)$,
and also in the local and conserved currents.
Thus, we introduce functions $g^{\rm cur}_+(q^2,a)$ and 
$g^{\rm cur}_0(q^2,a)$ to incorporate the lattice spacing dependence 
in fit forms given as
\begin{eqnarray}
f_+^{\rm cur}(q^2) &=& 
1 - \frac{4}{F_0^2} L_9 q^2 + K_+(q^2) + c_0 + c_2^+ q^4
+ g^{\rm cur}_+(q^2,a),
\label{eq:ato0_NLO_chpt_f+}\\
f_0^{\rm cur}(q^2) &=& 
1 - \frac{8}{F_0^2} L_5 q^2 + K_0(q^2) + c_0 + c_2^0 q^4
+ g^{\rm cur}_0(q^2,a),
\label{eq:ato0_NLO_chpt_f0}
\end{eqnarray}
where ${\rm cur} = {\rm loc, con}$ correspond to the local and 
conserved currents, respectively.
The same value of $F_0$ is adopted as in Sec.~\ref{sec:interpolate-each}.

From the observations of the finite lattice spacing effects
in each form factor discussed in the previous section, 
we empirically employ the following forms for
$g_{+,0}^{\rm cur}(q^2,a)$ as,
\begin{eqnarray}
{\rm fit\ A:}\ \ \ 
g_+^{\rm loc}(q^2,a) &=& 0,
\label{eq:gterm_f+-lcl_fitA}\\
g_0^{\rm loc}(q^2,a) &=& d_{21}^0 a^2 q^2,
\label{eq:gterm_f0-lcl_fitA}\\
g_+^{\rm con}(q^2,a) &=& e_{10} a + e_{11}^+ a q^2,
\label{eq:gterm_f+-cns_fitA}\\
g_0^{\rm con}(q^2,a) &=& e_{10} a + e_{11}^0 a q^2 ,
\label{eq:gterm_f0-cns_fitA}
\end{eqnarray}
where $d_{21}^0$, $e_{10}$, $e_{11}^{+,0}$ are free parameters.
Our calculation is performed with a nonperturbative 
$O(a)$-improved quark action, while we employ unimproved vector currents
in the calculation of the 3-point functions.
It means that $O(a)$ contributions could appear in our form factor data.
We observe that $f_+(0)$ with the conserved current
approximately behaves as a linear function of $a$ in 
Fig.~\ref{fig:fit-f+-q0-a0}.
In contrast to the conserved current data, 
$f_+(0)$ with the local current is reasonably flat
as shown in the same figure.
Furthermore, the lattice spacing dependence is invisible in $f_+(q^2)$
as shown in the left panel of Fig~\ref{fig:f+0-comp-lcl-cur}.
From the observations, we assume that $O(a)$ effects are well suppressed
in the local current data.
Thus, we choose the above forms for $g_{+,0}^{\rm loc}$ and $g_{+,0}^{\rm con}$
to explain the lattice spacing dependences for each form factor.
Note that $f_+(0) = f_0(0)$ is satisfied in the fit forms 
for each current and lattice spacing.
We will call this choice of the fit forms as ``fit A'' in the following.
The statistical error of the fit with the different lattice spacing data
is estimated by an extension of the jackknife method explained in 
Appendix B of Ref.~\cite{CP-PACS:2001vqx}.

\begin{figure}[ht!]
\includegraphics*[scale=.45]{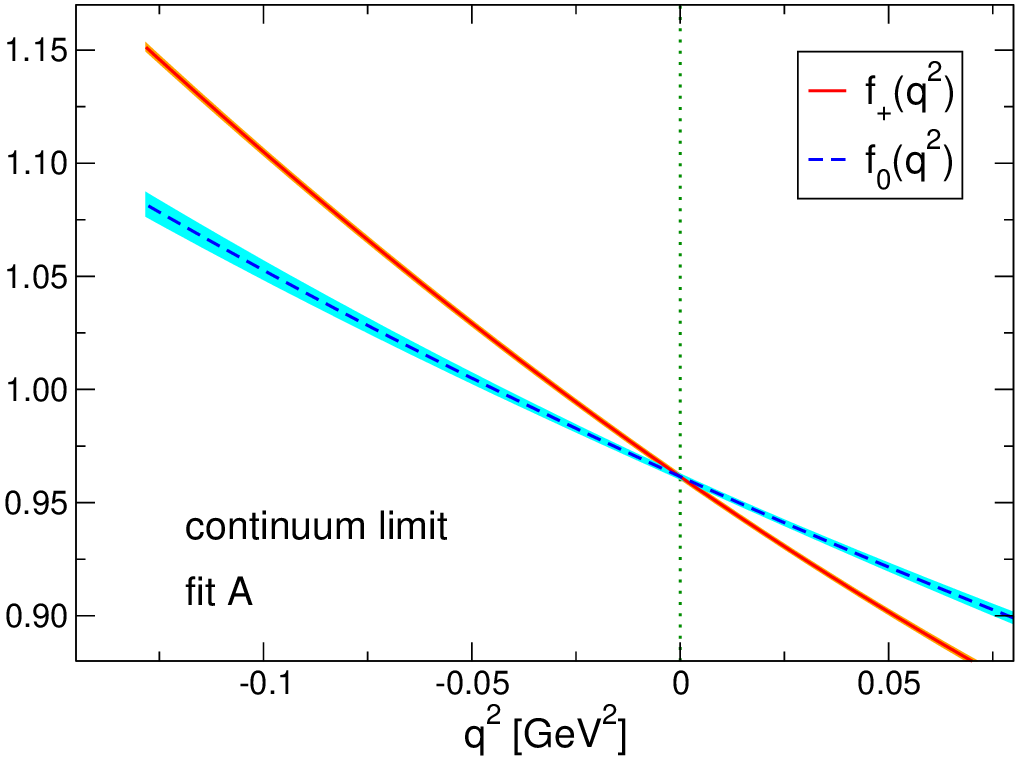}
\caption{The results for $f_+(q^2)$ and $f_0(q^2)$ in the continuum limit
and physical point obtained from fit A.
The solid and dashed curves denote $f_+(q^2)$ and $f_0(q^2)$, respectively.
Their bands correspond to the statistical errors.}
\label{fig:fit-f+0-a0}
\end{figure}

Figure~\ref{fig:fit-f+0-a0-lcl-cns-cur} shows that our data for both the form
factors are well fitted by the fit A form.
The form factors in the continuum limit at the physical point
are obtained from the fit result, where
the values at the physical point are evaluated in the same way
as explained in Sec.~\ref{sec:interpolate-each}.
The extrapolated results for $f_+(q^2)$ and $f_0(q^2)$ in the continuum limit
at the physical point are plotted in Fig.~\ref{fig:fit-f+0-a0} with 
only the statistical error.
The values for the fit parameters are given in 
Table~\ref{tab:fit_chpt-cont-limit} together with the
uncorrelated $\chi^2$/dof value.
It should be noted that while we perform uncorrelated fits,
the statistical error is properly estimated, because
the correlation among the data at the same $\beta$ is taken into
account in the jackknife analysis.
This fit gives $f_+(0) = 0.9615(10)$ in the continuum limit at
the physical point, which is well consistent with
the results of the local current at each lattice spacing listed in 
Table~\ref{tab:f0-slope-curvature-each-beta}.
The slopes and curvatures for the form factors defined in 
Eqs.~(\ref{eq:f_slope}) and (\ref{eq:f_curvature})
in the continuum limit are also obtained from the fit result, whose values are
summarized in Table~\ref{tab:fit_chpt-cont-limit}.
We choose this fit result as our main result in this study.

A more general choice of $g^{\rm cur}_{+,0}$ is examined by assuming
that finite lattice spacing effects start from $O(a^2)$ in
all the data.
The functional forms of $g^{\rm cur}_{+,0}$ with the assumption are given by
\begin{eqnarray}
{\rm fit\ B:}\ \ \ 
g_+^{\rm loc}(q^2,a) &=& d_{20} a^2 + d_{21}^+ a^2 q^2,
\label{eq:gterm_f+-lcl_fitB}\\
g_0^{\rm loc}(q^2,a) &=& d_{20} a^2 + d_{21}^0 a^2 q^2,
\label{eq:gterm_f0-lcl_fitB}\\
g_+^{\rm con}(q^2,a) &=& e_{20} a^2 + e_{21}^+ a^2 q^2,
\label{eq:gterm_f+-cns_fitB}\\
g_0^{\rm con}(q^2,a) &=& e_{20} a^2 + e_{21}^0 a^2 q^2.
\label{eq:gterm_f0-cns_fitB}
\end{eqnarray}
This choice is called ``fit B'' in the following.
The fit results are presented in Table~\ref{tab:fit_chpt-cont-limit}.
The fit curves in $q^2 = 0$ at the physical point for the fit A and B 
are compared in Fig.~\ref{fig:fit-f+-q0-a0} together with 
the data obtained from each current.
Those fit forms well explain the data,
while the result of $f_+(0) = 0.9660(35)$ in fit B is much higher than
that in fit A.
This discrepancy will be included in estimation of a systematic error
of $f_+(0)$ discussed later.

The lattice spacing dependences for
the slopes and curvatures at the physical point
evaluated from the fit A and B
are plotted in Figs.~\ref{fig:lam+0-a0} and \ref{fig:dlam+0-a0}.
The fit results of the curvatures are flat, because
in the above fits, it is assumed that the dominant lattice
artifact of the $q^2$ dependence is proportional to $q^2$ 
in all the form factors, except for $f_+(q^2)$ in the local current of fit A.
Thus, we employ another fit form under an assumption
that lattice artifact of $q^4$ is dominant, except
for $f_+(q^2)$ in the conserved current,
where a $q^2$ term is necessary to explain a linear behavior
of the slope as shown in the left panel of Fig.~\ref{fig:lam+0-a0}.
This fit form called ``fit C'' is given by
\begin{eqnarray}
{\rm fit\ C:}\ \ \ 
g_+^{\rm loc}(q^2,a) &=& d_{22}^+ a^2 q^4 ,
\label{eq:gterm_f+-lcl_fitC}\\
g_0^{\rm loc}(q^2,a) &=& d_{22}^0 a^2 q^4,
\label{eq:gterm_f0-lcl_fitC}\\
g_+^{\rm con}(q^2,a) &=& e_{10} a + e_{11}^+ a q^2 + e_{12}^+ a q^4,
\label{eq:gterm_f+-cns_fitC}\\
g_0^{\rm con}(q^2,a) &=& e_{10} a + e_{12}^0 a q^4,
\label{eq:gterm_f0-cns_fitC}
\end{eqnarray}
based on the fit A form.
The fit results for $f_+(0)$ and $\lambda_+^\prime$ agree with 
those of fit A as presented in Table~\ref{tab:fit_chpt-cont-limit} 
and Figs.~\ref{fig:fit-f+-q0-a0} and \ref{fig:lam+0-a0}.
On the other hand, the results for $\lambda_0^\prime$ and the curvatures
largely depend on the choice of the forms of $g^{\rm cur}_{+,0}$.
This is because we have only two different lattice spacings,
and those data are almost degenerate at each lattice spacing
for the three quantities.
For more precise determination for the quantities,
data at the third lattice spacing are important,
and we plan to carry out the calculation.

A reasonable fit can be performed with $F_0$ being a free parameter
in the fit forms Eqs.~(\ref{eq:ato0_NLO_chpt_f+}) and 
(\ref{eq:ato0_NLO_chpt_f0}) without the $c_0$ term.
The fit results with a free $F_0$ for fit A, B, and C are summarized in 
Appendix~\ref{sec:app_fit_results}.
While the fit result of $F_0$ is a little smaller than the fixed value 
used in the above fits, each result of $f_+(0)$ agrees with that using
the fixed $F_0$ fit.

We also carry out continuum extrapolations with monopole,
quadratic, and $z$-parameter expansion~\cite{Bourrely:2008za} 
fit forms using $g^{\rm cur}_{+,0}$ of fit A.
Furthermore, analyses using different data from the main analysis are performed:
the local and conserved current data at $\beta = 2.00$ or $\beta = 1.82$ 
are replaced by different dataset in a continuum extrapolation.
In an analysis, the data at $\beta = 2.00$ are replaced by
the local or smeared operator data described in Sec.~\ref{sec:result_finite_a}.
A similar analysis is also carried out with the A2 data 
instead of the A1 data at $\beta = 1.82$.
For the different data analyses, the NLO SU(3) ChPT fit formulas with fit A 
are employed using the fixed and free $F_0$.
Those results for the continuum extrapolations are summarized in
Appendix~\ref{sec:app_fit_results}, and will be used for 
systematic error estimations discussed below.

\begin{figure}[ht!]
\includegraphics*[scale=.45]{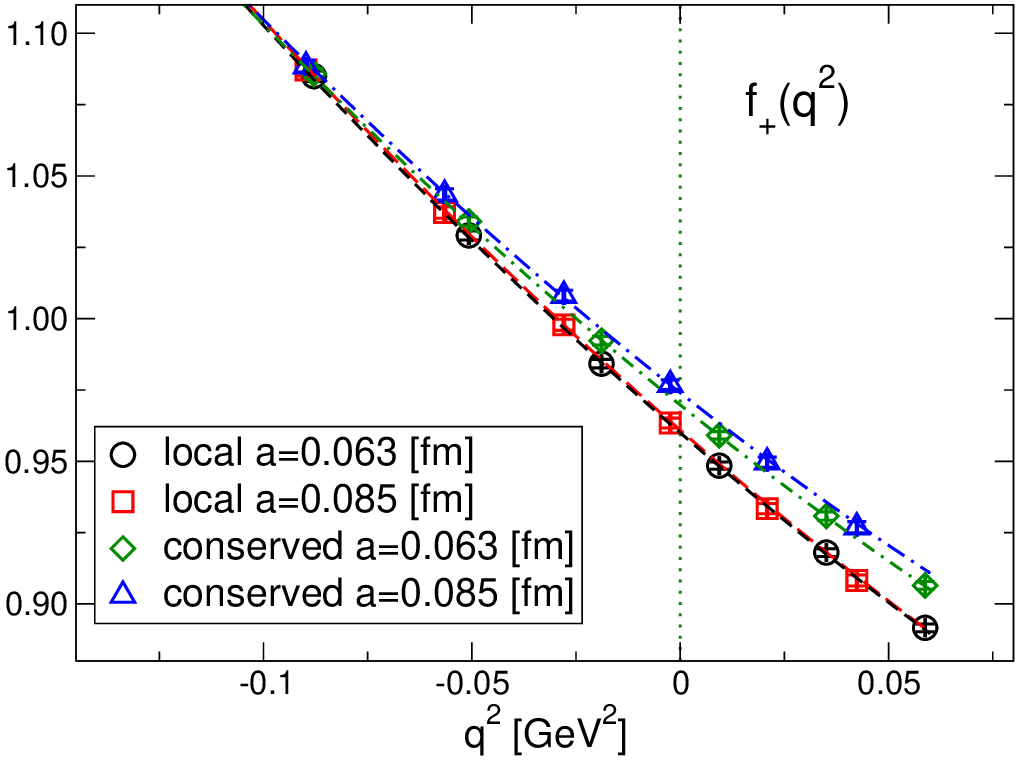}
\includegraphics*[scale=.45]{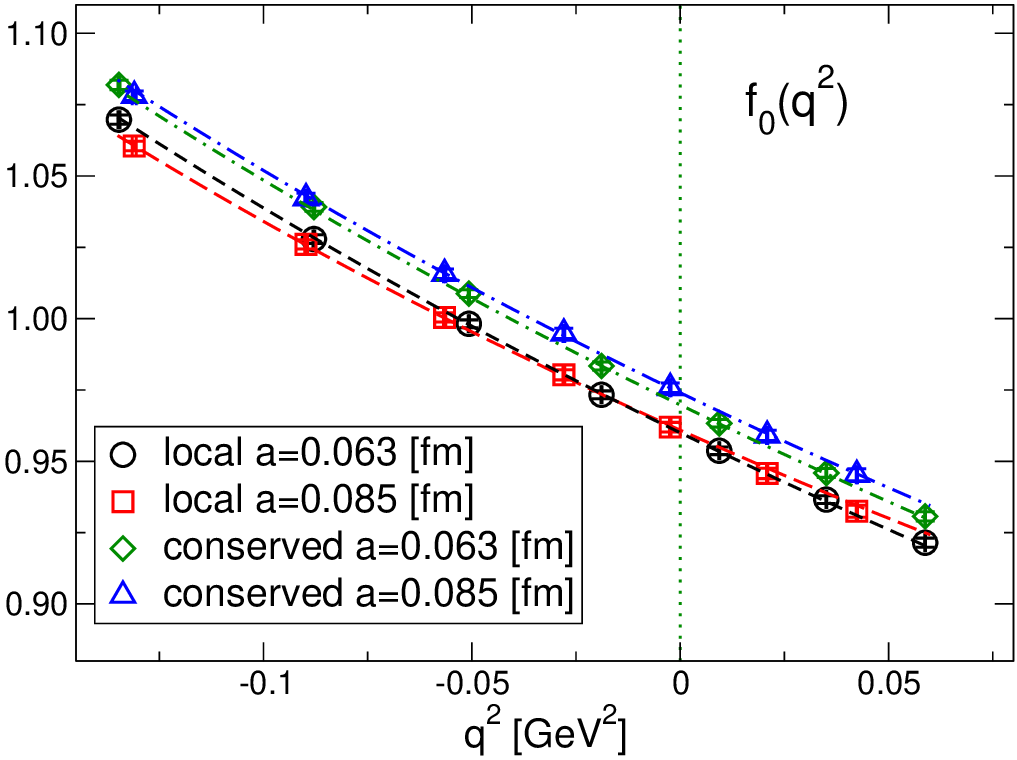}
\caption{Fit results of the continuum extrapolation with fit A,
defined in Eqs.~(\ref{eq:ato0_NLO_chpt_f+}), (\ref{eq:ato0_NLO_chpt_f0}), 
and (\ref{eq:gterm_f+-lcl_fitA})--(\ref{eq:gterm_f0-cns_fitA}), 
for each data of $f_+(q^2)$ (left) and $f_0(q^2)$ (right).
The different symbols express different data as explained in the legend.}
\label{fig:fit-f+0-a0-lcl-cns-cur}
\end{figure}

\begin{table}[ht!]
\caption{Fit results of the continuum extrapolation of $K_{\ell 3}$ form factors
based on the NLO SU(3) ChPT formulas in 
Eqs.~(\ref{eq:ato0_NLO_chpt_f+}) and (\ref{eq:ato0_NLO_chpt_f0}).
The coefficients $d_{nm}^s$ and $e_{nm}^s$ of $a^n q^{2m}$ terms for $s = +,0$
in $g_s^{\rm cur}$ are defined in text.
The values for $f_+(0)$, $\lambda_s^\prime$, and $\lambda_s^{\prime\prime}$ 
in the continuum limit at the physical point are also presented 
together with the value of the uncorrelated $\chi^2/$dof.
$F_0$ is fixed in all the fits.
}
\label{tab:fit_chpt-cont-limit}
\begin{tabular}{cccc}\hline\hline
                        & fit A         & fit B         & fit C         \\\hline
$L_9$ [$10^{-3}$]       & 3.878(32)     & 3.803(90)     & 3.880(32)     \\
$L_5$ [$10^{-4}$]       & 9.39(51)      & 8.90(30)      & 7.53(14)      \\
$c_2^+$ [GeV$^{-4}$]    & 1.435(96)     & 1.410(88)     & 2.26(88)      \\
$c_2^0$ [GeV$^{-4}$]    & $-0.194(89)$  & $-0.215(77)$  & $0.91(39)$    \\
$c_0$                   & $-0.0070(10)$ & $-0.0025(35)$ & $-0.0072(10)$ \\
$\chi^2/$dof            & 0.29          & 0.35          & 0.26          \\
$d_{20}$ [GeV$^2$]      & $\cdots$      & $-0.033(25)$  & $\cdots$      \\
$d_{21}^+$              & $\cdots$      & $-0.020(20)$  & $\cdots$      \\
$d_{21}^0$              & 1.00(27)      & 0.75(15)      & $\cdots$      \\
$d_{22}^+$ [GeV$^{-2}$] & $\cdots$      & $\cdots$      & $-5.5$(6.6)  \\
$d_{22}^0$ [GeV$^{-2}$] & $\cdots$      & $\cdots$      & $-$9.1(2.6)   \\
$e_{10}$ [GeV]          & 0.03032(81)   & $\cdots$      & 0.03099(80)   \\
$e_{11}^+$ [GeV$^{-1}$] & 0.2965(68)    & $\cdots$      & 0.284(12)    \\
$e_{11}^0$ [GeV$^{-1}$] & 0.32(10)      & $\cdots$      & $\cdots$      \\
$e_{12}^+$ [GeV$^{-3}$] & $\cdots$      & $\cdots$      & $-$2.5(2.3)   \\
$e_{12}^0$ [GeV$^{-3}$] & $\cdots$      & $\cdots$      & $-$3.0(1.0)   \\
$e_{20}$ [GeV$^2$]      & $\cdots$      & 0.048(25)     & $\cdots$      \\
$e_{21}^+$              & $\cdots$      & 0.61(20)      & $\cdots$      \\
$e_{21}^0$              & $\cdots$      & 0.62(15)      & $\cdots$      \\
$f_+(0)$                & 0.9615(10)    & 0.9660(35)    & 0.9613(10)    \\
$\lambda_+^\prime$ [$10^{-2}$] & 2.583(20) & 2.523(56) & 2.585(20) \\
$\lambda_0^\prime$ [$10^{-2}$] & 1.687(66) & 1.616(39) & 1.447(19) \\
$\lambda_+^{\prime\prime}$ [$10^{-3}$] & 1.253(76) & 1.228(60) & 1.90(70) \\
$\lambda_0^{\prime\prime}$ [$10^{-3}$] & 0.555(70) & 0.536(60) & 1.43(31) 
\\\hline\hline
\end{tabular}
\end{table}

\clearpage

\subsection{Result of $f_+(0)$}
\label{sec:res_f+0}

\begin{figure}[ht!]
\includegraphics*[scale=.45]{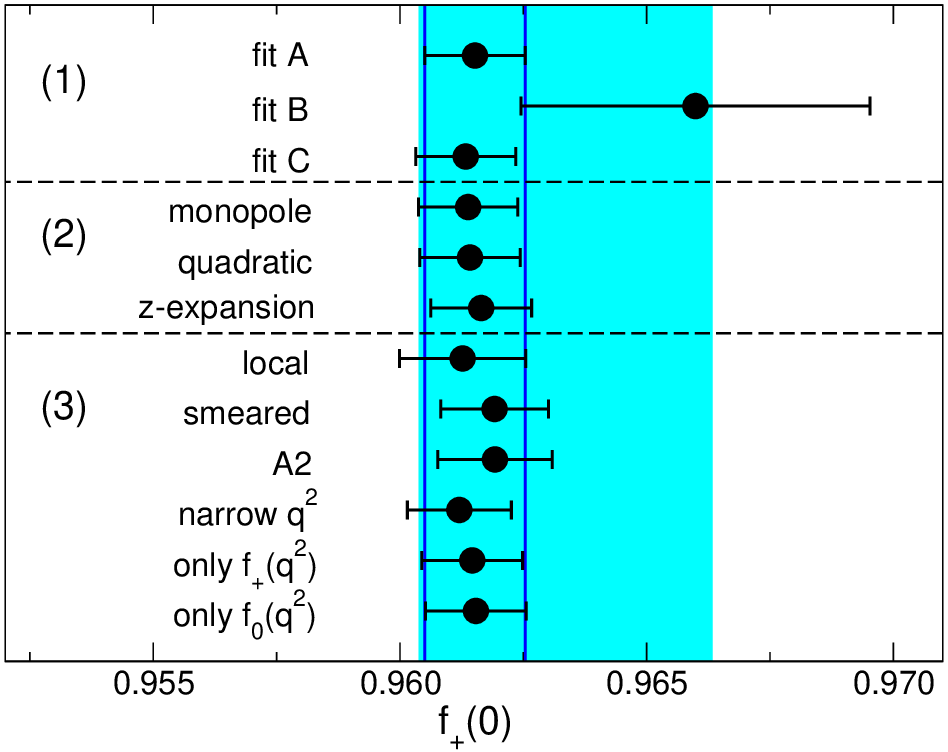}
\caption{Comparison of various fit results of $f_+(0)$.
The vertical solid lines express the statistical error of fit A as
our main result.
The light blue band represents a systematic error corresponding to
difference of fit results.
The (1), (2), and (3) regions express analyses using different fit forms
for continuum extrapolations, different fit forms for $q^2$
interpolations, and different dataset, respectively.
The analyses in the (2) and (3) regions employ the fit A form
for the continuum extrapolation.
All the analyses are described in Sec.~\ref{sec:continuum_extrapolation},
except for the last three, which are explained in Sec.~\ref{sec:res_f+0}.
}
\label{fig:comp_fit-f+-q0}
\end{figure}

Our main result of $f_+(0)$ explained in the last subsection
is compared with results from several analyses
presented in the above and also Appendix~\ref{sec:app_fit_results}.
Figure~\ref{fig:comp_fit-f+-q0} presents the comparison,
where we plot the results obtained from only the fixed $F_0$ analyses
in the NLO SU(3) ChPT fits denoted by fit A, B, C, local, smear, and A2.
The latter three analyses, which are presented in the (2) region of the figure, 
employ the fit A form as explained above.
For the comparison, we also carry out the same analysis as 
the main result (fit A), but with a narrower $q^2$ fit range of
$-0.03 \ {\rm GeV}^2 < q^2 < 0.04 \ {\rm GeV}^2 $ in 
the continuum extrapolation.
Furthermore, continuum extrapolations using only $f_+(q^2)$ or $f_0(q^2)$ data
are performed to study a stability of our result.
All the analyses are consistent with each other,
except for the one from fit B with the larger error.
This discrepancy is caused by the choice of the fit form
for the finite lattice spacing effects as explained in the last subsection.
Note that if we employ fit B instead of fit A in the analyses using
the different fit forms for $q^2$ interpolations and different dataset,
which are plotted in the (2) and (3) regions in Fig.~\ref{fig:comp_fit-f+-q0},
similar results are obtained to the fit B result in the (1) region.
Thus, we include the fit B result in the estimate of the systematic error.

A systematic error of $f_+(0)$ stemming from the choice of the fit forms and 
data is estimated by the maximum difference of the central values among
our main result and others.
Another systematic error from the isospin symmetry breaking effect is also
evaluated by the same analysis as in Ref.~\cite{PACS:2019hxd}.
In the evaluation the NLO ChPT functions $K_+$ and $K_0$ in 
Eqs.~(\ref{eq:ato0_NLO_chpt_f+})
and (\ref{eq:ato0_NLO_chpt_f0}) are replaced by
the ones for $f^{K^0\pi^-}_+$ and $f^{K^0\pi^-}_0$
in the NLO ChPT with the isospin breaking~\cite{Gasser:1984ux,Bijnens:2007xa}~\footnote{The $\pi$-$\eta$ mixing parameter $\epsilon = \sqrt{3}/(4R) = 0.01206$ is employed, which is calculated with the value of $R = 35.9(1.7)$ in the FLAG review~\cite{Aoki:2021kgd}.}.
The functions with the fit results from the main analysis
give $f_+^{K^0\pi^-}(0) = 0.96106$ in the continuum limit at the physical point.
The deviation from our main result is quoted as the systematic error,
although it is much smaller than the other systematic error.
This effect should be estimated by a nonperturbative calculation as 
in the $K_{\ell 2}$ decay~\cite{DiCarlo:2019thl}, but we leave it for
future work.

The physical volume in our calculation is more than $(10 \ {\rm fm})^3$
at both the lattice spacings, so that the systematic error from the 
finite volume effect is considered to be negligible.
A naive estimation of the effect, $e^{-m_\pi L}$, gives 0.08\%.
Since it is much smaller than our statistical error of $f_+(0)$,
we neglect the systematic error in this calculation.

\begin{figure}[ht!]
\includegraphics*[scale=.45]{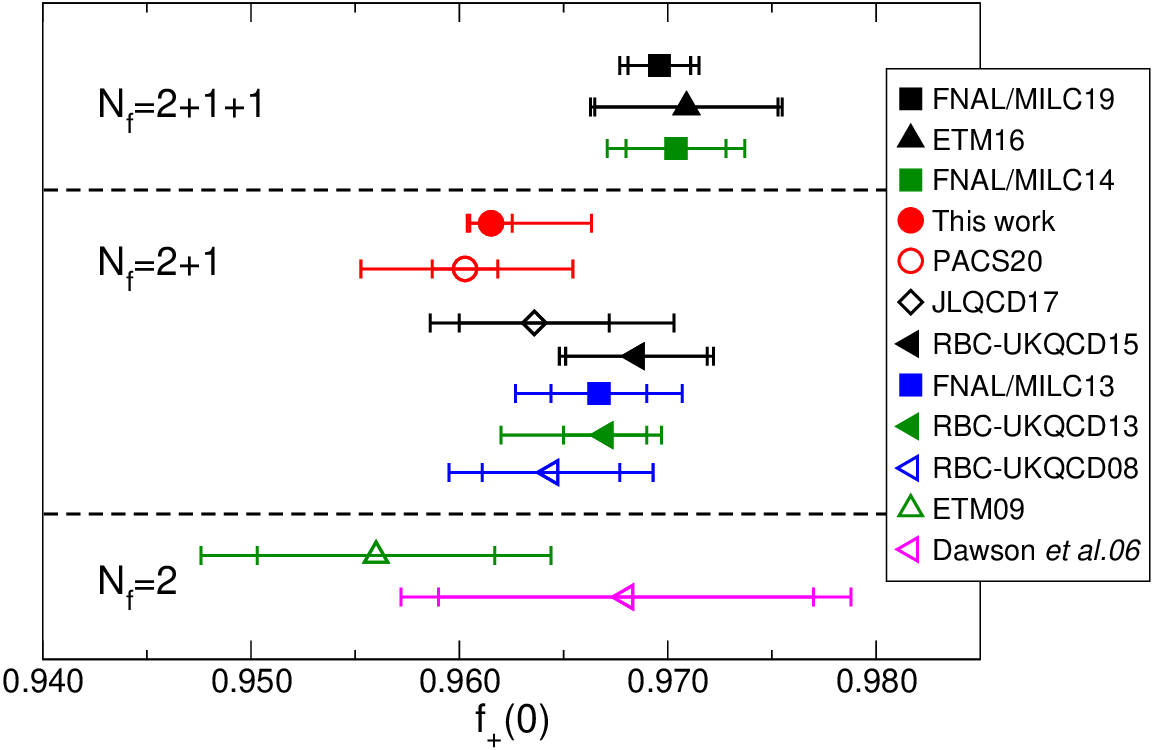}
\caption{Comparison of $f_+(0)$ obtained from this work with previous lattice QCD results in $N_f = 2$~\cite{Dawson:2006qc,Lubicz:2009ht},
$N_f = 2+1$~\cite{Bazavov:2012cd,Boyle:2007qe,Boyle:2013gsa,Boyle:2015hfa,Aoki:2017spo} including our previous work~\cite{PACS:2019hxd},
and $N_f = 2+1+1$~\cite{Bazavov:2013maa,Carrasco:2016kpy,Bazavov:2018kjg}.
The closed and open symbols express results 
in the continuum limit and at a finite lattice spacing, respectively.
The inner and outer errors express the statistical and total errors.
The total error is evaluated by adding the statistical and systematic
errors in quadrature.
}
\label{fig:comp-f+-q0}
\end{figure}

From the above analyses,
our result of $f_+(0)$ in this study is given by
\begin{equation}
f_+(0) = 0.9615(10)(^{+47}_{\ -3})(5) ,
\label{eq:res_f00}
\end{equation}
where the central value and statistical error (the first error) 
are determined from the main analysis with fit A.
The second and third errors are the systematic errors
from the fit forms and isospin breaking effect, respectively.
Figure~\ref{fig:comp-f+-q0} shows a comparison of our result
with those obtained in previous lattice calculations in $N_f = 2$~\cite{Dawson:2006qc,Lubicz:2009ht}, 2+1~\cite{Bazavov:2012cd,Boyle:2007qe,Boyle:2013gsa,Boyle:2015hfa,Aoki:2017spo} including our previous work~\cite{PACS:2019hxd}, and 2+1+1 QCD~\cite{Bazavov:2013maa,Carrasco:2016kpy,Bazavov:2018kjg}.
Our result in this work has smaller statistical and
systematic errors than the ones of our previous work~\cite{PACS:2019hxd},
which used a part of data in this calculation.
Especially, the lower systematic error is much reduced.

Due to our large upper systematic error,
our result is reasonably consistent with the ones from other groups
within 1.6 $\sigma$ in the total error, where the statistical
and systematic errors are added in quadrature.
The largest discrepancy comes from 
the result in Ref.~\cite{Bazavov:2018kjg}.
As discussed in our previous work~\cite{PACS:2019hxd},
reasons of the discrepancy are not clear at present.
There are several differences between the two calculations.
The calculation in Ref.~\cite{Bazavov:2018kjg} employed 
the HISQ action, and chiral and continuum extrapolations 
with data in $m_\pi < 0.3$ GeV 
using five different lattice spacings on the volumes of $m_\pi L > 3.2$
in $N_f = 2 + 1 + 1$ QCD.
The finite volume effect is corrected with NLO ChPT.
On the other hand, our calculation is performed with a nonperturbatively
improved Wilson quark action at
the physical point on the larger volumes than $(10\ {\rm fm})^3$
in $N_f = 2+1$ QCD, but we have only two different lattice spacings.
If our large systematic error would be significantly reduced
and the central value is not changed, 
serious investigations of the discrepancy will be necessary.
To decrease the systematic error for the continuum extrapolation, 
we plan to repeat the calculation with another set of the PACS10 configuration
at a finer lattice spacing.

\clearpage

\subsection{Shape of form factors}

\begin{figure}[ht!]
\includegraphics*[scale=.45]{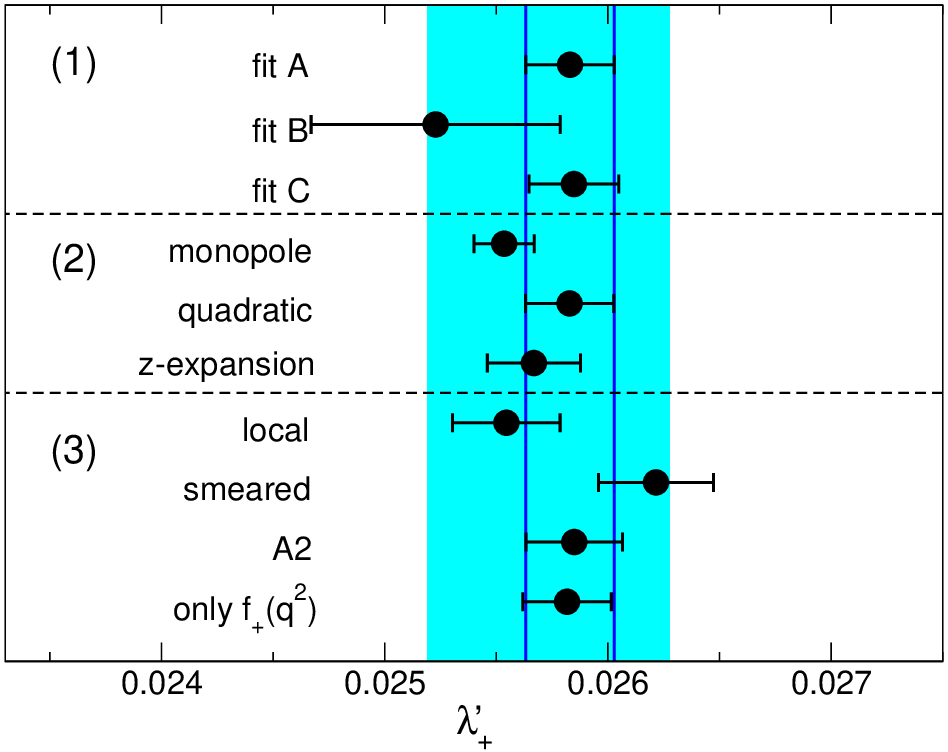}
\includegraphics*[scale=.45]{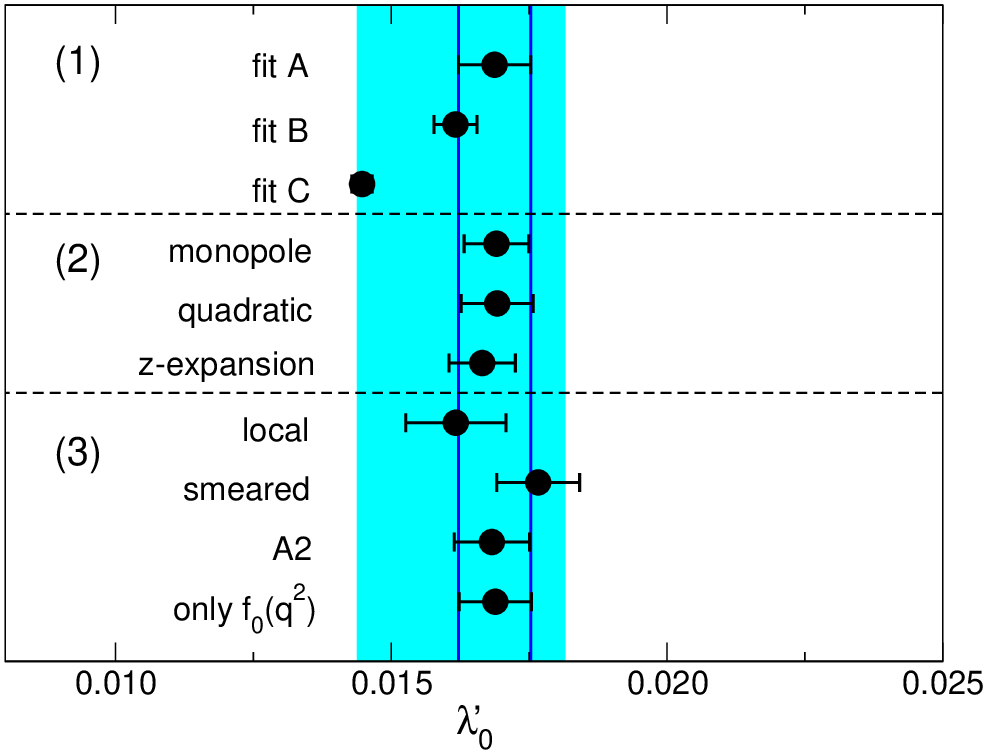}
\caption{The same figure as Fig.~\ref{fig:comp_fit-f+-q0},
but for $\lambda_+^\prime$ (left) and $\lambda_0^\prime$ (right).}
\label{fig:comp-fit_lam+0-a0}
\end{figure}

A comparison of the results of the slope for the form factors, 
$\lambda_+^\prime$ and $\lambda_0^\prime$ defined in Eq.~(\ref{eq:f_slope}), 
is presented in Fig.~\ref{fig:comp-fit_lam+0-a0}.
The format of the plot is the same as Fig.~\ref{fig:comp_fit-f+-q0},
but some analyses are not included in the comparison,
{\it e.g.,} the one from the narrower $q^2$ fit range analysis.
In both slopes, the choice of the fit form for the 
lattice spacing dependence causes the largest discrepancy from
our main result (fit A).
The central value and statistical error for the slopes 
are determined from the fit A result.
The systematic errors are evaluated in
the same way as in those of $f_+(0)$ explained in the last subsection.
Our results for the slopes are given as
\begin{eqnarray}
\lambda_+^{\prime} &=& 0.02583(20)(^{+39}_{-60})(9) ,
\label{eq:res_lam+_1}\\
\lambda_0^{\prime} &=& 0.0169(7)(^{+11}_{-24})(2) .
\label{eq:res_lam0_1}
\end{eqnarray}
The meanings of the second and third errors are the same as $f_+(0)$ 
in Eq.~(\ref{eq:res_f00}).

\begin{figure}[ht!]
\includegraphics*[scale=.45]{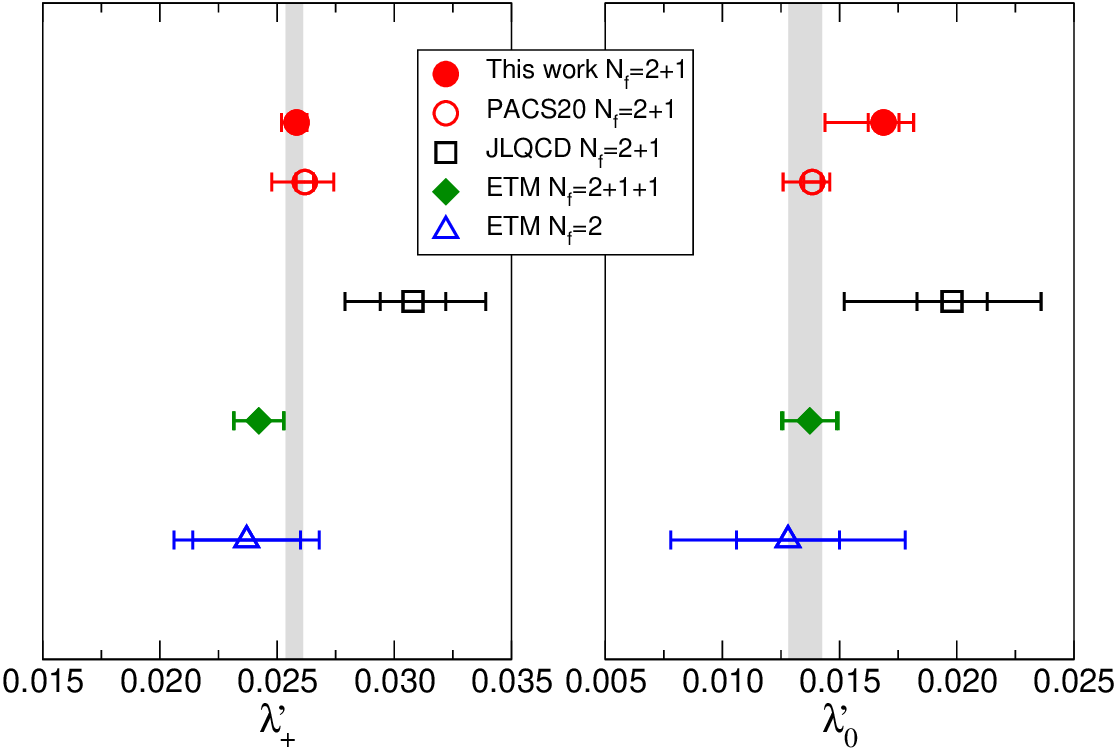}
\caption{Comparison of our result for 
$\lambda_+^\prime$ (left) and $\lambda_0^\prime$ (right) with
previous lattice QCD results~\cite{Lubicz:2009ht,Carrasco:2016kpy,Aoki:2017spo} including our previous work~\cite{PACS:2019hxd} and 
the experimental values~\cite{Moulson:2017ive}.
The different symbols denote different lattice QCD results,
and the gray band represents the experimental value with one standard
deviation.
The closed and open symbols express results 
in the continuum limit and at a finite lattice spacing, respectively.
The inner and outer errors express the statistical and total errors.
The total error is evaluated by adding the statistical and systematic
errors in quadrature.}
\label{fig:comp-lam+0-a0}
\end{figure}

Our results for the slopes are compared with 
previous lattice QCD results~\cite{Lubicz:2009ht,Carrasco:2016kpy,Aoki:2017spo} including our calculation~\cite{PACS:2019hxd}
in Fig.~\ref{fig:comp-lam+0-a0}.
Both the slopes are reasonably consistent with the previous results
and also the experimental values~\cite{Moulson:2017ive},
$\lambda_+^\prime = 0.02575(36)$ and
$\lambda_0^\prime = 0.01355(71)$\footnote{We employ $\lambda_0^\prime = m_{\pi^-}^2/(m_{K^0}^2 - m_{\pi^-}^2)(\log C - 0.0398(44))$~\cite{Bernard:2009zm} with $\log C = 0.1985(70)$.}.
Since the lattice spacing dependence of $\lambda_+^\prime$ is well
constrained in the continuum extrapolations,
its total uncertainty is smaller than our previous result,
and comparable to the experimental one.
On the other hand, the total error of $\lambda_0^\prime$
is larger than our previous result, because the lattice spacing
dependence is not constrained in our data as presented in 
the right panel of Fig.~\ref{fig:lam+0-a0}.

\begin{figure}[ht!]
\includegraphics*[scale=.45]{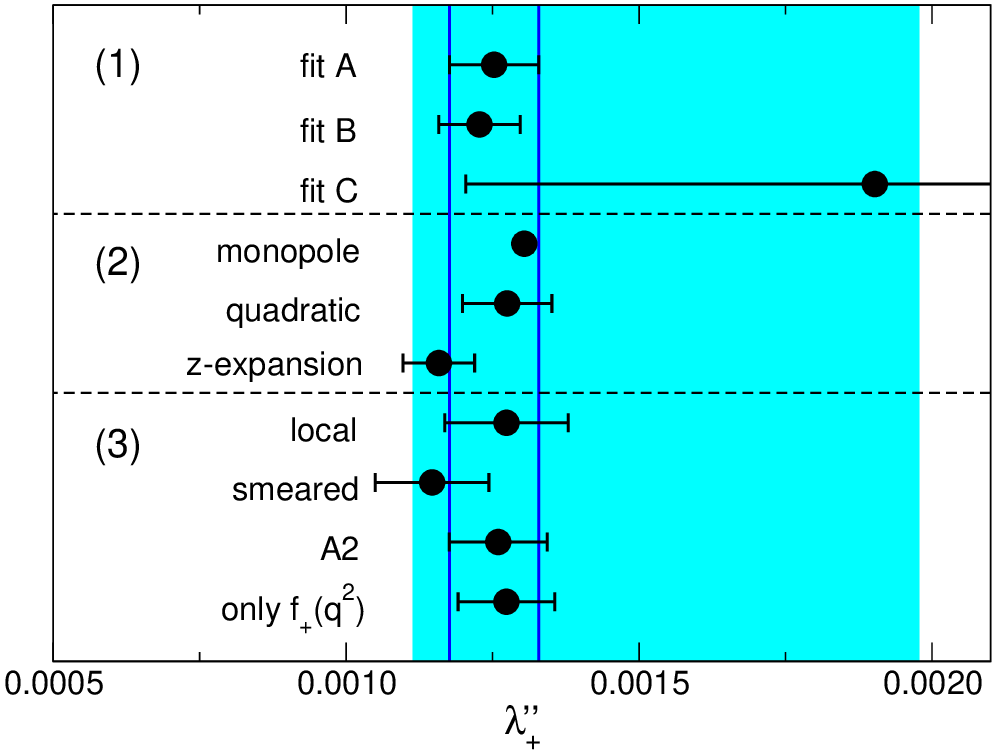}
\includegraphics*[scale=.45]{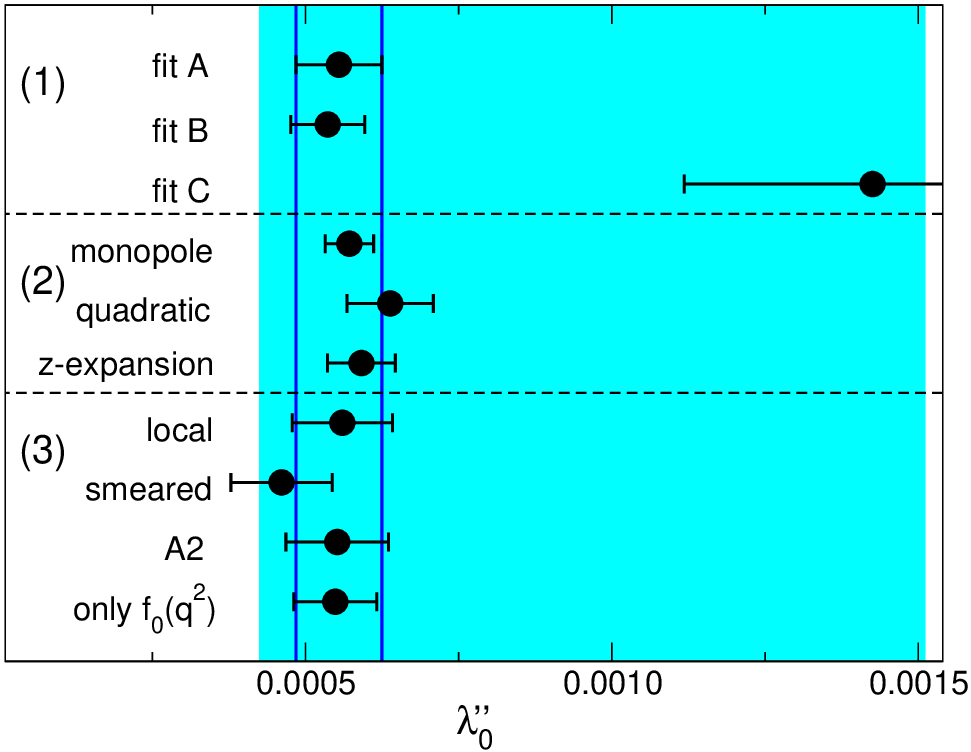}
\caption{The same figure as Fig.~\ref{fig:comp_fit-f+-q0},
but for $\lambda_+^{\prime\prime}$ (left) and 
$\lambda_0^{\prime\prime}$ (right).}
\label{fig:comp-fit_dlam+0-a0}
\end{figure}

\begin{figure}[ht!]
\includegraphics*[scale=.45]{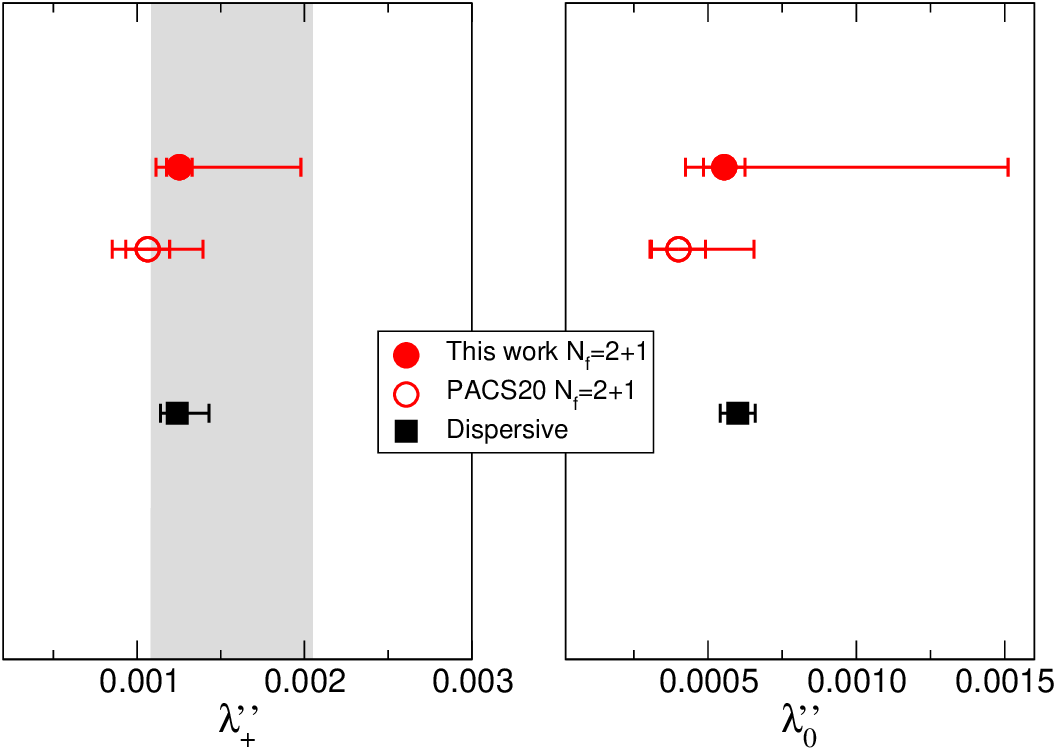}
\caption{Comparison of our result for 
$\lambda_+^{\prime\prime}$ (left) and $\lambda_0^{\prime\prime}$ (right)
with our previous work~\cite{PACS:2019hxd} and experimental values.
The square symbol is the experimental value evaluated with the dispersive
representation~\cite{Bernard:2009zm}, and
the gray band represents the average of 
the experimental results~\cite{Antonelli:2010yf}
with one standard deviation.
The inner and outer errors of the circles express 
the statistical and total errors.
The total error is evaluated by adding the statistical and systematic
errors in quadrature.}
\label{fig:comp-dlam+0-a0}
\end{figure}

The same comparison is shown in Fig.~\ref{fig:comp-fit_dlam+0-a0}
for the curvatures, $\lambda_+^{\prime\prime}$ and $\lambda_0^{\prime\prime}$ 
defined in Eq.~(\ref{eq:f_curvature}).
The largest difference from the fit A result is given by
fit C in both the cases.
The reason is the same as $\lambda_0^\prime$ due to
the unconstrained lattice spacing dependence in our data.
Using the same strategy of the estimation of the systematic errors
as in the slopes, 
we obtain the following results for the curvatures as
\begin{eqnarray}
\lambda_+^{\prime\prime} &=& 0.00125(8)(^{+72}_{-12})(0) ,
\label{eq:res_lam+_2}\\
\lambda_0^{\prime\prime} &=& 0.00055(7)(^{+95}_{-11})(1) .
\label{eq:res_lam0_2}
\end{eqnarray}
Figure~\ref{fig:comp-dlam+0-a0} shows that 
these values reasonably agree 
with our previous result~\cite{PACS:2019hxd},
the average of the experimental results, 
$\lambda_+^{\prime\prime} = 0.00157(48)$~\cite{Antonelli:2010yf},
and the experimental ones evaluated 
with the dispersive representation~\cite{Bernard:2009zm}.
The larger total errors in our calculation than those in our previous work 
come from the systematic error
of the choice of the fit forms in the continuum extrapolation.

Regarding the large systematic errors in the slopes and curvatures, 
if the third data at a smaller lattice spacing are calculated, 
these errors associated with the finite lattice spacing effect
can be significantly reduced.
Thus, it is an important future work in our calculation.

\clearpage

\subsection{Phase space integral}

The phase space factor $I_K^\ell$ in Eq.~(\ref{eq:decay_width_kl3}) 
is determined from an integral of 
the $q^2$ dependent form factors in a certain $q^2$ region,
$m_\ell^2 \le t \le t_{\rm max}$,
where $t = -q^2$, $t_{\rm max} = ( M_K - M_\pi )^2$,
and $m_\ell$ is the mass of the lepton $\ell$,
$m_e = 0.000511$ GeV, and $m_\mu = 0.10566$ GeV.
Since the $q^2$ dependence of the form factors is
obtained in our calculation, 
the phase space integral can be calculated using our result 
in the continuum limit.
The phase space integral~\cite{Leutwyler:1984je} is defined by
\begin{equation}
I^\ell_K = \int^{t_{\rm max}}_{m_\ell^2} dt
\frac{\lambda^{3/2}}{M_K^8}
\left( 1 + \frac{m_\ell^2}{2 t} \right)
\left( 1 - \frac{m_\ell^2}{t} \right)^2
\left(
\overline{F}_+^2(t) + 
\frac{3 m_\ell^2 \Delta_{K\pi}^2}{(2t + m_\ell^2)\lambda}\overline{F}_0^2(t)
\right),
\label{eq:phase_space_integral}
\end{equation}
where $\lambda = ( t - ( M_K + M_\pi )^2 ) ( t - t_{\rm max} )$,
$\Delta_{K\pi} = M_K^2 - M_\pi^2$,
$\overline{F}_s(t) = f_s(-t)/f_+(0)$ with $s = +$ and 0.
For $I^\ell_{K^0}$ related to the $K^0 \to \pi^- \ell^+ \nu_\ell$ process,
we employ $M_K = m_{K^0} = 0.497611$ GeV and
$M_\pi = m_{\pi^-} = 0.13957061$ GeV.
In the calculation of $I^\ell_{K^+}$ 
for the $K^+ \to \pi^0 \ell^+ \nu_\ell$ process,
$M_K = m_{K^+} = 0.493677$ GeV and
$M_\pi = m_{\pi^0} = 0.1349770$ GeV are used.

The four values of $I^\ell_K$ obtained from the fit results for
the fit A, B, C are tabulated in Table~\ref{tab:phase_space_integral}.
Using the results in the table and those from other fits shown 
in Appendix~\ref{sec:app_fit_results},
systematic errors are estimated in the same way as in $f_+(0)$.
In the four results, the systematic error of the isospin breaking effect 
is estimated using the NLO ChPT functions for 
$f_s^{K^0\pi^-}$~\cite{Gasser:1984ux,Bijnens:2007xa} with $s=+,0$, 
even in $I^\ell_{K^+}$.
In contrast to this evaluation of the phase space integrals,
a correction related to the difference between $f_+^{K^0\pi^-}$
and $f_+^{K^+\pi^0}$ is incorporated in the determination of $|V_{us}|$
from the phase space integrals discussed later.
From the estimations, we obtain results for $I_K^\ell$ as,
\begin{eqnarray}
I^e_{K^0} &=& 0.15477(7)(^{+55}_{-33})(5), \\
I^\mu_{K^0} &=& 0.1032(2)(^{\ +3}_{-20})(1),\\
I^e_{K^+} &=& 0.15919(8)(^{+58}_{-34})(1), \\
I^\mu_{K^+} &=& 0.1063(2)(^{\ +3}_{-21})(0).
\end{eqnarray}
These results agree well with the experimental values in
the dispersive representation of the form factors,
$I^e_{K^0} = 0.15476(18)$, $I^\mu_{K^0} = 0.10253(16)$,
$I^e_{K^+} = 0.15922(18)$, $I^\mu_{K^+} = 0.10559(17)$,
in Ref.~\cite{Antonelli:2010yf},
and also their updates,
$I^e_{K^0} = 0.15470(15)$, $I^\mu_{K^0} = 0.10247(15)$
$I^e_{K^+} = 0.15915(15)$, $I^\mu_{K^+} = 0.10553(16)$
in Ref.~\cite{Seng:2021nar}.

In the next subsection, we will determine $|V_{us}|$ using
our results of $I^\ell_K$.
For the determination, we also evaluate $f_+(0) \sqrt{I^\ell_{K}}$
in each process as
\begin{eqnarray}
f_+(0) \sqrt{I^e_{K^0}} &=& 0.3783(4)(^{+14}_{\ -3})(1), 
\label{eq:res_phase_space_integral_k0e}\\
f_+(0) \sqrt{I^\mu_{K^0}} &=& 0.3089(5)(^{\ +6}_{-17})(0),
\label{eq:res_phase_space_integral_k0mu}\\
f_+(0) \sqrt{I^e_{K^+}} &=& 0.3836(4)(^{+15}_{\ -3})(2), 
\label{eq:res_phase_space_integral_k+e}\\
f_+(0) \sqrt{I^\mu_{K^+}} &=& 0.3135(5)(^{\ +7}_{-17})(2).
\label{eq:res_phase_space_integral_k+mu}
\end{eqnarray}
The values obtained from each continuum extrapolation
are compiled in Table~\ref{tab:phase_space_integral}
and also summarized in Appendix~\ref{sec:app_fit_results}.

\begin{table}[ht!]
\caption{Results of phase space integrals
in the continuum limit at the physical point obtained
from each fit listed in Table~\ref{tab:fit_chpt-cont-limit}.}
\label{tab:phase_space_integral}
\begin{tabular}{cccc}\hline\hline
                           & fit A        & fit B       & fit C       \\\hline
$I^e_{K^0}$                & 0.154769(75) & 0.15444(29) & 0.15526(55) \\
$I^\mu_{K^0}$              & 0.10319(16)  & 0.10114(70) & 0.10335(49) \\
$I^e_{K^+}$                & 0.159186(77) & 0.15884(30) & 0.15969(56) \\
$I^\mu_{K^+}$              & 0.10630(17)  & 0.10418(72) & 0.10646(50) \\
$f_+(0)\sqrt{I^e_{K^0}}$   & 0.37827(42)  & 0.3796(15)  & 0.37879(83) \\
$f_+(0)\sqrt{I^\mu_{K^0}}$ & 0.30887(45)  & 0.3072(14)  & 0.30905(84) \\
$f_+(0)\sqrt{I^e_{K^+}}$   & 0.38363(43)  & 0.3850(15)  & 0.38416(85) \\
$f_+(0)\sqrt{I^\mu_{K^+}}$ & 0.31349(46)  & 0.3118(14)  & 0.31367(86) 
\\\hline\hline
\end{tabular}
\end{table}

\subsection{Result of $|V_{us}|$}

Combining our result of $f_+(0)$ in Eq.~(\ref{eq:res_f00}) and 
the experimental value,
$f_+(0)|V_{us}| = 0.21654(41)$~\cite{Moulson:2017ive},
we determine the value of $|V_{us}|$ as
\begin{equation}
|V_{us}| = 0.22521(24)(^{\ \ +6}_{-109})(11)(43).
\label{eq:vus_f0}
\end{equation}
The statistical (first) and systematical (second and third)
errors correspond to those in $f_+(0)$.
The additional fourth error comes from the experimental value.
When an updated value of $f_+(0)|V_{us}| = 0.21635(39)$~\cite{Seng:2021nar} 
is employed, the obtained value
is not largely changed as
$|V_{us}| = 0.22501(24)(^{\ \ +6}_{-109})(11)(41)$.

\begin{figure}[ht!]
\includegraphics*[scale=.45]{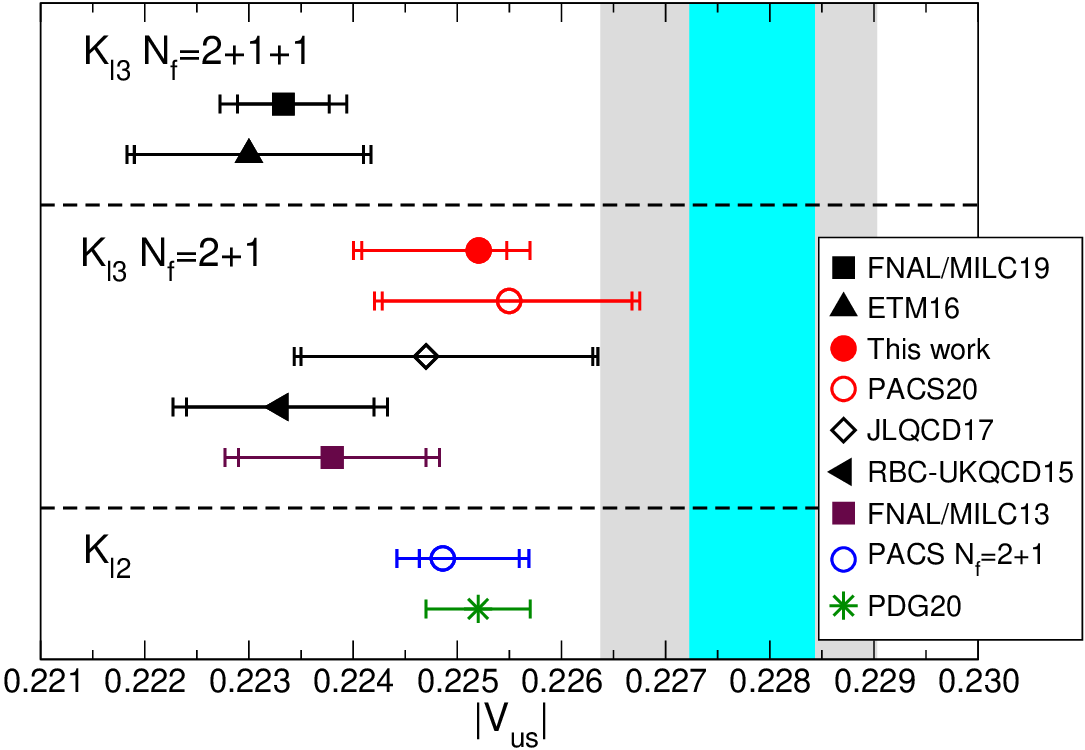}
\caption{Comparison of $|V_{us}|$ in this work with previous lattice QCD 
results obtained from the $K_{\ell 3}$ form factor~\cite{Boyle:2013gsa,Boyle:2015hfa,Aoki:2017spo,Carrasco:2016kpy,Bazavov:2018kjg} including our previous work~\cite{PACS:2019hxd}.
$|V_{us}|$ determined from the $K_{\ell 2}$ decay are also
plotted using $F_K/F_\pi$ in our calculation and 
PDG20~\cite{ParticleDataGroup:2020ssz}.
The inner and outer errors express the error of the lattice calculation 
and total error.
The total error is evaluated by adding the lattice QCD and experimental
errors in quadrature.
The closed and open symbols express results 
in the continuum limit and at a finite lattice spacing, respectively.
The unitarity values using $|V_{ud}|$ in Refs.~\cite{Seng:2018yzq} and \cite{Hardy:2020qwl} are presented by the light blue and gray bands, respectively.
}
\label{fig:vus-a0}
\end{figure}

In Fig.~\ref{fig:vus-a0}, our value of $|V_{us}|$ is 
compared with several previous results using $f_+(0)$
in the $N_f = 2+1$ and 2+1+1 lattice QCD calculations~\cite{Bazavov:2012cd,Boyle:2015hfa,Aoki:2017spo,Carrasco:2016kpy,Bazavov:2018kjg}
including our previous work~\cite{PACS:2019hxd}.
The inner error originates from the lattice calculation.
On the other hand,
the outer error corresponds to the total error,
where the errors in the lattice QCD calculation and experimental value
are added in quadrature.
Similar to the comparison of $f_+(0)$, our result is reasonably consistent 
with other lattice results, while our central value is a little larger than
most of all results.
The largest discrepancy, however, is only 1.4 $\sigma$,
so that it is not so significant in the current total error.

Our result also agrees with the value in PDG20~\cite{ParticleDataGroup:2020ssz}
determined through the $K_{\ell 2}$ processes,
and the one using our results of $F_K/F_\pi$ calculated with 
the PACS10 configurations at $\beta = 2.00$ and 1.82, $|V_{us}| = 0.22486(32)(^{+60}_{\ -0})(30)$, as plotted in Fig.~\ref{fig:vus-a0}.
Our central value and statistical (first) error
are determined from the result of $\beta = 2.00$, and 
the asymmetric systematic (second) error is estimated
from the difference between the results at the two lattice spacings.
The third error stems from the experimental values,
$|V_{us}|F_K/|V_{ud}|F_\pi = 0.27599(37)$~\cite{Moulson:2017ive}
and $|V_{ud}| = 0.97370(14)$~\cite{Seng:2018yzq}.
If an updated value of $|V_{us}|F_K/|V_{ud}|F_\pi = 0.27683(35)$~\cite{DiCarlo:2019thl} is used in the determination,
$|V_{us}|$ is shifted to a little larger value within the total uncertainty.

The value of $|V_{us}|$ is evaluated from the unitarity of 
the first row of the CKM matrix, 
$|V_{us}| \approx \sqrt{1-|V_{ud}|^2} = 0.2278(6)$ under the assumption
$|V_{ub}| \ll 1$, whose value with one standard deviation 
is shown by the light blue band in Fig.~\ref{fig:vus-a0}.
This value differs from our $K_{\ell 3}$ result by 3.4 $\sigma$.
We would need more precise studies of systematic errors in lattice calculations
and also $|V_{ud}|$~\cite{Hardy:2020qwl} to conclude
whether the difference suggests a signal of new physics or not.
The reason is that another determination of 
$|V_{ud}| = 0.97373(31)$~\cite{Hardy:2020qwl} 
with a larger error gives $|V_{us}| = 0.2277(13)$ through the unitarity,
as expressed by the gray band in the figure,
whose discrepancy from our result is 1.8 $\sigma$.

\begin{table}[ht!]
\caption{Results of $|V_{us}|$ obtained from our results
of the phase space integrals in Eqs.~(\ref{eq:res_phase_space_integral_k0e})--(\ref{eq:res_phase_space_integral_k+mu}). 
The errors of $|V_{us}|$ are explained in text.
The inputs for each decay process, $|V_{us}| f_+(0) \sqrt{I^\ell_K}$,
are also tabulated.}
\label{tab:Vus_phase_space_integral}
\begin{tabular}{cccc}\hline\hline
& $|V_{us}|$ & $|V_{us}|f_+(0)\sqrt{I^\ell_K}$ \\\hline
$K_L e$   & $0.22480(25)(^{+16}_{-86})(7)(53)     $ & 0.08503(20)  \\
$K_L \mu$ & $0.22439(33)(^{+121}_{\ -45})(2)(62)  $ & 0.06931(19)  \\
$K_S e$   & $0.22386(25)(^{+16}_{-85})(7)(127)    $ & 0.08468(48)  \\
$K_S \mu$ & $0.22026(32)(^{+119}_{\ -44})(2)(483) $ & 0.06803(149) \\
$K^+ e$   & $0.22581(25)(^{+16}_{-86})(10)(94)    $ & 0.08663(36)  \\
$K^+ \mu$ & $0.22486(33)(^{+122}_{\ -47})(13)(114)$ & 0.07049(36)  
\\\hline\hline
\end{tabular}
\end{table}

Using our results of the phase space
integrals in Eqs.~(\ref{eq:res_phase_space_integral_k0e})--(\ref{eq:res_phase_space_integral_k+mu}),
$|V_{us}|$ is determined through six kaon decay processes
as presented in Table~\ref{tab:Vus_phase_space_integral}.
For each decay process,
the experimental value of $|V_{us}| f_+(0) \sqrt{I^\ell_K}$ is
also tabulated in the table,
which is obtained using the experimental inputs and correction factors~\cite{Antonelli:2010yf,Seng:2021nar,Seng:2022wcw,Seng:2021boy,Seng:2021wcf} including the correction of 
the difference of $f^{K^0\pi^-}_+(0)$ and $f^{K^+\pi^0}_+(0)$.
A weighted average of the six processes gives
\begin{equation}
|V_{us}| = 0.22469(27)(^{+39}_{-77})(6)(40) .
\label{eq:vus_phase_space_integral}
\end{equation}
The results for the six processes and also
a weighted average are well consistent with 
the one determined from $f_+(0)$ in Eq.~(\ref{eq:vus_f0}).

\section{Conclusions}
\label{sec:conclusions}

We have updated our calculation of the $K_{\ell 3}$ form factors on
a more than (10 fm)$^3$ volume at the physical point by
adding the result at the second lattice spacing of 0.063 fm.
The calculation is performed using the local and conserved vector currents. 
We have observed that lattice spacing dependences in the $f_+(0)$ data 
for the two vector currents are clearly different.

Using the data of the form factors at the two different lattice spacings,
continuum extrapolation and $q^2$ interpolation are carried out
simultaneously using fit forms based on the NLO SU(3) ChPT.
The obtained result of $f_+(0)$ is stable against several fit forms
of $q^2$, while it largely depends on fitting function
of the lattice spacing.
This is because it is hard to constrain fit forms of the lattice spacing
dependence using our data at only two lattice spacings.
Similar trends of the fitting function dependence for the lattice spacing
effect are observed in the results for the slope and curvature
of the form factors.
Our final results are summarized as follows:
\begin{eqnarray}
f_+(0) &=& 0.9615(10)(^{+47}_{\ -3})(5) , \\
\lambda_+^{\prime} &=& 0.02583(20)(^{+39}_{-60})(9) , \\
\lambda_0^{\prime} &=& 0.0169(7)(^{+11}_{-24})(2) , \\
\lambda_+^{\prime\prime} &=& 0.00125(8)(^{+72}_{-12})(0) , \\
\lambda_0^{\prime\prime} &=& 0.00055(7)(^{+95}_{-11})(1) .
\end{eqnarray}
The first error is the statistical, 
the second one is the systematic error involving the fit form and data 
dependences, and the last one represents the systematic error from 
the isospin breaking effect, respectively.
As described above, the second error is the largest in our result.

Our $f_+(0)$ is reasonably consistent with recent lattice QCD results.
The relative difference from the most precise result is not
so significant, 1.6 $\sigma$, at present due to our large systematic error.
If our systematic error is largely decreased in a similar central value, 
we would need detailed investigations for the difference.
For the slopes and curvatures, our results are in good agreement
with other lattice results and the experimental values, 
while our systematic errors are much larger than those in the experiment, 
except for $\lambda_+^\prime$,
which has a comparable error to the experiment.

Using our $f_+(0)$ and the experimental input, 
we have obtained the value of $|V_{us}|$ as
\begin{equation}
|V_{us}| = 0.22521(24)(^{\ \ +6}_{-109})(11)(43).
\end{equation}
The errors from the first to the third correspond to the same ones
as in the above $f_+(0)$.
The fourth error comes from the experimental input.
This value can be expressed by $|V_{us}| = 0.2252(^{\ +5}_{-12})$.
Our result is reasonably consistent with those using the recent lattice 
results of $f_+(0)$, and also well agrees with the values of $|V_{us}|$ 
determined through the $K_{\ell 2}$ decay.
On the other hand, we have observed a discrepancy from that of
the unitarity of the first row of the CKM matrix.
Its significance, however, depends on the size of error of $|V_{ud}|$.
For the BSM search through $|V_{us}|$, it would be important to
decrease the uncertainties in both the lattice QCD and $|V_{ud}|$
in future.

We have also computed the phase space integrals from our $q^2$
dependent form factors as,
\begin{eqnarray}
I^e_{K^0} &=& 0.15477(7)(^{+55}_{-33})(5), \\
I^\mu_{K^0} &=& 0.1032(2)(^{\ +3}_{-20})(1),\\
I^e_{K^+} &=& 0.15919(8)(^{+58}_{-34})(1), \\
I^\mu_{K^+} &=& 0.1063(2)(^{\ +3}_{-21})(0).
\end{eqnarray}
Those results are well consistent with the experimental values
in the dispersive representation of the form factors.
Using our results for the phase space integrals, the experimental inputs, 
and the correction factors,
the values of $|V_{us}|$ are determined through 
six $K_{\ell 3}$ decay processes.
An average of the six decay processes given as,
\begin{equation}
|V_{us}| = 0.22469(27)(^{+39}_{-77})(6)(40) ,
\end{equation}
well agrees with that in the above with $f_+(0)$.
This agreement insists that the form factors calculated in the lattice QCD
are useful for the evaluation of the phase space integrals as well as 
for the determination of $f_+(0)$.

While in this work the lower uncertainty of our $f_+(0)$ is much
smaller than that in our previous work,
the upper one is still similar in size.
An additional data at a smaller lattice spacing could largely reduce
the uncertainty, because our main systematic error comes from
the choice of the fitting forms of the lattice spacing dependence
in the continuum extrapolation.
This is an important future direction of our calculation, and
we are generating the PACS10 configuration at the third lattice spacing.
Furthermore, the isospin breaking effect of the form factors 
is estimated with the NLO SU(3) ChPT formulas in this work, 
while this effect can be evaluated using a lattice calculation.
It is also an important future work for the indirect search for a BSM physics.

\section*{Acknowledgments}
Numerical calculations in this work were performed on Oakforest-PACS
in Joint Center for Advanced High Performance Computing (JCAHPC)
under Multidisciplinary Cooperative Research Program of Center for Computational Sciences, University of Tsukuba.
This research also used computational resources of Oakforest-PACS
by Information Technology Center of the University of Tokyo,
and of Fugaku by RIKEN CCS
through the HPCI System Research Project (Project ID: hp170022, hp180051, hp180072, hp180126, hp190025, hp190081, hp200062, hp200167, hp210112, hp220079).
The calculation employed OpenQCD system\footnote{http://luscher.web.cern.ch/luscher/openQCD/}.
This work was supported in part by Grants-in-Aid 
for Scientific Research from the Ministry of Education, Culture, Sports, 
Science and Technology (Nos. 18K03638, 19H01892).
This work was supported by the JLDG constructed over the SINET5 of NII.

\appendix

\section{Results with several fit forms}
\label{sec:app_fit_results}

In this appendix results for continuum extrapolations using
various fit forms and different data are summarized.
These results are used to estimate systematic errors of our main result
as discussed in the main text.

The same continuum extrapolations using the formulas based on 
the NLO SU(3) ChPT explained in Sec.~\ref{sec:continuum_extrapolation}
are performed, but the decay constant $F_0$ 
is a free parameter and $c_0$ is omitted in the fit functions
shown in Eqs.~(\ref{eq:ato0_NLO_chpt_f+}) and (\ref{eq:ato0_NLO_chpt_f0}).
These fit results for the fit A, B, C, whose functional forms are described 
in Sec.~\ref{sec:continuum_extrapolation}, are summarized in
Table~\ref{tab:fit_chpt-cont-limit-free_F0}.
The phase space integral calculated by Eq.~(\ref{eq:phase_space_integral})
for each fit is shown in Table~\ref{tab:phase_space_integral-free_F0}.

We also compare results with a different choice of data.
In a continuum extrapolation, the form factor data at $\beta = 2.00$ 
are replaced by those obtained from the analysis with only the local 
or smeared operator data, whose values are presented in 
Table~\ref{tab:kl3_form_factor_b200}
for the local current and in Table~\ref{tab:kl3_form_factor_b200-cns-cur}
for the conserved current in Sec.~\ref{sec:result_finite_a}.
Furthermore, we perform a continuum extrapolation with
the replaced data at $\beta = 1.82$ by
the ones obtained from the A2 analysis shown in
Table~\ref{tab:kl3_form_factor_b182-cns-cur} for the conserved current
in Sec.~\ref{sec:result_finite_a} and Table~II of Ref.~\cite{PACS:2019hxd}
for the local current.
In all the extrapolations, the fit A form is employed for the lattice 
spacing dependence.
Those fit results and phase space integrals are tabulated in
Tables~\ref{tab:fit_chpt-cont-limit-diff_data-fixed_F0} and
\ref{tab:phase_space_integral-diff_data-fixed_F0}
for a fixed $F_0$, 
and Tables~\ref{tab:fit_chpt-cont-limit-diff_data-free_F0}
and \ref{tab:phase_space_integral-diff_data-free_F0}
for a free $F_0$.

\begin{table}[ht!]
\caption{Fit results of the continuum extrapolation of $K_{\ell 3}$ form factors
based on the NLO SU(3) ChPT formulas in 
Eqs.~(\ref{eq:NLO_chpt_f+}) and (\ref{eq:NLO_chpt_f0})
together with the value of the uncorrelated $\chi^2/$dof.
$F_0$ is a free parameter in all the fits.
}
\label{tab:fit_chpt-cont-limit-free_F0}
\begin{tabular}{cccc}\hline\hline
                        & fit A        & fit B        & fit C       \\\hline
$L_9$ [$10^{-3}$]       & 3.178(96)    & 3.54(38)     & 3.164(97)   \\
$L_5$ [$10^{-4}$]       & 7.30(55)     & 8.1(1.1)     & 5.54(27)    \\
$c_2^+$ [GeV$^{-4}$]    & 1.390(94)    & 1.395(88)    & 2.32(88)    \\
$c_2^0$ [GeV$^{-4}$]    & $-0.404(92)$ & $-0.29(13)$  & 0.82(40)    \\
$F_0$ [GeV]             & 0.1017(13)   & 0.1081(52)   & 0.1015(13)  \\
$\chi^2/$dof            & 0.37         & 0.35         & 0.32        \\
$d_{20}$ [GeV$^2$]      & $\cdots$     & $-0.034$(24) & $\cdots$    \\
$d_{21}^+$              & $\cdots$     & $-0.20$(20)  & $\cdots$    \\
$d_{21}^0$              & 1.11(27)     & 0.77(16)     & $\cdots$    \\
$d_{22}^+$ [GeV$^{-2}$] & $\cdots$     & $\cdots$     & $-6.3$(6.6) \\
$d_{22}^0$ [GeV$^{-2}$] & $\cdots$     & $\cdots$     & $-$10.1(2.6) \\
$e_{10}$ [GeV]          & 0.03015(89)  & $\cdots$     & 0.03084(88) \\
$e_{11}^+$ [GeV$^{-1}$] & 0.2972(66)   & $\cdots$     & 0.285(13)   \\
$e_{11}^0$ [GeV$^{-1}$] & 0.36(10)     & $\cdots$     & $\cdots$    \\
$e_{12}^+$ [GeV$^{-3}$] & $\cdots$     & $\cdots$     & $-$2.7(2.3) \\
$e_{12}^0$ [GeV$^{-3}$] & $\cdots$     & $\cdots$     & $-$3.39(95) \\
$e_{20}$ [GeV$^2$]      & $\cdots$     & 0.048(24)    & $\cdots$    \\
$e_{21}^+$              & $\cdots$     & 0.62(20)     & $\cdots$    \\
$e_{21}^0$              & $\cdots$     & 0.64(16)     & $\cdots$    \\
$f_+(0)$                & 0.96184(98)  & 0.9662(32)   & 0.96164(98) \\
$\lambda_+^\prime$ [$10^{-2}$] & 2.584(20) & 2.525(55) & 2.587(20) \\
$\lambda_0^\prime$ [$10^{-2}$] & 1.718(66) & 1.624(42) & 1.451(18) \\
$\lambda_+^{\prime\prime}$ [$10^{-3}$] & 1.243(74) & 1.224(69) & 1.97(69) \\
$\lambda_0^{\prime\prime}$ [$10^{-3}$] & 0.540(69) & 0.530(60) & 1.51(31) 
\\\hline\hline
\end{tabular}
\end{table}

\begin{table}[ht!]
\caption{Results of phase space integrals
in the continuum limit at the physical point obtained
from each fit listed in Table~\ref{tab:fit_chpt-cont-limit-free_F0}.}
\label{tab:phase_space_integral-free_F0}
\begin{tabular}{cccc}\hline\hline
                           & fit A        & fit B       & fit C       \\\hline
$I^e_{K^0}$                & 0.154766(75) & 0.15445(29) & 0.15532(54) \\
$I^\mu_{K^0}$              & 0.10325(16)  & 0.10121(73) & 0.10344(48) \\
$I^e_{K^+}$                & 0.159192(77) & 0.15886(30) & 0.15977(56) \\
$I^\mu_{K^+}$              & 0.10639(17)  & 0.10426(76) & 0.10658(50) \\
$f_+(0)\sqrt{I^e_{K^0}}$   & 0.37839(41)  & 0.3797(13)  & 0.37899(82) \\
$f_+(0)\sqrt{I^\mu_{K^0}}$ & 0.30907(43)  & 0.3074(12)  & 0.30928(83) \\
$f_+(0)\sqrt{I^e_{K^+}}$   & 0.38376(41)  & 0.3851(14)  & 0.38437(83) \\
$f_+(0)\sqrt{I^\mu_{K^+}}$ & 0.31372(44)  & 0.3120(12)  & 0.31394(84) 
\\\hline\hline
\end{tabular}
\end{table}

\begin{table}[ht!]
\caption{The same table as Table~\ref{tab:fit_chpt-cont-limit-free_F0},
but for using the local and smeared operator data at $\beta = 2.00$,
and the A2 analysis data at $\beta = 1.82$.
$F_0$ is fixed in all the fits.
}
\label{tab:fit_chpt-cont-limit-diff_data-fixed_F0}
\begin{tabular}{cccc}\hline\hline
                        & local         & smeared       & A2           \\\hline
$L_9$ [$10^{-3}$]       & 3.833(38)     & 3.940(40)     & 3.883(35)    \\
$L_5$ [$10^{-4}$]       & 8.85(71)      & 10.01(58)     & 9.36(53)     \\
$c_2^+$ [GeV$^{-4}$]    & 1.46(13)      & 1.30(12)      & 1.44(11)     \\
$c_2^0$ [GeV$^{-4}$]    & $-0.19(10)$   & $-0.31(10)$   & $-0.20(11)$  \\
$c_0$                   & $-0.0073(13)$ & $-0.0066(11)$ & $-0.0066(12)$ \\
$\chi^2/$dof            & 0.26          & 0.45          & 0.21         \\
$d_{21}^0$              & 0.80(33)      & 1.17(27)      & 1.01(30)     \\
$e_{10}$ [GeV]          & 0.03064(91)   & 0.03032(85)   & 0.03043(77)  \\
$e_{11}^+$ [GeV$^{-1}$] & 0.2879(63)    & 0.3103(73)    & 0.2984(84)   \\
$e_{11}^0$ [GeV$^{-1}$] & 0.25(13)      & 0.39(10)      & 0.32(11)     \\
$f_+(0)$                & 0.9613(13)    & 0.9619(11)    & 0.9619(12)   \\
$\lambda_+^\prime$ [$10^{-2}$] & 2.545(24) & 2.622(26) & 2.585(22) \\
$\lambda_0^\prime$ [$10^{-2}$] & 1.617(91) & 1.766(75) & 1.682(68) \\
$\lambda_+^{\prime\prime}$ [$10^{-3}$] & 1.27(11)  & 1.147(97) & 1.260(84) \\
$\lambda_0^{\prime\prime}$ [$10^{-3}$] & 0.560(82) & 0.461(83) & 0.552(84)
\\\hline\hline
\end{tabular}
\end{table}

\begin{table}[ht!]
\caption{The same table as Table~\ref{tab:phase_space_integral-free_F0},
but for each fit result listed in Table~\ref{tab:fit_chpt-cont-limit-diff_data-fixed_F0}.}
\label{tab:phase_space_integral-diff_data-fixed_F0}
\begin{tabular}{cccc}\hline\hline
                           & local        & smeared      & A2       \\\hline
$I^e_{K^0}$                & 0.154635(85) & 0.154893(87) & 0.154784(83) \\
$I^\mu_{K^0}$              & 0.10296(22)  & 0.10340(19)  & 0.10319(17)  \\
$I^e_{K^+}$                & 0.159048(88) & 0.159314(90) & 0.159202(86) \\
$I^\mu_{K^+}$              & 0.10606(22)  & 0.10651(20)  & 0.10630(18) \\
$f_+(0)\sqrt{I^e_{K^0}}$   & 0.37801(51)  & 0.37858(44)  & 0.37845(48) \\
$f_+(0)\sqrt{I^\mu_{K^0}}$ & 0.30845(56)  & 0.30931(45)  & 0.30900(53) \\
$f_+(0)\sqrt{I^e_{K^+}}$   & 0.38336(52)  & 0.38394(44)  & 0.38381(49) \\
$f_+(0)\sqrt{I^\mu_{K^+}}$ & 0.31306(56)  & 0.31393(46)  & 0.31362(54) 
\\\hline\hline
\end{tabular}
\end{table}

\begin{table}[ht!]
\caption{The same table as Table~\ref{tab:fit_chpt-cont-limit-diff_data-fixed_F0},
but for fits with a free $F_0$.
}
\label{tab:fit_chpt-cont-limit-diff_data-free_F0}
\begin{tabular}{cccc}\hline\hline
                        & local       & smeared     & A2           \\\hline
$L_9$ [$10^{-3}$]       & 3.11(11)    & 3.26(10)    & 3.22(11)     \\
$L_5$ [$10^{-4}$]       & 6.78(73)    & 7.92(60)    & 7.38(64)     \\
$c_2^+$ [GeV$^{-4}$]    & 1.42(13)    & 1.26(12)    & 1.40(10)     \\
$c_2^0$ [GeV$^{-4}$]    & $-0.41(11)$ & $-0.51(10)$ & $-0.39(11)$ \\
$F_0$ [GeV]             & 0.1014(16)  & 0.1023(14)  & 0.1023(15)   \\
$\chi^2/$dof            & 0.30        & 0.55        & 0.23         \\
$d_{21}^0$              & 0.90(33)    & 1.27(28)    & 1.11(30)     \\
$e_{10}$ [GeV]          & 0.0305(10)  & 0.03011(95) & 0.03030(84)  \\
$e_{11}^+$ [GeV$^{-1}$] & 0.2884(62)  & 0.3103(71)  & 0.2993(82)   \\
$e_{11}^0$ [GeV$^{-1}$] & 0.30(13)    & 0.43(10)    & 0.36(11)     \\
$f_+(0)$                & 0.9615(12)  & 0.9622(11)  & 0.9622(11)   \\
$\lambda_+^\prime$ [$10^{-2}$] & 2.554(24) & 2.622(26) & 2.586(22) \\
$\lambda_0^\prime$ [$10^{-2}$] & 1.649(93) & 1.796(76) & 1.711(67) \\
$\lambda_+^{\prime\prime}$ [$10^{-3}$] & 1.27(10)  & 1.136(95) & 1.251(82) \\
$\lambda_0^{\prime\prime}$ [$10^{-3}$] & 0.544(79) & 0.445(82) & 0.539(82) 
\\\hline\hline
\end{tabular}
\end{table}

\begin{table}[ht!]
\caption{The same table as Table~\ref{tab:phase_space_integral-free_F0},
but for each fit result listed in Table~\ref{tab:fit_chpt-cont-limit-diff_data-free_F0}.}
\label{tab:phase_space_integral-diff_data-free_F0}
\begin{tabular}{cccc}\hline\hline
                           & local        & smeared      & A2       \\\hline
$I^e_{K^0}$                & 0.154630(86) & 0.154888(87) & 0.154784(84) \\
$I^\mu_{K^0}$              & 0.10302(22)  & 0.10345(19)  & 0.10325(17)  \\
$I^e_{K^+}$                & 0.159052(89) & 0.159317(90) & 0.159211(86) \\
$I^\mu_{K^+}$              & 0.10615(22)  & 0.10659(19)  & 0.10638(17) \\
$f_+(0)\sqrt{I^e_{K^0}}$   & 0.37811(50)  & 0.37869(42)  & 0.37857(46) \\
$f_+(0)\sqrt{I^\mu_{K^0}}$ & 0.30863(55)  & 0.30949(43)  & 0.30919(50) \\
$f_+(0)\sqrt{I^e_{K^+}}$   & 0.38348(50)  & 0.38407(43)  & 0.38395(47) \\
$f_+(0)\sqrt{I^\mu_{K^+}}$ & 0.31328(55)  & 0.31415(44)  & 0.31385(51) 
\\\hline\hline
\end{tabular}
\end{table}

In order to carry out continuum extrapolations with
other fitting forms for a $q^2$ interpolation, {\it e.g.,} a monopole form,
estimate of the form factors at the physical
point is necessary, though effects of the chiral extrapolation are 
considered to be tiny.
We estimate the values of the form factors at the physical point
in each lattice spacing by using the $q^2$ fit results with
the NLO SU(3) ChPT formula 
in each current data and lattice spacing
presented in Sec.~\ref{sec:interpolate-each}.
At each $q^2_{n_p}$, a difference of $f_s(q^2)$ for $s = +,0$ in between
the measure meson masses and the physical point,
\begin{equation}
\Delta f_s(q^2) = \left.f_s(q^2)\right|_{m_{\pi^-},m_{K^0}}
- \left.f_s(q^2)\right|_{m_\pi,m_K} ,
\end{equation}
is evaluated from the fit results, where 
$\left.f_s(q^2)\right|_{m_{\pi^-},m_{K^0}}$ is 
estimated in the same way as explained in Sec.~\ref{sec:interpolate-each}.
Each data is shifted by adding $\Delta f_s(q^2)$ to estimate
the value at the physical point.
The values of the shifted data are tabulated in 
Tables~\ref{tab:kl3_form_factor_lcl-cur-shifted} and
\ref{tab:kl3_form_factor_cns-cur-shifted}.
The original values of the form factors are summarized in 
Tables~\ref{tab:kl3_form_factor_b200} and 
\ref{tab:kl3_form_factor_b200-cns-cur} labeled by ``combined'', 
Table~\ref{tab:kl3_form_factor_b182-cns-cur}, and
Table~II in Ref.~\cite{PACS:2019hxd} labeled by ``A1''.
Comparing to those original values, it is found that the shifts are almost
the same size as or less than the statistical errors.

Using the shifted data, we perform continuum extrapolations with
a monopole form,
\begin{eqnarray}
f_+^{\rm cur}(q^2) &=& \frac{f_+(0)}{1 + c_1^+ q^2} 
+ g^{\rm cur}_+(q^2,a), \label{eq:ato0_mono_f+}\\
f_0^{\rm cur}(q^2) &=& \frac{f_+(0)}{1 + c_1^0 q^2}
+ g^{\rm cur}_0(q^2,a), \label{eq:ato0_mono_f0}
\end{eqnarray}
a quadratic form,
\begin{eqnarray}
f_+^{\rm cur}(q^2) &=& f_+(0) + c_1^+ q^2 + c_2^+ q^4
+ g^{\rm cur}_+(q^2,a), \label{eq:ato0_quad_f+}\\
f_0^{\rm cur}(q^2) &=& f_+(0) + c_1^0 q^2 + c_2^0 q^4
+ g^{\rm cur}_0(q^2,a), \label{eq:ato0_quad_f0}
\end{eqnarray}
and the second order of the $z$-parameter expansion~\cite{Bourrely:2008za},
\begin{eqnarray}
f_+^{\rm cur}(q^2) &=& f_+(0) + c_1^+ z^2(q^2) + c_2^+ z^4(q^2)
+ g^{\rm cur}_+(q^2,a), \label{eq:ato0_zexp_f+}\\
f_0^{\rm cur}(q^2) &=& f_+(0) + c_1^0 z^2(q^2) + c_2^0 z^4(q^2)
+ g^{\rm cur}_0(q^2,a), \label{eq:ato0_zexp_f0}
\end{eqnarray}
where 
\begin{equation}
z(q^2) = \frac{\sqrt{(m_{K^0}+m_{\pi^-})^2+q^2} - (m_{K^0}+m_{\pi^-})}
{\sqrt{(m_{K^0}+m_{\pi^-})^2+q^2} + (m_{K^0}+m_{\pi^-})} .
\end{equation}
Our choice of $z(q^2)$ corresponds to the one with $t_0 = 0$ in the general representation of $z(q^2)$~\cite{Bourrely:2008za}.
In the three fits, 
the functional form of $g_s^{\rm cur}$ is fixed to the ones in fit A
described in Sec.~\ref{sec:continuum_extrapolation}.
The fit results are tabulated in Table.~\ref{tab:fit_mono-poly-zexp-cont-limit}
together with the results for the slopes and curvatures.
The phase space integrals evaluated using the fit results
are presented in Table~\ref{tab:phase_space_integral-mono-poly-zexp}.

\begin{table}[ht!]
\caption{Form factors $f_+(q^2)$ and $f_0(q^2)$
at the physical point in each $q^2$ at $\beta = 2.00$ and 1.82 
using the local vector current for the combined and A1 analyses, respectively.}
\label{tab:kl3_form_factor_lcl-cur-shifted}
\begin{tabular}{cccccc}\hline\hline
& \multicolumn{2}{c}{$\beta = 2.00$} && \multicolumn{2}{c}{$\beta = 1.82$} \\
$q^2$ & $f_+(q^2)$   & $f_0(q^2)$ && $f_+(q^2)$ & $f_0(q^2)$ \\\hline
$q^2_0$ & $\cdots$   & 1.0741(17) && $\cdots$   & 1.0602(18) \\
$q^2_1$ & 1.0871(19) & 1.0312(16) && 1.0876(22) & 1.0260(17) \\
$q^2_2$ & 1.0310(16) & 1.0007(14) && 1.0377(21) & 1.0007(17) \\
$q^2_3$ & 0.9859(15) & 0.9752(14) && 0.9984(20) & 0.9807(18) \\
$q^2_4$ & 0.9499(13) & 0.9550(13) && 0.9640(17) & 0.9626(17) \\
$q^2_5$ & 0.9193(13) & 0.9375(13) && 0.9340(18) & 0.9465(18) \\
$q^2_6$ & 0.8928(13) & 0.9220(16) && 0.9089(19) & 0.9333(21) 
\\\hline\hline
\end{tabular}
\end{table}

\begin{table}[ht!]
\caption{The same table as Table~\ref{tab:kl3_form_factor_lcl-cur-shifted},
but for the conserved current data.}
\label{tab:kl3_form_factor_cns-cur-shifted}
\begin{tabular}{cccccc}\hline\hline
& \multicolumn{2}{c}{$\beta = 2.00$} && \multicolumn{2}{c}{$\beta = 1.82$} \\
$q^2$ & $f_+(q^2)$   & $f_0(q^2)$ && $f_+(q^2)$ & $f_0(q^2)$ \\\hline
$q^2_0$ & $\cdots$   & 1.0863(17) && $\cdots$   & 1.0778(18) \\
$q^2_1$ & 1.0879(19) & 1.0424(16) && 1.0889(21) & 1.0423(17) \\
$q^2_2$ & 1.0359(16) & 1.0112(14) && 1.0440(20) & 1.0160(17) \\
$q^2_3$ & 0.9939(15) & 0.9852(14) && 1.0086(20) & 0.9953(18) \\
$q^2_4$ & 0.9605(13) & 0.9646(13) && 0.9774(17) & 0.9763(17) \\
$q^2_5$ & 0.9321(15) & 0.9468(16) && 0.9503(18) & 0.9598(18) \\
$q^2_6$ & 0.9076(14) & 0.9312(16) && 0.9276(19) & 0.9461(21) 
\\\hline\hline
\end{tabular}
\end{table}

\begin{table}[ht!]
\caption{Fit results of the continuum extrapolation of $K_{\ell 3}$ form factors
at the physical point using monopole, quadratic, and z-parameter expansion
fit forms, defined in Eqs.~(\ref{eq:ato0_mono_f+})--(\ref{eq:ato0_zexp_f0}), 
together with the value of the uncorrelated $\chi^2/$dof.
}
\label{tab:fit_mono-poly-zexp-cont-limit}
\begin{tabular}{cccc}\hline\hline
                     & monopole    & quadratic    & z-parameter  \\\hline
$c_1^+$ [GeV$^{-2}$]    & 1.3108(69)  & $-$1.275(10) & $-$2.058(17) \\
$c_1^0$ [GeV$^{-2}$]    & 0.868(30)   & $-$0.835(32) & $-$1.335(49) \\
$c_2^+$ [GeV$^{-4}$]    & $\cdots$    & 1.615(96)    & $-$0.24(23)  \\
$c_2^0$ [GeV$^{-4}$]    & $\cdots$    & 0.809(89)    & $-$0.69(16)  \\
$\chi^2/$dof            & 0.32        & 0.28         & 0.31         \\
$d_{21}^0$              & 1.01(26)    & 1.01(27)     & 1.39(37)     \\
$e_{10}$ [GeV]          & 0.03032(81) & 0.03033(81)  & 0.0300(88)   \\
$e_{11}^+$ [GeV$^{-1}$] & 0.2919(70)  & 0.2965(68)   & 0.2949(66)   \\
$e_{11}^0$ [GeV$^{-1}$] & 0.323(97)   & 0.32(10)     & 0.311(97)    \\
$f_+(0)$                & 0.9614(10)  & 0.9614(10)   & 0.9616(10)   \\
$\lambda_+^\prime$ [$10^{-2}$] & 2.554(14) & 2.583(20) & 2.567(21) \\
$\lambda_0^\prime$ [$10^{-2}$] & 1.691(59) & 1.692(65) & 1.665(60) \\
$\lambda_+^{\prime\prime}$ [$10^{-3}$] & 1.304(14) & 1.275(76) & 1.158(61) \\
$\lambda_0^{\prime\prime}$ [$10^{-3}$] & 0.572(40) & 0.638(70) & 0.591(55) 
\\\hline\hline
\end{tabular}
\end{table}

\begin{table}[ht!]
\caption{The same table as Table~\ref{tab:phase_space_integral-free_F0},
but for each fit result listed in Table~\ref{tab:fit_mono-poly-zexp-cont-limit}.}
\label{tab:phase_space_integral-mono-poly-zexp}
\begin{tabular}{cccc}\hline\hline
                           & monopole     & quadratic    & z-parameter  \\\hline
$I^e_{K^0}$                & 0.154744(83) & 0.154781(75) & 0.154710(77) \\
$I^\mu_{K^0}$              & 0.10320(15)  & 0.10322(16)  & 0.10313(16)  \\
$I^e_{K^+}$                & 0.159118(85) & 0.159155(77) & 0.159519(80) \\
$I^\mu_{K^+}$              & 0.10621(16)  & 0.10623(17)  & 0.10653(17) \\
$f_+(0)\sqrt{I^e_{K^0}}$   & 0.37818(42)  & 0.37824(42)  & 0.37825(43) \\
$f_+(0)\sqrt{I^\mu_{K^0}}$ & 0.30885(44)  & 0.30889(45)  & 0.30883(45) \\
$f_+(0)\sqrt{I^e_{K^+}}$   & 0.38349(42)  & 0.38355(43)  & 0.38408(43) \\
$f_+(0)\sqrt{I^\mu_{K^+}}$ & 0.31331(45)  & 0.31335(46)  & 0.31387(46) 
\\\hline\hline
\end{tabular}
\end{table}

\clearpage

\bibliographystyle{apsrev4-1}
\bibliography{reference}

\end{document}